\documentclass[preprint]{elsarticle}
\usepackage{float,graphicx,amsmath,changepage,amsthm,amssymb,subcaption,caption,bbold,lineno,hyperref}
\usepackage{multirow}
\usepackage{cleveref,natbib}
\usepackage{autonum}
\usepackage{xcolor}
\modulolinenumbers[1]
\newtheorem{example}{Example}
\newcommand{\bx}{\mathbf{x}}
\newcommand{\X}{\Psi}
\newcommand{\be}{\mathbf{e}}
\newcommand{\by}{\mathbf{y}}

\newcommand{\bp}{\mathbf{p}}
\newcommand{\bc}{\mathbf{c}}
\newcommand{\bs}{\mathbf{s}}

\newcommand{\bu}{\mathbf{u}}

\newcommand{\bv}{\mathbf{v}}
\newcommand{\dd}{\mathrm{d}}
\newcommand{\bF}{\mathbf{F}}
\newcommand{\bbf}{\mathbf{f}}
\newcommand{\bff}{\mathbf{f}}

\newcommand{\bom}{\boldsymbol{\omega}}

\newcommand{\trans}{^{\mathrm{T}}}

\newcommand{\bdelta}{\boldsymbol{\delta}}
\usepackage[margin=1in]{geometry}

\journal{}
\begin{document}
\begin{frontmatter}
	\title{\textbf{Computational geometric methods for preferential clustering of particle suspensions}.}

\author[add1]{Benjamin K Tapley\corref{cor1}}
\ead{Benjamin.Tapley@ntnu.no}
\author[add2]{Helge I Andersson}
\ead{Helge.I.Andersson@ntnu.no}
\author[add1]{Elena Celledoni}
\ead{Elena.Celledoni@ntnu.no}
\author[add1]{Brynjulf Owren}
\ead{Brynjulf.Owren@ntnu.no}

\cortext[cor1]{Corresponding author}

\address[add1]{Department of Mathematical Sciences, The Norwegian University of Science and Technology, 7491 Trondheim, Norway}

\address[add2]{Department of Energy and Process Engineering, The Norwegian University of Science and Technology, 7491 Trondheim, Norway}

\begin{abstract}
	A geometric numerical method for simulating suspensions of spherical and non-spherical particles with Stokes drag is proposed. The method combines divergence-free matrix-valued radial basis function interpolation of the fluid velocity field with a splitting method integrator that preserves the sum of the Lyapunov spectrum while mimicking the centrifuge effect of the exact solution. We discuss how breaking the divergence-free condition in the interpolation step can erroneously affect how the volume of the particulate phase evolves under numerical methods. The methods are tested on suspensions of $10^4$ particles evolving in discrete cellular flow field. The results are that the proposed geometric methods generate more accurate and cost-effective particle distributions compared to conventional methods.
\end{abstract}
\end{frontmatter}

\section{Introduction}
Since the influential work of Maxey and Riley \cite{maxey1983equation} in deriving the equations of motion of an inertial spherical particle immersed in viscous flow, there have been a multitude of studies exploring the collective behavior of suspensions of particles. In particular, the remarkable phenomenon of preferential concentration of inertial particles in turbulence has attracted the attention of many authors. This phenomenon, sometimes referred to as the ``centrifuge effect", is also attributed to Maxey \cite{maxey_1987} who showed that particles disperse in regions where the fluid velocity strain rate is low compared to the vorticity. 
The theoretical mechanisms for particle clustering has since been further explored by means of Lyapunov exponent analysis \cite{bec2003fractal}, caustics \cite{wilkinson2005caustics} and perturbative methods to name a few. Sophisticated numerical simulations \cite{squires1991preferential} have also advanced and verified our understanding of this phenomenon for a variety of flows and extended such observations to non-spherical particles \cite{mortensen2008dynamics}. As the need for large-scale simulations increase, the demand for cost-effective numerical methods is growing. However, despite the fact that numerical simulations are so well documented, there have been few studies that explore the extent to which the numerical methods used in simulations accurately reproduce the geometric properties that explain the preferential clustering of particles. In this paper we discuss some features of the equations of motion that influence the preferential concentration of particles and determine to what extent these features can be replicated by well designed numerical methods. In doing so, we propose an efficient numerical algorithm that is designed to replicate these features. The method combines matrix-valued radial basis functions for the divergence-free interpolation of the discrete fluid field with a splitting method that is designed specifically for the equations of motion under study. 

Interpolation methods are necessary for simulating suspensions of particles as the flow field is usually generated by a direct numerical simulation of the Navier-Stokes equations and is therefore only available at discrete points in space, meaning that it must be approximated at the location of the particle. To achieve this in a simple and efficient manner many authors use a variant of a tri-polynomial interpolant, for example \cite{squires1991preferential,Bernard, Portela, rouson2001preferential, challabotla2015orientation, uijttewaal1996particle, van2000dynamics, PanBanerjee, wang1996large, Deardorff}. Previous studies \cite{yeung_pope_1989, BALACHANDA} have explored the extent to which these interpolation methods accurately reproduce statistical properties of the turbulent flow field. However, all the interpolation methods considered in the aforementioned references are based on polynomials that create an approximation to the fluid velocity field that is not divergence-free. One major consequence is that the hydrodynamic Stokes force that determines the particle path lines is instead calculated from a non-conservative fluid velocity field. These non-conservative interpolation methods are still used in practice today despite the fact that the theoretical mechanisms that explain the preferential concentration are derived with the assumption of incompresibility. Furthermore, it is often argued (e.g., \cite{uijttewaal1996particle}) that interpolation errors are ``averaged out" and it is concluded that one can acheive statistically similar results using a fast low-order interpolation method. This claim is supported by the fact that linear interpolation produces similar statistics to simulations using cubic interpolation \cite{yeung_pope_1989}. However, neither linear nor cubic interpolation preserves the divergence-free condition of the fluid field and therefore it is not truly understood whether or not errors to the divergence of the fluid field are averaged out in the same way that standard truncation errors are. The implications of these divergence-errors have not been studied in detail, however there is numerical evidence suggesting that breaking this condition can lead to erroneous clustering in PDF methods, first presented in \cite{meyer2004conservative} and also in \cite{gobert2006lagrangian}. Divergence free interpolation has been used to good effect in particle-laden flow simulations \cite{Esmaily-Moghadam2016analysis,gobert2010numerical} as well as in other particle simulation problems, such as in geodynamic modelling \cite{wang2015advantages} and magnetospheric physics \cite{mackay2006divergence}, for example. One of the goals of this study is to explain, from a numerical analysis point of view, the consequences of breaking the divergence-free condition in the flow field. We also show the benefit of divergence-free interpolation using in simulations of suspensions of inertial particles as well as show how these errors affect the numerical time integration. 

In addition, we study some numerical integration methods and how their errors affect the preferential concentration of particles. Two popular classes of methods are explicit Runge-Kutta \cite{Bernard,rouson2001preferential} and Adams-Bashforth methods \cite{Portela, challabotla2015orientation, van2000dynamics, wang1996large, PanBanerjee}. As the accuracy of the time stepping algorithm is limited by the interpolation error we consider only explicit order one and two methods. Such methods often do a reasonably good job at integrating the ODEs under study as they are efficient and easy to implement. However, the exact solution to the ODEs that govern the dynamics of particles with Stokes drag possess a number of physical features that can be exploited to increase the accuracy of the time stepping methods without increasing its order or cost. Such features include constant contractivity of phase space volume, the centrifuge effect, rigid body motion, linear dissipation and, in some cases, perturbative forces. These features are able to be exploited by a carefully designed splitting method. In this work we propose, as an alternative to Runge-Kutta and Adams-Bashforth methods, a splitting method that is especially designed to reduce the error in the centrifuge effect, which combined with divergence free interpolation techniques allow us to obtain a higher lever of accuracy in the distribution of particles in viscous flows.

\subsection{Main contributions and summary of paper}
We now highlight the main contributions and give a brief outline of the paper. We begin by outlining the equations of motion and the centrifuge effect in section \ref{sec:EoM}. In section \ref{sec:integration} we develop and analyze a contractive splitting method whose flow preserves the sum of the Lyapunov spectrum of the exact solution and show that conventional methods cannot do this. The splitting method is then applied to the equations of motion for spherical particles and the so-called ``centrifuge-preserving" methods are presented, which are constructed to minimize the error of the centrifuge effect. 

Section \ref{sec:interp} presents the use and implementation of matrix-valued radial basis function interpolation to construct a divergence-free interpolation of the discrete flow field. We show that a vector field approximated by matrix-valued radial basis functions are compatible with the Stokes equations due to the fact they they are identical to the method of regularized Stokeslets. This results in a more physically realistic approximation to the underlying Navier-Stokes equations.

In section \ref{sec:spherical} we focus our attention to how physical volume of the particle phase $\X$ evolves over a small time $h$. Upon expanding $\X$ in $h$ under the exact solution, we recover the centrifuge effect at $O(h^4)$. When expanding $\X$ under the numerical solution, we find that errors to the divergence of the fluid velocity field appear at $O(h^2)$, overshadowing the centrifuge effect. However, when a divergence-free interpolation method is used, all the numerical methods under consideration replicate the \textit{qualitative} behavior of the centrifuge effect. That is, physical volumes of particles will contract in regions where the vorticity is lower than the strain rate and vice versa, however, they do so at a slightly erroneous rate. To account for this error, we show that the centrifuge-preserving methods contract physical volume at the same rate as the exact solution to leading order in $h$, hence also preserving the \textit{quantitative} behavior of the centrifuge-effect. 

Section \ref{sec:numerical tests} is dedicated to simulations of particle suspensions evolving in a discrete cellular flow field where we compare the proposed geometric methods against conventional methods. What we observe is that a computationally inexpensive combination of divergence-free interpolation and centrifuge-preserving splitting methods yield far more accurate spatial distributions of particles compared to standard methods of higher cost. We present many examples where our geometric algorithm produced distributions of particles that are more similar to the ``exact" distribution despite having higher error per particle than distributions produced by slow conventional methods. The main conclusion here is that numerical solutions that preserve the sum of the Lyapunov spectrum, the contractivity of phase space volume, the divergence-free condition and the centrifuge effect in simulations is of great benefit. 

Section \ref{sec:conclusions} is dedicated to conclusions.

\section{The equations of motion}\label{sec:EoM}
The translational dynamics of a small particle immersed in a viscous fluid is governed by the rigid body equations with a Stokes force term
\begin{align}
\dot{\bv} =& \alpha K (\bu(\bx)-\bv)\label{sphericalv}\\
\dot{\bx} =& \bv\label{sphericalx}
\end{align}
where $\bu(\bx)$ is the fluid velocity at the particle's location $\bx$, $\bv$ the velocity, $K$ is a positive definite resistance tensor and $\alpha=1/St$ is the inverse particle Stokes number, which is a dimensionless measure of particle inertia. Note that unless mentioned we will assume that $\bu(\bx)$ does not explicitly depend on $t$. Doing so improves the readability and presentation of the paper and does not affect the forthcoming results. 

For spherical particles, $K = I$ is the identity, the rotational variables are constant and the above ODEs uniquely specify the dynamics of each particle. For non-spherical particles, the resistance tensor $K = Q^TK_bQ$, where $K_b$ is the diagonal positive definite body frame resistance tensor and $Q\in SO(3)$ is a rotation matrix that transforms a vector in the body frame to one in the inertial frame. The angular velocity $\bom$ evolves via
\begin{equation}\label{eq:rotation}
J\dot{\bom}=J\bom\times\bom-\mathbf{T},
\end{equation}
where $J$ is the diagonal body frame moment of inertia tensor and $\mathbf{T}$ is the hydrodynamic torque. The rotation matrix $Q$ is calculated by solving the matrix ODE
\begin{equation}\label{eq:Qdot}
\dot{Q} = Q\widehat{\bom}, 
\end{equation}
where $\widehat{\cdot}:\mathbb{R}^3 \rightarrow \mathfrak{so}(3)$ is defined by
\begin{equation}
\left(
\begin{array}{c}
\omega_1\\
\omega_2\\
\omega_3\\
\end{array}
\right)
\mapsto 
\widehat{\bom} = \left( 
\begin{array}{ccc}
0 & -\omega_1 &  \omega_2 \\
\omega_1 & 0  & -\omega_3 \\
-\omega_2  & \omega_3 & 0   \\
\end{array} 
\right),
\end{equation} 
such that $\widehat{\bom}\bv = \bom\times\bv$. The expressions for $K_b$ and $\mathbf{T}$ for spheroidal particles are given in \ref{model}. 
\subsection{The centrifuge effect}
Here we will outline the centrifuge effect of the particle equations of motion, which is one of the mechanisms for particle clustering that is referred to throughout the paper. In \cite{maxey_1987}, Maxey assumes $\alpha\gg 1$ and expands the the spherical particle ODEs \eqref{sphericalv} and \eqref{sphericalx} in powers of $\alpha^{-1}$ to derive a first-order ODE expression for $\bx$
\begin{equation}\label{xdot}
\dot\bx = \bu(\bx) - \alpha^{-1}\left(\frac{\partial \bu}{\partial t} + \bu\cdot\nabla\bu\right) + O(\alpha^{-2})
\end{equation}
where we have ignored the effect of gravity. Taking the divergence gives
\begin{equation}\label{divV}
\nabla\cdot\bv = \frac{\partial u_i}{\partial x_i} - \frac{1}{\alpha}\left(\frac{\partial}{\partial t}  \frac{\partial u_i}{\partial x_i} + \frac{\partial u_i}{\partial x_j}\frac{\partial u_j}{\partial x_i} + u_i\frac{\partial }{\partial x_i}\frac{\partial u_j}{\partial x_j}\right)+ O(\alpha^{-2})
\end{equation}
where there is an implied summation over repeated indices, which is the convention that is assumed throughout the paper. Assuming that the fluid field is divergence-free, we arrive at the familiar relationship between the fluid field rate of strain, rate of rotation and the divergence of the particle velocity field
\begin{equation}\label{MaxeyCentrifuge}
\nabla\cdot\bv = - \frac{1}{\alpha}\frac{\partial u_i}{\partial x_j}\frac{\partial u_j}{\partial x_i} = - \frac{1}{\alpha} \left( \|S\|^2_F - \|\Omega\|^2_F\right)+ O(\alpha^{-2})
\end{equation}
where the rate of strain and rotation tensors $S$ and $\Omega$ are given by 
\begin{equation}
S_{ij} = \frac{1}{2}\left(\frac{\partial u_i}{\partial x_j} + \frac{\partial u_j}{\partial x_i}\right)\quad\text{and}\quad	\Omega_{ij} = \frac{1}{2}\left(\frac{\partial u_i}{\partial x_j} - \frac{\partial u_j}{\partial x_i}\right)
\end{equation}
and $\|\cdot\|_F$ is the Frobenius matrix norm. In other words, the divergence of the particle velocity field $\nabla\cdot\bv$ is positive when the vorticity is large compared to the strain rate tensor meaning that the particulate phase disperses in these regions. Conversely, particles concentrate in regions where the strain rate is large compared to the vorticity. This phenomenon is the ``centrifuge effect" of the exact solution to \eqref{sphericalv} and \eqref{sphericalx}. 

Finally, we remark that while the centrifuge effect was derived for spherical particles, one can make similar observations for non-spherical particles. In this scenario, the resistance tensor can be decomposed into a spherical part and a non-spherical part, e.g., for a spheroidal particle with rotational symmetry (see \ref{model}) we can write $K_b = a\,I + b\,\be_z\be_z^T$, where $\be_z = (0,0,1)^T$ and $b\rightarrow a$ in the spherical limit. In other words, the centrifuge effect still plays a central role in the preferential clustering of non-spherical particles in addition to the non-spherical effects due to $b\be_z\be_z^T$ term in the resistance tensor. 

 \section{Numerical integration of dissipative vector fields}\label{sec:integration}
 The dynamics of small inertial particles (both spherical and non-spherical) can be modeled as the flow of a vector field with linear dissipation. Such vector fields arise due to the fact that for low Reynolds number flow the drag forces are linear in the slip velocity, for example the Stokes drag force for small ellipsoids, spheres or rigid slender particles \cite{andersson2020integral}. We begin this section with a discussion of such linearly dissipative vector fields and their contractive properties of phase space volume. We then discuss the application of some conventional explicit  methods for integrating such ODEs. In particular, we show that conventional methods cannot preserve the contractivity of phase space volume. A splitting scheme is then shown to preserve the exact contractivity of phase space volume. The section concludes with the application of the splitting scheme to spherical particles. 

 \subsection{Linearly dissipative vector fields and contractivity of phase space volume}
 A linearly dissipative vector field in $n$ dimensions is given in general by
 \begin{equation}\label{dissipativeODE}
 \dot{\by} = \bff(\by)-A\by
 \end{equation}
 where $A$ is a positive definite matrix and $\bff(\by)$ is volume preserving, that is, it satisfies $\nabla\cdot\bbf(\by) = 0$ (e.g., any Hamiltonian vector field). Note that the ODEs of both non-spherical and spherical particles can be cast in this form, where $\bff(\by)$ represents the free rigid-body vector field plus the conservative part of the Stokes force and $-A\by$ represents the dissipative part of the Stokes force.
 
 The quantitative behavior of particle clustering can be explained in part by analyzing the Lyapunov exponents $\lambda_i$ of the ODE, see for example \cite{bec2003fractal}. It is therefore desirable that the numerical solution of the ODE reproduces similar Lyapunov exponent characteristics. Whilst there do not currently exist numerical methods that preserve individual Lyapunov exponents we can however construct a numerical method that preserves the sum of the Lyapunov spectrum $\sum_{i=1}^{n}\lambda_i$. From a backward error analysis point of view, a numerical method that preserves the Lyapunov spectrum is one that is the exact solution to an ODE with the same Lyapunov spectrum sum as the ODE being solved. For the equations of motion for spherical particles (equations \eqref{sphericalv} and \eqref{sphericalx}), the sum of the first three Lyapunov exponents characterizes the divergence of the velocity field and the sum of the spatial Lyapunov exponents characterizes the rate at which particle clouds contract or expand \cite{Esmaily-Moghadam2016analysis}. Generally speaking, the sum of the Lyapunov spectrum describes the rate at which phase space volume exponentially contracts or expands \cite{hoover1988negative}. That is, by letting $\by(t)$ denote the exact solution of \eqref{dissipativeODE} with initial conditions $\by(0) = \by_0$ and 
\begin{equation}\label{Y1}
 Y = \det\left(	\frac{\partial \by(t)}{\partial \by_0}\right)
\end{equation}
the $n$ dimensional phase space volume, then 
\begin{equation}\label{lyapsum}
 Y = \prod_{i=1}^{n} e^{t\lambda_i}.
\end{equation}
It is also known that linearly dissipative systems contract phase space volume at a constant rate, as we will now show. Taking the Jacobian of $\by(t)$ with respect to $\by_0$ gives
 \begin{equation}
 \frac{\dd}{\dd t}\frac{\partial \by(t)}{\partial \by_0} = (\bff'-A)\frac{\partial \by(t)}{\partial \by_0}.
 \end{equation}
 We now recall Jacobi's formula, which relates the derivative of the determinant of a square matrix $M(t)$ by the following
 \begin{equation}\label{JacobiFormula}
 \frac{\dd}{\dd t}\det(M(t)) = \det(M(t))\, \mathrm{Tr}\left(M(t)^{-1}	\frac{\dd}{\dd t}M(t)\right).
 \end{equation} 
 Differentiating \eqref{Y1} with respect to time and applying \eqref{JacobiFormula} gives 
 \begin{equation}\label{detAODE}
 \frac{\dd}{\dd t}Y = -Y\, \mathrm{Tr}\left(A\right),
 \end{equation}
 as $\bff'$ has zero trace. This is solved by
 \begin{equation}\label{detA}
 Y=e^{-t\,\mathrm{Tr}(A)}.
 \end{equation}
 By equating this with \eqref{lyapsum} we obtain the relation
 \begin{equation}\label{L-P}
 \sum_{i=1}^{n}\lambda_i = - \mathrm{Tr}(A).
 \end{equation}
 As the trace of $A$ is by definition positive, the phase space volume $Y$ is strictly monotonically contracting in time. Equation \eqref{L-P} implies that a numerical integration method that preserves phase space volume $Y$ also preserves the sum of the Lyapunov exponents of the underlying ODE. It is therefore logical to imply that a numerical flow of \eqref{dissipativeODE} that preserves the contractivity of phase space volume will better reproduce the clustering properties of the exact solution than one that doesn't. The rest of this section is dedicated to analysing to what extent some common numerical integration methods for particle dynamics can preserve this constant contractivity of phase space volume. 
 
 \subsection{Preservation of the contractivity of phase space volume by numerical methods}
Denote by $\Phi_h$ a numerical method for solving \eqref{dissipativeODE} such that ${\Phi_h(\by_0)\approx \by(h)}$ for time step $h\ll1$. For $\Phi_h$ to be called \textit{contractivity preserving} when applied to \eqref{dissipativeODE}, we require that $\det\left(\frac{\partial\Phi_h(y)}{\partial y}\right) = e^{-h\,\mathrm{Tr}(A)}$ \cite{grimm2008geometric}. It is known that no standard methods (e.g., one with a B-series \cite{GNI}) can preserve phase space volume for divergence free vector fields \cite{Iserles2007}. The same is also expected when it comes to preserving contractivity of phase space volume for dissipative vector fields \cite{mclachlan2000numerical}. Instead, a weaker requirement is that they contract phase space volume when the exact solution does so, that is, $\det\left(\frac{\partial\Phi_h(y)}{\partial y}\right) < 1$ when applied to \eqref{dissipativeODE}. Such a numerical method that possesses this property is called \textit{contractive}.   
 
 We now consider some popular numerical methods for particle dynamics. We consider explicit methods used in the literature, namely, explicit Runge-Kutta methods, Adams-Bashforth methods and a splitting method scheme. 
 
 \subsubsection{Runge-Kutta methods and phase volume contractivity}
 Take an order-$p$ Runge-Kutta method $\Phi^{[RK]}_h(\by_0)$ with stability function $R(z)$ applied to an ODE of the form \eqref{dissipativeODE} with linear $\bff(\by)$, say
 \begin{equation}\label{lindiss}
 \dot{\by} = B\by - A\by,
 \end{equation} 
 where $B$ is a square and traceless matrix. Then the numerical solution by the Runge-Kutta method is given by 
 \begin{equation}
 \Phi^{[RK]}_h(\by_0) = R(h(B-A))\by_0.
 \end{equation}
 Recall that $R(z)$ is an order-$p$ Pad{\'e} approximation to the exponential function. We therefore have
 \begin{equation}\label{rkexp}
 \Phi^{[RK]}_h(\by_0) = \exp\left(h(B-A)\right)\by_0 + O(h^{p+1})
 \end{equation}
 which means that over one time-step, the phase space volume contracts via
 \begin{equation}\label{RKphasevol}
 \det\left(\frac{\partial \Phi^{[RK]}_h(\by_0)}{\partial \by_0}\right) = e^{-h\mathrm{Tr}(A)} + O(h^{p+1}). 
 \end{equation}
 That is, for such a linear system, a Runge-Kutta method will only preserve the phase volume contractivity up to the order of the method. So for a non-linear dissipative ODE of the form \eqref{dissipativeODE}, one can hardly expect a Runge-Kutta method to preserve phase space volume exactly. In fact, due to this error explicit Runge-Kutta methods usually have a time-step restriction on $h$ to even be contractive at all \cite{grimm2008geometric}. This is illustrated by the following examples of some low order explicit Runge-Kutta methods applied to equations \eqref{sphericalv} and \eqref{sphericalx}.  
 
 \begin{example}\label{FEcontraction}
 	We apply the forward Euler method $\Phi^{[F\!E]}_h$ to the ODEs \eqref{sphericalv} and \eqref{sphericalx}. Note that these ODEs are non-linear due to $\bu(\bx)$. Setting $\by:=(\bv,\bx)$, we have for the contractivity of phase space volume under the forward Euler method
 	\begin{equation}\label{jdetFE}
 	\det\left(\frac{\partial\Phi^{[F\!E]}_h(\by_0)}{\partial \by_0}\right) = 1-3\alpha h + \alpha\,h^2\,\left(3\alpha-\frac{\partial u_i}{\partial x_i}\right) + O(h^3),
 	\end{equation}
 	which is an order one approximation to the exact contractivity 
 	\begin{equation}\label{exactContractivity}
 	\det\left(\frac{\partial \by(h)}{\partial \by_0}\right) = e^{-3\alpha h}.
 	\end{equation} 
 	The Forward Euler method must therefore satisfy the following time-step restriction for it to be contractive
 	\begin{equation}
 	h\lessapprox\frac{3}{3\alpha-\frac{\partial u_i}{\partial x_i}}.
 	\end{equation}
 	Violating this restriction means that the forward Euler method will expand phase space volume despite the ODE dictating that it is always contracting. Furthermore, it can be seen that large values of $|\frac{\partial u_i}{\partial x_i}|$ will place further restrictions on the size of $h$.
 \end{example}
 
 \begin{example}
 	Consider the following second order explicit Runge-Kutta method 
 	\begin{equation}\label{RK2}
 	\Phi^{[RK]}_h(\by_0) = \by_0 + h\bigl( (1-\tfrac1{2\theta}) \bff(\by_0) + \tfrac1{2\theta} \bbf(\by_0 + \theta h \bbf(\by_0)),
 	\end{equation}
	where $\theta=\frac{1}{2},\frac{2}{3}$ and $1$ correspond to the explicit midpoint method, Ralston's method and Heun's method, respectively. We apply this to the ODEs \eqref{sphericalv} and \eqref{sphericalx}. Setting $\by:=(\bv,\bx)$, we have for the contractivity of phase space volume under $\Phi^{[RK]}_h$
 	\begin{align}\label{jdetRK}
 	\det\left(\frac{\partial\Phi^{[{RK}]}_h(\by_0)}{\partial \by_0}\right) &= 1\,-\,3\alpha h\,+\, 9\alpha^2\frac{h^2}{2!}\\
 	&\quad \,+\,\left( 3\alpha(\theta - 1)\frac{\partial^2 u_i}{\partial x_i\partial x_j}v_j\,+\,3\alpha^2\frac{\partial u_i}{\partial x_i}\,-\,24\alpha^3\right)\frac{h^3}{3!}\,+\,O(h^4)
 	\end{align}
 	which is an order two approximation to the exact contractivity \eqref{exactContractivity}, which is expected for an order two method. The time-step $h$ must be chosen small enough such that the $O(h^3)$ error term does not violate the contractivity condition. Violating this restriction means that the method will expand phase space volume despite the ODE dictating that it is always contracting. Furthermore, it can be seen that large values of $|\frac{\partial u_i}{\partial x_i}|$ and $v_i$ will place further restrictions on the size of $h$.
\end{example}
We remark that we can make similar observations for the above methods applied to the ODEs for non-spherical particles. That is, the phase space volume is conserved only to the order of the method. 

 \subsubsection{Multi-step methods and phase volume contractivity}\label{sec:multistep_contractivity}
 Another popular numerical method used for particle dynamics are multi-step methods. Consider the explicit $k$-step Adams-Bashforth methods. Such methods are of global order-$k$ and can be seen as a map $\Phi_h^{[AB]}:(\by_0,...,\by_{k-1})\rightarrow(\by_1,...,\by_{k})$ such that 
 \begin{align}\label{ABk}
 	\by_k &= \by_{k-1} + h\sum_{i=0}^{k-1}b_i\bbf(\by_i),\\
 	\by_i &= \by_{i-1},\quad\text{for}\quad i=1,...,k-1
 \end{align}
 where the coefficients $b_i$ satisfy $\sum_{i=0}^{k-1}b_i=1$. That is, $\Phi_h^{[AB]}$ takes a point in a $kn$ dimensional phase space to another point in the same space. Due to this and the fact that the initial vectors in the domain $\by_i$ for $i=0,...,k-1$ are independent of one another, it's less clear how to define the notion of numerical phase space volume that relates to that of the underlying ODE. However, in practice these initial vectors $\by_k$ for $i=0,...,k-1$, are usually computed by an order-$k$ one step method, for example a Runge-Kutta method. Therefore, the vectors $\by_i$ depend on the vectors $\by_j$ for $j<i$. This implies that each $\by_i$ has the same series expansion as the exact solution up to $O(h^{k})$. While a detailed analysis of multi-step methods and the preservation of phase volume lie outside the scope of the paper, we can illustrate this concept by the following example, which considers the phase volume properties of a second order Adams-Bashforth using a second order Runge-Kutta method to compute the initial vectors. 
 
\begin{example}\label{AB2example}
 Consider the second-order Adams-Bashforth method $\Phi_h^{[AB]}:(\by_1,\by_0)\rightarrow(\by_2,\by_1)$ where 
 \begin{align}
\by_2  =& \by_1 + \frac{h}{2}\left(3\bff(\by_1)-\bff(\by_0)\right)
 \end{align}
 Now define $\by_1=\Phi^{[RK]}_h(\by_0)$ in the domain by the second order Runge-Kutta method \eqref{RK2}. Applying $\Phi_h^{[AB]}$ to the ODEs \eqref{sphericalv} and \eqref{sphericalx} and taking the Jacobian determinant of $\by_2$ with respect to $\by_0$ gives 
 \begin{align}
\det\left(\frac{\partial \by_2}{\partial \by_0}\right) &= 1\,-\,6\alpha h\, +\, 36\alpha^2\frac{h^2}{2!}\\
&\quad \,+\,
\left(\alpha\left(3\theta - \frac{21}{2}\right)\frac{\partial^2 u_i}{\partial x_i\partial x_j}v_j\,+\,\frac{21\alpha^2}{2}\frac{\partial u_i}{\partial x_i}\,-\,24\alpha^3\right)\frac{h^3}{3!}\,+\,O(h^4)
 \end{align}
 which is an $O(h^2)$ approximation to the exact contractivity \eqref{exactContractivity}. Note that here we have taken the the contractivity over two time-steps $2h$. Like the previous example, we observe that the contractivity is affected by non-zero values of $\frac{\partial u_i}{\partial x_i}$ and $v_i$.
 \end{example}
 
 
 \subsection{A splitting scheme that preserves the contractivity of phase space volume} \label{sec:contractivitysplitting}
 We now analyze the splitting method based on the following splitting of equation \eqref{dissipativeODE}
 \begin{equation}\label{splitDissipativeODE}
 \dot{\by} = \bff(\by)-\mathbf{b}(\by) \quad\text{and}\quad \dot{\by} = -A\by+\mathbf{b}(\by)
 \end{equation}
 where $\mathbf{b}(\by)$ is any vector that is constant along the flow of the second vector field. A similar splitting was proposed in \cite{tapley2019novel} for non-spherical particle dynamics. Denote their exact flow operators by $\psi^{[1]}_h$ and $\psi^{[2]}_h$, respectively. In the context of small particles immersed in a viscous fluid, the first vector field represents the free rigid body equations and the second is due to the Stokes viscous drag forces. The free rigid body vector field can be solved exactly. That is, by a forward Euler step for the spherical case or otherwise using trigonometric or Jacobi elliptic functions depending  whether or not the body is axially symmetric \cite{celledoni2008exact}. Due to the existence of an exact solution, we immediately have volume preservation
 \begin{equation}
 \left|	\frac{\partial \psi^{[1]}_h (\by_0)}{\partial \by_0}\right| = 1.
 \end{equation}
 The second vector field is solved by the variation of parameters formula
 \begin{equation}
 \psi^{[2]}_h (\by_0) = e^{-hA}(\by_0 + A^{-1}\mathbf{b})+A^{-1}\mathbf{b}.
 \end{equation}
 Taking the Jacobian determinant gives
 \begin{equation}
 \left|	\frac{\partial \psi^{[2]}_h (\by_0)}{\partial \by_0}\right| = e^{-h\mathrm{Tr}(A)},
 \end{equation}
 which is consistent with the exact solution \eqref{detA}. As the Jacobian of the composition of two or more maps is the product of the Jacobians of the maps, any splitting method based on the alternating compositions of the flows $\psi^{[1]}_h$ and $\psi^{[2]}_h$ will be contractivity preserving. 
 
 In forthcoming numerical experiments, we will consider only order one and two methods including the order one Lie-Trotter method
 \begin{equation}\label{LieTrotter}
 \Phi^{[LT]}_h = \psi^{[1]}_h\circ\psi^{[2]}_h,
 \end{equation}
and the order two Strang method
 \begin{equation}\label{strang}
\Phi^{[SS]}_h =  \Phi^{[LT]}_{\frac{h}{2}}\circ  \Phi^{[LT]*}_{\frac{h}{2}}.
 \end{equation}
Here, we denote by $\Phi^{[LT]*}_{h} = \psi^{[2]}_h\circ\psi^{[1]}_h$ the conjugate of the $\Phi^{[LT]}_{h}$. 
  
  \subsection{Application to spherical particle dynamics and the centrifuge-preserving methods}
  To construct a contractivity preserving splitting method for spherical particles, we split the ODEs \eqref{sphericalv} and \eqref{sphericalx} in to the following two vector fields 
  \begin{equation}\label{ode11}
  \left(\begin{array}{c}
  \dot{\bv}\\\dot{\bx}\\
  \end{array}\right)
  =
  \left(\begin{array}{c}
  0\\\bv\\
  \end{array}\right)
  ,\quad\text{and}\quad
  \left(\begin{array}{c}
  \dot{\bv}\\\dot{\bx}\\
  \end{array}\right)
  =
  \left(\begin{array}{c}
  \alpha (\bu(\bx)-\bv)\\0\\
  \end{array}\right).
  \end{equation}
  Their exact flow operators are
  \begin{equation}
  \psi^{[1]}_h\left(\begin{array}{c}
  \bv\\\bx\\
  \end{array}\right) = \left(\begin{array}{c}
  0\\\bx + h\bv\\
  \end{array}\right)  \quad\mathrm{and}\quad \psi^{[2]}_h\left(\begin{array}{c}
  \bv\\\bx\\
  \end{array}\right) 
  = 
  \left(\begin{array}{c}
  e^{-\alpha\,h}(\bv-\bu(\bx)) + \bu(\bx)\\\bx\\
  \end{array}\right).
  \end{equation}
  Indeed, letting $\by_0 = (\bv_0^T,\bx_0^T)^T$ we see that
  \begin{equation}
  \left|	\frac{\partial \psi^{[1]}_h (\by_0)}{\partial \by_0}\right| = 1\quad\mathrm{and}\quad \left|	\frac{\partial \psi^{[1]}_h (\by_0)}{\partial \by_0}\right| = e^{-3\alpha h}
  \end{equation} 
  hence any composition of the above flows will preserve contractivity. For the construction of the splitting method for non-spherical particle dynamics, we refer the reader to \cite{tapley2019novel}.
  
  A draw back of splitting methods is that composing methods of order higher than two requires the use of negative time steps. As the ODEs in question are dissipative, such higher order methods would therefore require strict time step restrictions, which is preferably avoided. The idea behind geometric numerical integrators is that preserving relevant properties of the exact solution upon discretisation can lead to better qualitative and long time numerical solutions. With this in mind, instead of improving the accuracy of the method in a conventional sense by increasing the order, we propose as an alternative the following composition methods 
 \begin{align}
 \Phi^{[C\!P_1]}_h =&  \Phi^{[LT]}_{(1-\frac{\sqrt{6}}{6})h}\circ  \Phi^{[LT]*}_{\frac{\sqrt{6}}{6}h} \label{CP1}
\\
  \Phi^{[C\!P_2]}_h =&  \Phi^{[LT]}_{\frac{3h}{12}}\circ  \Phi^{[LT]*}_{\frac{5h}{12}}\circ \Phi^{[LT]}_{\frac{4h}{12}} \label{CP2}
 \end{align}
 which are order one and order two methods, respectively. We propose that the above splitting methods are particularly well suited to calculation of particle dynamics as their numerical solution preserves the centrifuge effect of the exact solution when considering the contraction of physical volume of the particle field. We will therefore refer to the methods \eqref{CP1} and \eqref{CP2} as the ``centrifuge-preserving" methods. This favorable property is discussed in more detail in section \ref{sec:num int sph}. In section \ref{sec:clusters} we show through numerical simulations that integrators possessing this property predict more accurately the spatial distribution of particles (both spherical and non-spherical) compared to methods without this property.

\section{Divergence-free interpolation with matrix-valued radial basis functions}\label{sec:interp}
To construct a divergence-free approximation to the discrete fluid field, we propose using matrix-valued radial basis functions (MRBFs). In this section, we will give a brief outline on their use and implementation. We then further motivate their use by showing that the interpolated vector field generated by MRBFs is a solution to the Stokes equation. 

The interpolation problem is as follows. Given a set of vector-valued data $\{\bu_{i},\bx_{i}\}_{i=1}^{n^3}$ generated by an accurate direct numerical simulation to the Navier-Stokes equations, construct a divergence-free vector field that locally interpolates the data. In our context, $\bu_{i} = \bu(\bx_{i})$ is the fluid velocity vector at the grid node located at $\bx_{i} = (x_i,y_i,z_i)\trans$. When implementing a polynomial interpolation method, one usually chooses the $ n\times n\times n$ cube of data points neighboring the particle, where $n=2,3$ or $4$. This is because polynomial interpolation of degree $n-1$ requires $n$ data points in each dimension to specify a unique interpolating polynomial. MRBFs are not restricted by this particular choice of data points, however to keep the interpolation methods comparable we will adopt this convention. The MRBF interpolating vector field $\bs(\bx)$ is then constructed by
\begin{equation}\label{MRBFsurf}
\bs(\bx) = \sum_{i=1}^{n^3} \Theta_i(\bx) \bc_i,
\end{equation}
where
\begin{equation}
\Theta_i(\bx) = (\nabla\nabla^T - \nabla^2 I)\theta(r_i(\bx))\in\mathbb{R}^{3 \times 3}
\end{equation}
is called an MRBF, $\theta(r_i(\bx))$ is a (scalar-valued) radial basis function, $r_i(\bx) = \|\bx_i-\bx\|$ is the distance from the point $\bx_i$ and $I$ is the identity matrix in three dimensions. The $n^3$ vector-valued coefficients $\bc_i\in\mathbb{R}^{3}$ are chosen such that $\bs(\bx_i) = \bu(\bx_i)$, which amounts to solving the following $3 n^3$ dimensional linear system
\begin{equation}
\left(\begin{array}{ccc}
\Theta_1(\bx_1)&\cdots&\Theta_n(\bx_1)\\
\vdots &\ddots & \vdots \\
\Theta_1(\bx_n)&\cdots&\Theta_n(\bx_n)\\
\end{array}\right)
\left(\begin{array}{c}
\bc_1\\ \vdots \\ \bc_n\\
\end{array}\right) 
=
\left(\begin{array}{c}
\bu_1\\ \vdots \\ \bu_n\\
\end{array}\right)\in\mathbb{R}^{3n^3}.
\end{equation}
The particular RBF we use in the forthcoming experiments is the Gaussian $\theta(r) = \exp(-\epsilon^2r^2)$, where $\epsilon$ is some user defined parameter that controls the flatness of the RBF. In general, one should choose $\epsilon$ as small as possible as this leads to less interpolation error, although more ill conditioned systems.

It can be easily seen that $\bs(\bx)$ is divergence free. Using the double curl identity in $\mathbb{R}^3$ we have
\begin{equation}
\nabla\cdot\bs = \sum_{i=1}^{3n}\nabla\cdot\left((\nabla\nabla^T - \nabla^2 I)\theta(r_i)\bc_i\right)=\sum_{i=1}^{3n}\nabla\cdot\left(\nabla\times\nabla\times\left(\theta(r_i)\bc_i\right)\right)=0.
\end{equation} 
Finally, we list some advantages of MRBF interpolation over standard tri-polynomial interpolation: (1) they work equally well on scattered data points, meaning that they are just as well suited to interpolate data generated by a direct numerical simulation involving complex geometries on unstructured grids; (2)  they have faster convergence of their derivatives \cite{buhmann2003radial,wendland2004scattered}, compared to tri-polynomial interpolation \cite{carlson1973error}; and (3), they are compatible with the Stokes equations, meaning that they construct a more physically realistic fluid field for fluid simulations. We will discuss point (3) in the next section. 

\subsection{MRBFs as regularised Stokeslet solutions to the Stokes equations}\label{sec:interpNS}
In addition to the fact that the underlying flow field should be divergence-free, we are given extra knowledge that can be exploited; namely that the data is a numerical solution to the incompressible Navier-Stokes equations
\begin{equation}\label{NSeq}
\rho\left(\frac{\partial \bu}{\partial t} + (\bu\cdot\nabla)\bu\right) - \mu\nabla^2 \bu ~=~ - \nabla p +\bF \quad \mathrm{and}\quad \nabla \cdot \bu ~=~0.
\end{equation}
We are only interpolating in space and hence approximating steady-state solutions to \eqref{NSeq} and as the grid-spacing $\Delta x$ is comparable to the smallest length scales of the flow (e.g., the Kolmogorov scale for turbulent flows), the Reynolds number is small and the non-linear terms of equations \eqref{NSeq} can be ignored. Under the above assumptions, a good approximation for the local flow in a grid-cell can be given by the steady Stokes equations, which reads
\begin{equation}\label{StokesEq}
\mu \nabla^2 \bu - \nabla p ~=~ -\bF   \quad\mathrm{and}\quad \nabla \cdot \bu ~=~0,
\end{equation}
where we have set $\mu=1$. Cortez \cite{cortez2001method} presents what's called the regularised Stokeslet solution to the Stokes equation, which is an approximation to Green's function of the Stokes equation for body force $\bF = \phi_\epsilon(\bx)\,\bbf_0$, where $\bff_0\in\mathbb{R}^3$ is constant. Here, $\phi_\epsilon(\bx)$ is the so-called ``blob" function, which is a radially symmetric smooth approximation of the Dirac delta function $\delta(\bx)$ that decays to zero at infinity whilst satisfying
\begin{equation}
\int \phi_\epsilon (\bx)\,\mathrm{d}\bx = 1 \quad\mathrm{and}\quad \lim\limits_{\epsilon\rightarrow0}(\phi_{\epsilon}(\bx))=\delta(\bx).
\end{equation}
Now define the functions $G_\epsilon(\bx)$ and $B_\epsilon(\bx)$ as the solutions to
\begin{equation}
\nabla^2  G_\epsilon(\bx) = \phi_\epsilon(\bx) \quad\mathrm{and}\quad
\nabla^2  B_\epsilon(\bx) = G_\epsilon(\bx),
\end{equation}
which are smooth approximations to Green's function and the biharmonic equation $\nabla^4B(\bx)=\delta(\bx)$, respectively. Then Cortez's regularised Stokeslet solution reads
\begin{equation}\label{rs}
\bu_\epsilon(\bx) = (\bbf_0\cdot\nabla)\nabla B_\epsilon(\bx) - \bbf_0 G_\epsilon(\bx)
\end{equation}
with pressure term
\begin{equation}
p_\epsilon(\bx) = \bbf_0\cdot\nabla G_\epsilon(\bx).
\end{equation}
Using the definition for $G_\epsilon(\bx)$, we can rewrite the regularised Stokeslet \eqref{rs} as
\begin{equation}
\bu_\epsilon(\bx) = \,(\nabla \nabla^T - \nabla^2I)(B_\epsilon(\bx)\bbf_0),
\end{equation}
which is identical to an MRBF element if we can identify $B_\epsilon(\bx)$ with a positive-definite RBF $\psi(||\bx||)$ (e.g., the Gaussian $\psi(||\bx||) = \exp\left(-\epsilon^2||\bx||^2\right)$) and the force vectors are identified with the interpolation coefficient vectors $\bc_i$ from equation \eqref{MRBFsurf}. This means that a vector field that is constructed from a linear combination of MRBFs, (i.e., equation \eqref{MRBFsurf}) corresponds to a linear combination of regularised Stokeslet solutions, with force $\bbf_i=\mu \bc_i$. This leads to the following solution to the Stokes equation, now written in terms of MRBFs
\begin{equation}
\bs(\bx) = \bu_\epsilon(\bx) = \sum_{i=0}^{N}\,\Theta_i(\bx)\bbf_i, ~~\mathrm{and}~~ p_\epsilon(\bx) =\sum_{i=0}^{N} \bbf_i\cdot\nabla(\nabla^2\theta(r_i)).
\end{equation}
One implication of this is that the interpolated background fluid field is related to the gradient of a scalar pressure field, when MRBF interpolation is used. The benefit of this can be illustrated by inserting equation \eqref{NSeq} into equation \eqref{xdot} to derive an expression for $\nabla\cdot\bv(\bx)$ in terms of the pressure field \cite{elperin1996turbulent} 
\begin{equation}
\nabla\cdot\bv(\bx) = \nabla\cdot\bu(\bx) + \alpha^{-1}(\nabla^2 p_\epsilon(\bx))+O(\alpha^{-2}).
\end{equation}
This equation tells us that the pressure field is also related to the preferential concentration of particles. Moreover, this suggests that particles cluster in regions of maximum pressure ($\nabla^2 p_\epsilon(\bx) <0$) \cite{elperin1996turbulent}. Indeed, in \cite{luo2007pressure}, numerical evidence is found to support the correlation between the Laplacian of the pressure field $\nabla^2 p_\epsilon(\bx)$ and the spatial distribution of the particles. However, if the background fluid field is interpolated by a standard polynomial method, then there is no background scalar pressure field, which could erroneously influence the particle path lines.

\section{Numerical errors and preferential concentration of spherical particles} \label{sec:spherical}
In what follows we will consider how volumes of inertial spherical particles evolve under the flow of the ODEs \eqref{sphericalv} and \eqref{sphericalx}. The goal of this section is to relate the numerical interpolation and integration errors to the clustering mechanisms of the exact solution. This is done by first expanding the exact solution into its elementary differentials. We then discuss the effects of integration errors and interpolation errors on the evolution of volumes of particles. In what follows, we will initially assume that $\bu(\bx)$ is an arbitrary vector field that is not necessarily divergence-free until explicitly mentioned. 

\subsection{Expanding the exact solution}\label{sec:expansion}
We start this section by defining the notion of volume of the particle suspension. Given an open and bounded set $D_t\subset\mathbb{R}^3$ at time $t$, then its volume at $t=0$ is given by 
\begin{equation}
\mathrm{vol}(D_0) := \int_{D_0}\!\dd \bx_0
\end{equation}
where $\bx(0)=\bx_0$. This can be thought of as the volume occupied by a suspension of inertial particles confined to the region $D_0$. The idea is to consider how this volume expands or contracts in time. Consider now the same volume after evolving under the ODEs \eqref{sphericalv} and \eqref{sphericalx} for time $t$
\begin{equation}
\mathrm{vol}(D_t) = \int_{D_t}\!\!\dd\bx(t)= \int_{D_t}\!\!\det \left(\frac{\partial\bx(t)}{\partial \bx_0}\right) \dd \bx_0.
\end{equation}
Hence, the quantity 
\begin{equation}\label{X}
\X := \det \left(\frac{\partial \bx(t)}{\partial \bx_0}\right)
\end{equation}
determines how volumes of particles contract or expand over time. That is, given a volume of particles, if $\X>1$ the volume is expanding, $\X<1$ the volume is contracting and $\X=1$ the volume is preserved. These three cases correspond to the particulate phase dispersing, concentrating or remaining a constant density, respectively. Note that we will refer to $\X$ as the \textit{physical} volume, to distinguish between phase space volume. 

To illustrate the connection between $\nabla\cdot\bv$ and $\X$, we can take the Jacobian of equation \eqref{sphericalx} with respect $\bx_0$ \cite{ijzermans_meneguz_reeks_2010}
\begin{equation}
\frac{\partial \dot{\bx}}{\partial \bx_0} = \frac{\partial \bx}{\partial \bx_0}\frac{\partial \bv}{\partial \bx}.
\end{equation}
Applying Jacobi's formula \eqref{JacobiFormula} yields a differential equation for $\X$
\begin{equation}\label{dXdt}
\frac{\partial}{\partial t}\X =   \left(\nabla \cdot \bv\right)\X.
\end{equation}
It is clear that if $\nabla \cdot \bv<1$, then $\X$ is decreasing and if $\nabla \cdot \bv>1$ then $\X$ is increasing. 

We will now show this more concretely, by expanding the exact solution into its elementary differentials, which we now recall. Denote by $\by(t)$ the exact solution of an ODE
\begin{equation}
\dot{y}_i(t) = f_i(\by(t)), \quad\text{for}\quad i=1,...,n
\end{equation}
For some small time $0<h\ll1$, $y_i(h)$ has the following elementary differential expansion \cite{GNI}
\begin{align}
y_i(h) =\,& y_i(0) + h f_i\big|_{t=0} + \frac{h^2}{2}\left(\frac{\partial f_i}{\partial y_j}f_j\right)\bigg|_{t=0} \\\quad
& +\frac{h^3}{3!}\left(\frac{\partial^2 f_i}{\partial y_j\partial y_k}f_jf_k+\frac{\partial f_i}{\partial y_j}\frac{\partial f_j}{\partial y_k}f_k\right)\bigg|_{t=0} \!+ \frac{h^4}{4!}\bigg(
\frac{\partial^3 f_i}{\partial y_j\partial y_k\partial y_l}f_jf_kf_l \\
&
+
3\frac{\partial^2 f_i}{\partial y_j\partial y_k}\frac{\partial f_l}{\partial y_l}f_lf_k
+
\frac{\partial f_i}{\partial y_j}\frac{\partial^2 f_j}{\partial y_k\partial y_l}f_kf_l
+
\frac{\partial f_i}{\partial y_j}\frac{\partial f_j}{\partial y_k}\frac{\partial f_k}{\partial y_l}f_l\bigg)\bigg|_{t=0} + ...
\end{align}
for $i=1,...,n$. We note that the expansion is convergent if $h$ is small compared to $\|\bff\|$. 

The elementary differentials of the ODEs \eqref{sphericalv} and \eqref{sphericalx} are calculated and the terms up to $O(h^3)$ are presented 
\begin{align}
v_i(h) =& v_i + h\alpha(u_i-v_i) + \frac{h^2}{2} \left(-\alpha^2(u_i-v_i) + \alpha \frac{\partial u_i}{\partial x_j}v_j\right)  \\ 
& + \frac{h^3}{3!}\left(\alpha v_jv_k\frac{\partial^2u_i}{\partial x_j \partial x_k} + \alpha^3 (u_i-v_i) - \alpha^2 \frac{\partial u_i}{\partial x_j} (u_j-2v_j)\right) + ...\label{expansionv}	\\
x_i(h) =& x_i + hv_i + \frac{h^2}{2} \left(\alpha(u_i-v_i)\right) + \frac{h^3}{3!}\left(\alpha\frac{\partial u_i}{\partial x_j}v_j - \alpha^2(u_i-v_i)\right) + ... \label{expansionx}	
\end{align}
where the variable appearing on the right hand side are evaluated at $t=0$. Here we assumed nothing about the size of $\alpha$, but instead take $h\ll\alpha$ such that the series converges. Taking the determinant of the Jacobian of $\bx(h)$ from equation \eqref{expansionx} with respect to $\bx_0$ yields an expansion for $\X$ with repect to $h$
\begin{equation}\label{vol expansion}
\X = 1 + h\X_1 + h^2\X_2 + h^3\X_3 + h^4\X_4 + O(h^5) 
\end{equation}
where 
\begin{gather}\label{Xexpand}
\X_1 =\, 0,\quad \X_2 =\, {\frac {\alpha}{2}}\,\chi_3 ,\quad
\X_3 =\, {\frac {\alpha}{6}}\left(\chi_5-\alpha\chi_3\right) \\
\X_4 =\, \frac{\alpha}{24}\,\left(\chi_1 + \alpha \chi_2 + \alpha^2 \chi_3 + 3\alpha \chi_3^2 -2 \alpha \chi_4 - 2\alpha\chi_5\right).
\end{gather}
The elementary differentials $\chi_i$ are given by
\begin{gather}
\chi_1 = v_iv_j\frac{\partial^3u_k}{\partial x_i\partial x_j\partial x_k},\quad	\chi_2 =  u_i\frac{\partial^2u_j}{\partial x_i\partial x_j},\quad 
\chi_3 =  \frac{\partial u_i}{\partial x_i}\\ 
\chi_4 = \frac{\partial u_j}{\partial x_i}\frac{\partial u_i}{\partial x_j} = \|S\|^2_F - \|\Omega\|^2_F, \quad
\chi_5 = v_j\frac{\partial^2 u_i}{\partial x_i\partial x_j}.
\end{gather}
This can be verified by expansion of \eqref{dXdt} into its Taylor series. If we insist that the fluid field is divergence-free then the $\X_i$ and $\chi_i$ all vanish except for $\X_4$ and $\chi_4$. We are then left with
\begin{align}\label{vol centrifuge effect}
\X\big|_{\nabla\cdot\bu=0}= 1-\frac{\alpha^2}{12}\,{h}^{4}
\left( \|S\|^2_F - \|\Omega\|^2_F\right)+O \left( {h}^{5} \right)
\end{align}
which relates the fluid rate of strain and rotation with the contractivity of physical volume in the same way as the centrifuge effect \eqref{MaxeyCentrifuge}. That is, if the rate of vorticity is greater than the rate of strain, physical volumes of particles will contract and vice versa. 

\subsection{Expanding the numerical solution and numerical errors}\label{sec:num int sph}\label{sec:interp errors}
In this section, we perform a similar analysis to that of section \ref{sec:expansion} but instead of the exact flow of the ODE, we consider now how physical volumes of particles evolve under the \textit{numerical} flow. That is, we will look at how errors to the divergence of the fluid field affect the evolution of volumes of particles under the numerical solution to the equations of motion. We do so by expanding the numerical methods into their elementary differentials and comparing the expansions with the exact solution. We then look at how errors to the divergence of the fluid field affect the evolution of physical volume under the numerical flow. The results are that if a divergence-free interpolation method is used, the numerical methods preserve the same qualitative behavior of the centrifuge effect. Moreover, we show here that the centrifuge-preserving methods replicate the centrifuge-effect from equation \eqref{vol centrifuge effect} up to the accuracy of the interpolation method when the fluid field is divergence-free. 

Consider the map $\Phi^{[n]}_h:(\bv_0,\bx_0)\rightarrow(\bv_1,\bx_1)$, where the superscript $[n]$ denotes the numerical method in consideration. We calculate $\X^{[n]} = \det\left(\frac{\partial \bx_1}{\partial \bx_0}\right)$ and expand the solution in $h$ yielding an expression of the form 
\begin{equation}\label{num expansion}
\X^{[n]} = 1 + h\X^{[n]}_1+h^2\X^{[n]}_2+h^3\X^{[n]}_3+h^4\X^{[n]}_4+O(h^5)
\end{equation}
The values of $\X^{[n]}_i$ for $i = 2,3,4$ for the Forward Euler (FE1), Lie-Trotter (LT1), order one centrifuge-preserving (CP1), Ralston (RK2), Adams-Bashforth two-step (AB2), order two centrifuge-preserving (CP2) and Strang splitting (SS2) methods are presented in table \ref{table:num expansion}. Note that $\X^{[n]}_1=0$ for all the above methods. 

\renewcommand{\arraystretch}{1.5}
\begin{table}
	\centering
	\begin{tabular}{c|ccc}
		Method & $\X_2^{[n]}$ & $\X_3^{[n]}$ & $\X_4^{[n]}$  \\
		\hline
		\begin{tabular}{c}
			Exact\\
			solution\\
		\end{tabular} & $\frac{\alpha}{2}\chi_3$&${\frac {\alpha}{6}}\left(\chi_5-\alpha\chi_3\right)$ & \begin{tabular}{c}
			$ \frac{\alpha}{24}\,\big(\chi_1 + \alpha\chi_2 + \alpha^2 \chi_3+ 3\alpha\chi_3^2 $ \\
			$\qquad -2 \alpha\chi_4 - 2\alpha\chi_5\big)$  \\
		\end{tabular} \\
	\hline
		FE1 &0 & 0 & 0 \\
		\hline
		LT1 & $\alpha\chi_3$ & $\frac{\alpha^2}{2}\chi_3$ &$-\frac {{\alpha}^{2}}{6} \left(3\,\chi_{{4}}- \alpha\,\chi_{{3}}-3\,{\chi_{{3}}}^{2} \right)$ \\
		\hline
		CP1 & $\frac{\alpha\sqrt{6}}{6}\chi_3$ &${\frac {\alpha\,
				\left( \sqrt {6}-1 \right) }{6}}\chi_{{5}}-{\frac {{\alpha}^{2}\sqrt {6}}{12}}\chi_{{3}}
		$&\begin{tabular}{c}
			$\frac {\alpha}{36} \big(  ( {\alpha}^{2}\chi_{{3}}-3\,\alpha\,
			\chi_{{5}}+{\frac {7}{2}\chi_{{1}}} ) \sqrt {6}$\\
			$+ \big( 3\,{
				\chi_{{3}}}^{2}-3\,\chi_{{4}}+3\,\chi_{{5}} \big) \alpha-6\,\chi_{{1}} \big) 
			$
		\end{tabular}\\
		\hline
		AB2 & $\frac{3\alpha}{16\theta}\chi_3$ & ${\frac {3\alpha}{32}}\left(\chi_5-\alpha\chi_3\right)$ & $\frac{\alpha}{128}\left(3\theta\chi_1+16\alpha\chi_3^2-16\alpha\chi_4\right)$ \\
		\hline
		RK2 & $\frac{\alpha}{2}\chi_3$& 0 & $\frac{\alpha^2}{8}\left(\chi_3^2 - \chi_4\right)$ \\
		\hline
		CP2 & $\frac{\alpha}{2}\chi_3$ &$\frac{3\alpha^2}{16}\chi_3+\frac{\alpha}{6}\chi_5$ & 
		\begin{tabular}{c}
			$\frac{\alpha}{576}\, \big( 33\,{\alpha}^{2}\chi_{{3}}+72\,\alpha\,{\chi_{{3}}}^{2}+24\,\alpha\,\chi_{{2}}$\\
			$\quad -48\,\alpha\,\chi_{{4}}-60\,\alpha\,\chi_{{5}}+32\,\chi_{{1}} \big)$\\
		\end{tabular} 
		\\
		\hline
		SS2 & $\frac{\alpha}{2}\chi_3$ &${\frac {\alpha}{4}}\left(\chi_5-\alpha\chi_3\right)$ & $\frac{\alpha}{48}\,\left(3\chi_1 + 4\alpha^2 \chi_3 + 6\alpha \chi_3^2 -6 \alpha \chi_4 - 6\alpha\chi_5\right)$ \\
	\end{tabular}
	\caption{The terms in the series expansion \eqref{num expansion} for the physical volume $\Psi^{[n]}$ under various numerical methods. Note that $\Psi^{[n]}_1 = 0$ for all the methods. }
	\label{table:num expansion}
\end{table}
\renewcommand{\arraystretch}{1}

We make a number of observations from this table. First, the divergence of the fluid field affects $\X^{[n]}$ at $O(h^2)$ for each method. The one exception to this is FE1, which satisfies $\X^{[F\!E]}=1$ and therefore erroneously preserves physical volume. When the divergence of the fluid field is zero, all the $\chi_i=0$ except $\chi_4$. For example, setting $\nabla\cdot\bu=0$ gives for the SS2 method
\begin{equation}\label{strangCentrifuge}
\X^{[SS]}\big|_{\nabla\cdot\bu=0} = 1- \frac{\alpha^2}{8}\,{h}^{4}
\left( \|S\|^2_F - \|\Omega\|^2_F\right)+O \left( {h}^{5} \right).
\end{equation}
This means that the numerical solution generated by the Strang splitting method \eqref{strang} reproduces the \textit{qualitative} nature of the centrifuge effect, in the sense that $\X^{[SS]}\big|_{\nabla\cdot\bu=0}>1$ when $\left.\|\Omega\|_F>\|S\|_F\right.$.  This qualitative centrifuge effect is seen by all the methods (other than FE1) by setting $\nabla\cdot\bu=0$ in table \ref{table:num expansion}. However, we note here that the coefficient of the $O(h^4)$ term in equation \eqref{strangCentrifuge} is different to that of the exact solution \eqref{vol centrifuge effect}. Meaning that, while the method contracts physical volume when the exact solution does, it does so at an erroneous rate. This issue is circumvented by the centrifuge-preserving methods (CP1 and CP2), where we have chosen the time-step coefficients in such a way such that they yield the exact same expansion as \eqref{vol centrifuge effect} up to $O(h^4)$ and hence contracts physical volume at the same rate as the exact solution to leading order. 

We now discuss the effect of interpolation errors in simulations of spherical particles in numerically calculated flows. Say that $\bu_e(\bx)$ is the true solution to the underlying Navier-Stokes equations that satisfies $\nabla\cdot\bu_e(\bx)=0$. As this exact solution is generally not available, we consider the following three cases
\begin{enumerate}
	\item Case (a): the fluid field has interpolation errors $\bdelta(\bx)$ that are not divergence free $\bu(\bx)= \bu_e(\bx) + \bdelta(\bx)$, where $\nabla\cdot\bdelta(\bx) \ne 0$ (e.g., using standard polynomial interpolation)
	\item Case (b): the fluid field has interpolation errors $\bdelta(\bx)$ and is divergence free $\bu(\bx)= \bu_e(\bx) + \bdelta(\bx)$, where $\nabla\cdot\bdelta(\bx) = 0$ (e.g., using MRBF interpolation)
	\item Case (c): the fluid field is free of errors $\bu(\bx) = \bu_e(\bx)$ and $\nabla\cdot\bu(\bx)=0$ (e.g., when the velocity field is available in closed form, also referred to as ``exact" interpolation.)
\end{enumerate} 

We pay particular attention to how errors resulting in $\nabla\cdot\bdelta(\bx)\ne0$ affect how the numerical methods evolve physical volume. To quantify this we define the physical volume error by 
\begin{equation}
\Delta \X^{[n]} = \X-\X^{[n]}.
\end{equation}
Here, $\X$ is used to denote the physical volume over time $h$ of the exact solution with fluid field corresponding to case (c), that is, the physical volume of the true solution in the absence of any errors, whereas $\X^{[n]}$ is the physical volume of the numerical solution with fluid field corresponding to one of the three cases given below. The results are presented in table \ref{table:vol errors}. We see here that the physical volume errors in case (a) are $O(h^2)$ and proportional to $\nabla\cdot\bdelta(\bx)$ for all the methods. In case (b), the physical volume errors are proportional to $O(h^4)$ except for the centrifuge-preserving methods, which have physical volume error proportional to $O(h^4\delta_4)$, where $\delta_4 = \left| \chi_4(\bu) - \chi_4(\bu+{\boldsymbol \delta})\right| = O(|\bdelta|) \ll |\chi_4(\bu)|$ is the error of $\chi_4$ from the interpolation method, which we assume is small. In case (c), $\delta_4=0$ and the centrifuge-preserving methods have physical volume error proportional to $O(h^5)$, while the other methods are $O(h^4)$. It is due to this behavior that we expect all the methods to more accurately evolve physical volume, when a divergence-free interpolation method is implemented such as with MRBFs. In this case, we expect the centrifuge-preserving methods to perform especially well due to table \ref{table:vol errors}.

\renewcommand{\arraystretch}{1.5}
\begin{table}[h!]
	\centering
	\begin{tabular}{c|ccc}
		& \multicolumn{3}{c}{$|\Delta\X^{[n]}|$}\\\hline
		Method &  \renewcommand{\arraystretch}{1.1}\begin{tabular}{c}
		Case (a) \\
		($\nabla\cdot\bu(\bx)\ne0$)\\
	\end{tabular}
 & \renewcommand{\arraystretch}{1.1}\begin{tabular}{c}
 	Case (b)\\
 	($\nabla\cdot\bu(\bx)=0$)\\
 \end{tabular}&\renewcommand{\arraystretch}{1.1}\begin{tabular}{c}
 Case (c)\\
 ($\bu(\bx)=\bu_e(\bx)$)\\
\end{tabular}\\ 
		\hline 
		FE1 & $\frac{\alpha}{2}h^2|\chi_3|$&$\frac{\alpha^2}{12}\,{h}^{4}
		|\chi_4|$ &$\frac{\alpha^2}{12}\,{h}^{4}
		|\chi_4|$ \\
		LT1 & $\alpha h^2|\chi_3|$ &$\frac{5\alpha^2}{12}\,{h}^{4}
		|\chi_4+\frac{\delta_4}{2}|$ &$\frac{5\alpha^2}{12}\,{h}^{4}
		|\chi_4|$ \\
		CP1 & $\frac{\alpha\sqrt{6}}{6}h^2|\chi_3|$ &$\frac{\alpha^2}{12}\,{h}^{4}
		|\delta_4|$ &$O(h^5)$ \\
		AB2 & $\frac{3\alpha }{16\theta}h^2|\chi_3|$&$\frac{\alpha^2}{24}\,{h}^{4}|\chi_4+\frac{\delta_4}{8}|$ &$\frac{\alpha^2 h^2}{24}\,{h}^{4}
		|\chi_4|$ \\
		RK2 &$\frac{\alpha}{2}h^2|\chi_3|$ &$\frac{\alpha^2}{24}\,{h}^{4}|\chi_4+\frac{\delta_4}{8}|$ &$\frac{\alpha^2}{24}\,{h}^{4}
		|\chi_4|$ \\
		CP2 & $\frac{\alpha}{2}h^2|\chi_3|$ & $\frac{\alpha^2}{12}\,{h}^{4}
		|\delta_4|$ &$O(h^5)$ \\
		SS2 & $\frac{\alpha}{2}h^2|\chi_3|$& $\frac{\alpha^2}{24}\,{h}^{4}|\chi_4+\frac{\delta_4}{8}|$ & $\frac{\alpha^2}{24}\,{h}^{4}
		|\chi_4|$\\
	\end{tabular}
	
	\caption{The errors of the physical volume after one time step for the numerical methods under consideration. }
	\label{table:vol errors}
\end{table}
\renewcommand{\arraystretch}{1}

In addition to the erroneous contraction of physical volume, we note from examples \ref{FEcontraction} - \ref{AB2example} that large divergence errors impose more stringent restrictions on the time step for the numerical methods to be contractive.

\section{Numerical simulations}\label{sec:numerical tests}

In this section we test our numerical methods for simulating suspensions of particles in viscous flows. The section begins by outlining the flow field and summarizing the methods and numerical parameters. We then outline the computational cost and verify the convergence of the methods. Next we simulate suspensions of $10^4$ particles in Taylor-Green vortices. This is the most important part of the section and is comprised of three experiments. The first compares the integration methods with exact evaluation of the fluid field. The second compares the effect of different interpolation errors with the CP2 integration. The third and final experiment explores how a combination of the proposed interpolation and integration methods can be used to generate cost-effective accurate particle distributions compared to conventional methods. 

\subsection{Preliminaries}\label{preliminaries}
Here we will briefly outline the numerical methods that are under consideration in the forthcoming numerical experiments, the fluid field and finally the particle models.

The integration methods under consideration and their properties are summarized in table \ref{table:integration properties}. 
\begin{table}[h!]
	\centering
	\begin{tabular}{c|ccccccc}
		& FE1 & LT1 & CP1 & AB2 & RK2 & SS2 & CP2 \\
		\hline
		Order & 1 & 1 & 1 & 2 & 2 & 2 & 2 \\
		Contractivity-preserving & No & Yes & Yes & No & No & Yes & Yes \\
		Centrifuge-preserving & No & No & Yes & No & No & No & Yes \\
	\end{tabular}
	\caption{Summary of the properties of the integration methods under consideration} \label{table:integration properties}
\end{table} 

We will abbreviate the divergence-free MRBF interpolation with the nearest $(n+1)\times (n+1)\times (n+1)$ data points by MRBF$n$ and the non-divergence-free order $n$ tripolynomial interpolation by TP$n$. The MRBF shape parameters are set to $\epsilon_1 = 0.31$, $\epsilon_2 = 0.23$ and $\epsilon_3 = 0.16$ corresponding to the MRBF1, MRBF2 and MRBF3 schemes, respectively, and are chosen empirically. We will compare the methods against a reference solution that uses exact evaluation of the analytic fluid field and the classical fourth order Runge-Kutta method for time integration with a time step that is 10 times smaller then that of the other methods. Note that such a reference solution is only available in the case that the flow field is known in closed form.

The discrete fluid field is generated by evaluating a closed form solution to the Navier Stokes equation on a regularly spaced grid with uniform sampling in each direction $\Delta x = \Delta y = \Delta z=1/10$. We use a stationary Taylor-Green vortex solution that was proposed in \cite{taylor1937mechanism} and has been used by other authors to study the behaviour of particles in cellular flow fields \cite{maxey1987motion,ruan2020structural,bergougnoux2014motion,jayaram2020clustering}. The particular Taylor-Green flow field used in the experiments is given by $\bu(\bx) = (u(\bx),v(\bx),w(\bx))^T$ where
\begin{align}
u(\bx) =& \,2\cos(2\pi x)\sin(2\pi y)\sin(2\pi z),\\
v(\bx) =& -\sin(2\pi x)\cos(2\pi y)\sin(2\pi z),\label{TGV}\\
w(\bx) =& -\sin(2\pi x)\sin(2\pi y)\cos(2\pi z).
\end{align}
We will perform experiments on both spherical and non-spherical particles. Denoting by $\lambda$ the aspect ratio of the particle, then $\lambda=1$ corresponds to spherical particles, $\lambda>1$ corresponds to a prolate spheroid and $\lambda<1$ corresponds to an oblate spheroid. For $\lambda=1$, the equations of motion are given by equations \eqref{sphericalv} and \eqref{sphericalx}, while for $\lambda\ne1$ the equations of motion are \eqref{sphericalv}, \eqref{sphericalx}, \eqref{eq:rotation} and \eqref{eq:Qdot}. For details about the moment of inertia tensor $J$, torque term $\mathbf{T}$, resistance tensor $K$ for the $\lambda\ne1$ cases we refer to \ref{model}. Finally, we note that for all of the following experiments, the particles are given a random initial location within in a box of width $0.01$ centered at the point $x_0 = (1/3,1/5,1/7)^T$ in the domain and a random initial orientation for non-spherical particles. 

\subsection{Computational cost}
Here, we outline the main computational costs associated with the methods. The two main steps in the algorithm are the interpolation step and the time integration step, which we examine separately. The wall clock times $T_w$ for $10^4$ time steps of the considered integration methods using exact evaluation of the fluid field are measured and presented in table \ref{integration times} and $T_w$ for $10^4$ time steps of the various interpolation methods using the FE1 method are presented in table \ref{interpolation times}. 

We note that the centrifuge-preserving methods are slightly more costly due to extra evaluations of the $\Phi_{ah}^{[LT]}$ operator. However, we note that one could speed up many of these splitting methods by observing that they are conjugate to a lower stage faster method, for example $$\left(\Phi_h^{[SS]}\right)^N = \psi^{[1]}_{\frac{h}{2}}\circ\left(\Phi_h^{[LT]*}\right)^N \circ\psi^{[1]}_{\frac{-h}{2}},$$ hence repeated evaluations of the operator $\Phi_h^{[SS]}$ when implemented in this way effectively has the same cost as $\Phi_h^{[LT]*}$. Similar observations are made for the centrifuge preserving methods. 

For the interpolation step, there are two main calculations that contribute the most to the computational cost. The first being the solution of a linear system of size $3n^3\times3n^3$ to find the interpolation coefficients. Guassian elimination is used for this purpose due to simplicity and the fact that the systems are not so large (at most $192\times192$ for the MRBF3 and TP3 methods). However, we note the existence of the exact matrix inverses for the coefficient matrices of these linear systems. This can be found in \cite{lekien2005tricubic} for the TP method and \cite{akaike1973block} for the MRBF method, the latter being due to the fact that the coefficient matrix has a block toeplitz structure for MRBF interpolation on Cartesian grids. The next most significant cost is evaluation of the sums of basis functions, that is, the sum in \eqref{MRBFsurf} and a similar equation for the TP methods. MRBF interpolation involves evaluation of more complex basis functions (i.e., matrix-vector products containing exponentials of polynomials), which is more costly than evaluating sums of monomials for the TP interpolation. This cost is more of a burden for the MRBF2 and MRBF3 methods as seen in table \ref{interpolation times}. 

\begin{table}[h]
	\centering
	\begin{tabular}{c|ccccccc}
& FE1 & LT1 & CP1 & AB2 & RK2 & CP2 & SS2  \\
\hline
$T_w\,(s)$ & 1.1817 & 1.3610 & 1.6577 & 2.0524 & 2.0376 & 2.5001 & 1.6455  \\
	\end{tabular}
	\caption{The wall clock times for $10^4$ time steps using different integration methods and exact interpolation.} \label{integration times}
	\vspace{0.5cm}
	\begin{tabular}{c|cccccc}
& MRBF1 & MRBF2 & MRBF3 & TP1 & TP2 & TP3\\
\hline
$T_w\,(s)$ & 3.3796 & 5.0776 & 12.0333 & 3.7263 & 4.2226 & 5.7551 \\
\end{tabular}
	\caption{The wall clock times for $10^4$ time steps using the FE1 method and different integration methods.} \label{interpolation times}
\end{table}

We see here that the MRBF1 and TP1 methods are roughly equal in cost. The MRBF2 method is about double that of TP1 and MRBF1 and is more expensive than the TP2 method. The MRBF3 method is double the cost of the TP3 method. We recall that we are not constrained to these three choices of MRBF methods and one is free to use any number of data points to achieve an optimum balance of accuracy and cost. This freedom is due to the fact that MRBF interpolation was designed for interpolation on scattered data points \cite{wendland2004scattered}. This option is not available for the TP$n$ methods, where $n+1$ grid points in each dimension are required to ensure the existence of a unique degree $n$ interpolating polynomial.

\subsection{Convergence}
In this section, we will verify the convergence of the integration methods first with exact interpolation then with various combinations of the interpolation methods for spherical ($\lambda = 1$) and non-spherical ($\lambda = 10$). In these experiments, we set $St=1$ and compute the particles' dynamics for time $T=1$.

\begin{figure}[!h]
	\centering
		\begin{subfigure}{0.4\textwidth}
			\includegraphics[width=\linewidth]{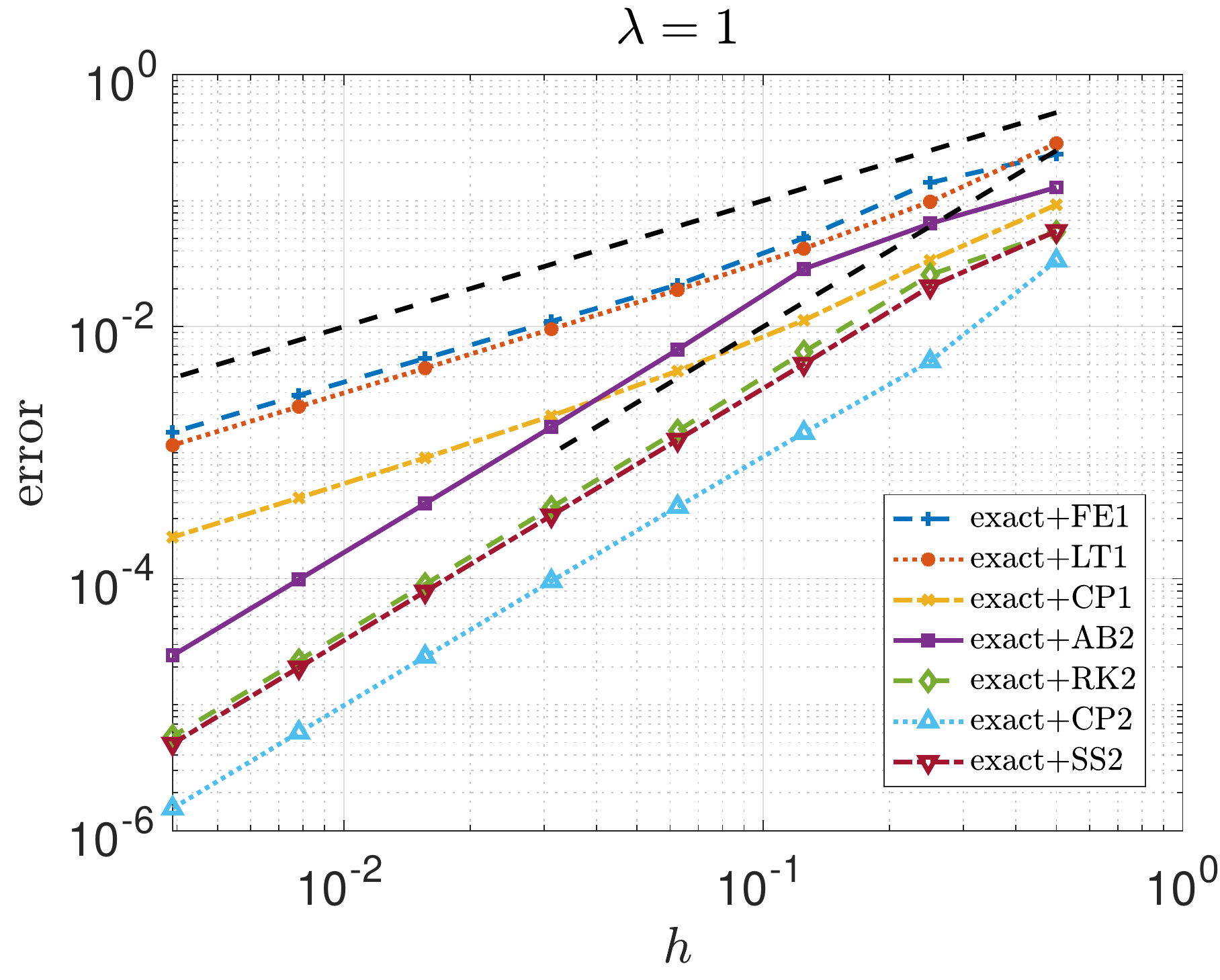}
			\caption{}
			\label{fig:conv sph}
		\end{subfigure}
	\begin{subfigure}{0.4\textwidth}
		\includegraphics[width=\linewidth]{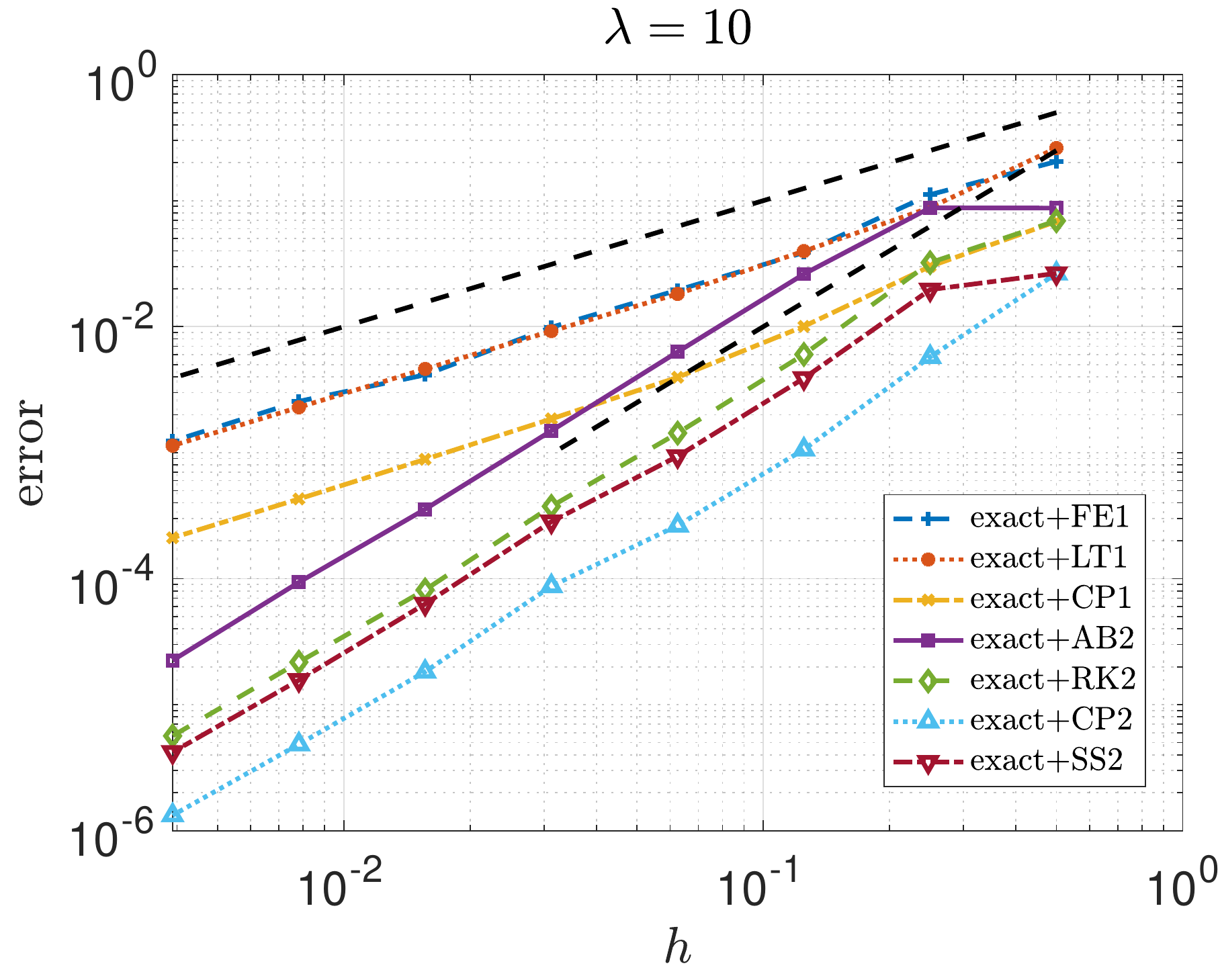}
		\caption{}
		\label{fig:conv rod}
	\end{subfigure}

\begin{subfigure}{0.4\textwidth}
	\includegraphics[width=\linewidth]{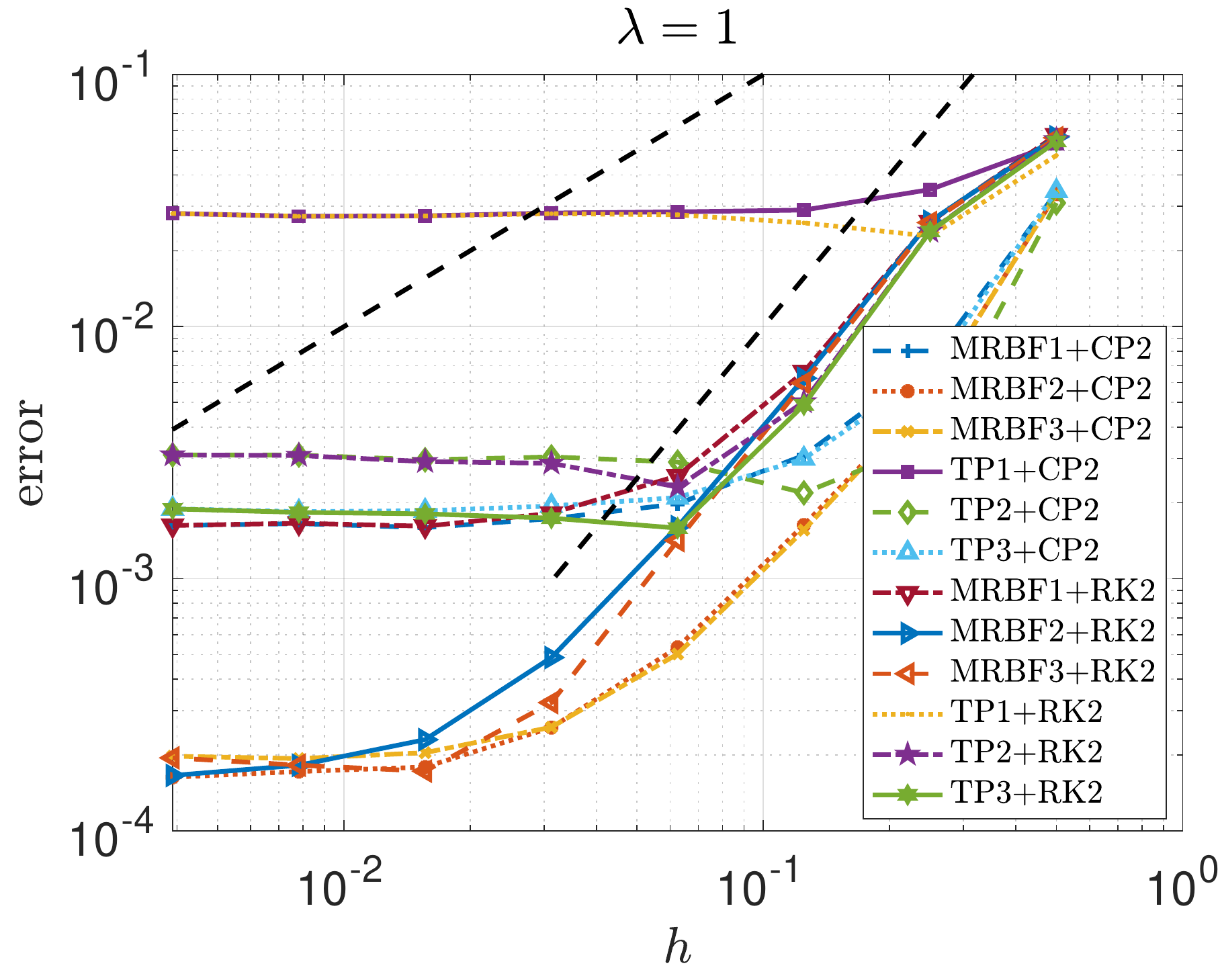}
	\caption{}
	\label{fig:conv2 sph}
\end{subfigure}
\begin{subfigure}{0.4\textwidth}
	\includegraphics[width=\linewidth]{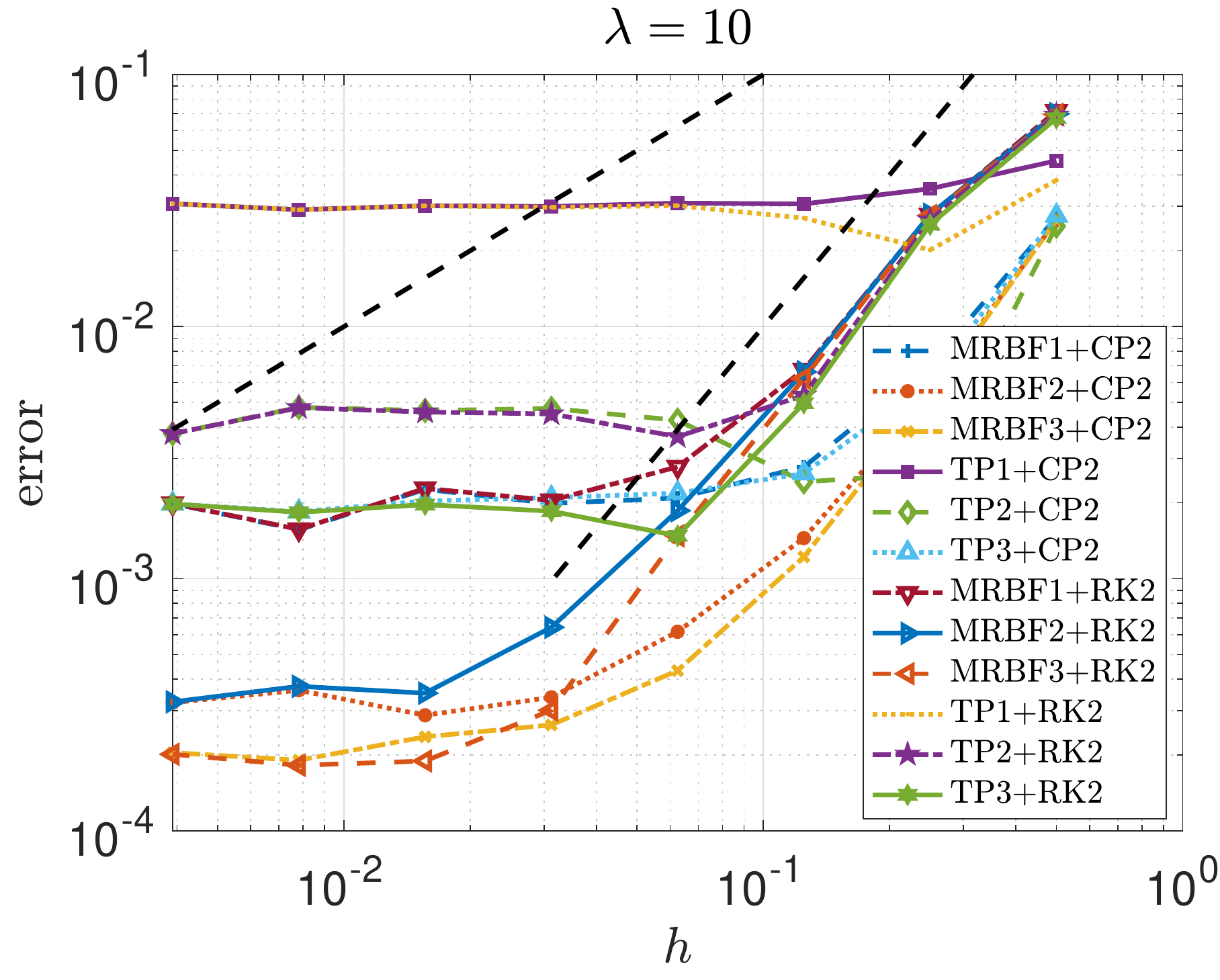}
	\caption{}
	\label{fig:conv2 rod}
\end{subfigure}
	\caption{Figures (a) and (b) show the convergence of the numerical integration methods using exact interpolation for the $\lambda = 1$ and $\lambda = 10$ equations, respectively. Figures (c) and (d) show the convergence of the CP2 and RK2 methods with the six interpolation methods for the $\lambda = 1$ and $\lambda = 10$ equations, respectively. The dashed lines are $O(h)$ and $O(h^2)$ }
	\label{}
\end{figure}

The convergence of the error, measured in the 2-norm, of the integration methods are presented in figures \ref{fig:conv sph} and \ref{fig:conv rod}. We observe here that the FE1 and LT1 methods have similar accuracy as do the RK2 and SS2 methods. It is noted that the benefits of preserving contractivity in the various splitting methods are expected to be seen after longer times. One remarkable observation here is that for this Stokes number the first order CP1 method is competitive with the second order RK2 and AB2 methods at large time steps, furthermore, the CP2 method is the most accurate by a factor of about 5 in both scenarios. 

Figures \ref{fig:conv2 sph} and \ref{fig:conv2 rod} show the convergence of the CP2 and RK2 methods with different interpolation methods. We see that the methods initially converge at their expected order, but as $h$ goes to zero  we see that the integration error becomes overshadowed by the $h$-independent interpolation error. We observe that the MRBF solutions yield more accurate solutions than the TP solutions using the equal number of data points. However, the TP3 solution is expected to perform better for longer simulations where particles cross grid cells. This is because the piece-wise fluid field constructed from the TP3 method is globally $C^1(\mathbb{R}^3)$, meaning that the spatial derivatives of the fluid velocity are everywhere continuous. This is not true for the other methods.

Finally, we remark that the centrifuge-preserving methods perform equally well for non-spherical particles. 


\subsection{Simulating suspensions of particles}\label{sec:clusters}

%

Up until now we have mainly focused on the average error in the positions of individual particles. It is well known that standard methods such as polynomial interpolation and Adams-Bashforth integration do a good job at minimizing this truncation error in some norm.  However, while it is indeed important that this conventional measure of error is kept at a minimum, in practice one is usually more interested in properties of distributions of many indistinguishable particles meaning that the individual error of each particle is less important. Due to this, it is more desirable that an algorithm reproduces accurately the spatial statistical properties of many particles rather than minimizing the absolute errors of each individual particle. With this in mind, the main goal here is to test to what extent the aforementioned errors affect suspensions of particles when viewed as a single discrete probability distribution. We will show that distributions of particles calculated by our proposed geometric methods will more closely resemble that of the exact solution, despite sometimes having higher average error per particle in the conventional sense.

In our context, a distribution $P=\{ (\bx_i,w_i)\}_{i=1}^{n_c}$ is a set of $n_c$ non-empty equally sized cells, where $\bx_i$ is the location of the cell center and $w_i$ is a weight that is equal to the number of particles in that cell. We let $P_n$ denote a distribution where the particle locations are calculated by a numerical method, $P_{\mathrm{ref}}$ refers to the distribution obtained by the reference solution and we use $300\le n_c\le400$ depending on the spread of particles. We will determine the accuracy of $P_n$ using three measures which we now outline.


The first is the first Wasserstein distance, which is a natural metric to compare the distance between two discrete probability distributions of equal size (also known as the Earth Mover Distance). The first Wasserstein distance between two probability distributions is denoted by $W_1(P_1,P_2)$ and is a measure of the cost of transporting the distribution $P_1$ into $P_2$ in the cheapest way possible. The cost is measured as the distance between cell centers, measured in the 2-norm and weighted by the number of particles being transported. For mathematical details about the first Wasserstein distance, we refer the reader to \cite{rubner2000earth} and the numerical computation of the first Wasserstein distances are computed using a publicly available MATLAB code \cite{ulmaz}. We denote by $W_1(P_n) = W_1(P_n,P_{\mathrm{ref}})$ the first Wasserstein distance between $P_n$ and $P_{\mathrm{ref}}$.

The second is the relative entropy (also known as the Kullback-Leibler divergence) \cite{kullback1951information}, which is a measure of how much information is lost from a reference distribution $P_2$ when an approximate distribution $P_1$ is used. The relative entropy is calculated by
\begin{equation}
E(P_1,P_2) = \sum_{\bx_i\in\Omega_P} P_1(\bx_i) \log\left(\frac{P_1(\bx_i)}{P_2(\bx_i)}\right),
\end{equation}
where $P(\bx_i) = w_i$ is the number of particles in the cell at $\bx_i$ and $\Omega_P$ is the support of the two distributions. If there is an empty cell in one distribution and not the other, say at $\bx=\bx_0$ we use $P(\bx_0)=10^{-1}$, to avoid singularities. This modestly penalizes the approximate solution for predicting a non-zero probability of having a particle in a cell that should have zero particles according to the reference distribution. We denote by $E(P_n) = E(P_n,P_{\mathrm{ref}})/n_p$ the relative entropy between $P_n$ and $P_{\mathrm{ref}}$ scaled by the number of particles $n_p = 10^4$.

Finally, the third means of determining the accuracy of the distribution is by the average error of the particle positions $\overline{\Delta \bx_n}$. This conventional measure of error is calculated by taking the difference between the final position of the numerical and reference solution starting from the same initial conditions and averaging over all the $n_p = 10^4$ particles, that is
\begin{equation}
	\overline{\Delta \bx_n} = \frac{1}{n_p}\sum_{i=1}^{n_p}\|\bx_{n,i}-\bx_{\mathrm{ref},i}\|_2,
\end{equation}
where $\bx_{n,i}$ is the $i$th particle calculated by the numerical method and $\bx_{\mathrm{ref},i}$ is the $i$th particle under the reference solution. As the rotational variables are strongly coupled with the translational variables, errors in the rotational dynamics will also influence the final positions of the particles, hence this is a reasonable measure of the error of the algorithms' overall accuracy in computing the dynamics of a single particle. We recall that this error is that which the conventional methods are designed to reduce when referring the the global order of accuracy of a method. 

In the forthcoming experiments, we will use various combinations of integration and interpolation methods to compute the paths of $10^4$ particles in the discrete Taylor-Green vortices starting with random positions and orientations within a cube of width $1/100$ centered at the point $(1/3,1/5,1/7)^T$. We perform three experiments. The first compares how the various numerical integration methods and their errors affect the spatial distribution of suspensions of particles in the absence of interpolation errors. The second experiment investigates how interpolation errors affect the spatial distribution of particles using the CP2 method. Finally, we look at how a combination of MRBF interpolation and centrifuge-preserving integration can be used to calculate fast and accurate suspensions of particles compared to the conventional AB2+TP$n$ methods, similar to the methods used in \cite{Portela, challabotla2015orientation, van2000dynamics, PanBanerjee, wang1996large}, for example.

\subsubsection{Comparison of integration methods}\label{sec:spherical clusters}
\begin{figure}
	\centering
	
	\begin{subfigure}{0.32\textwidth}
		\includegraphics[width=\linewidth]{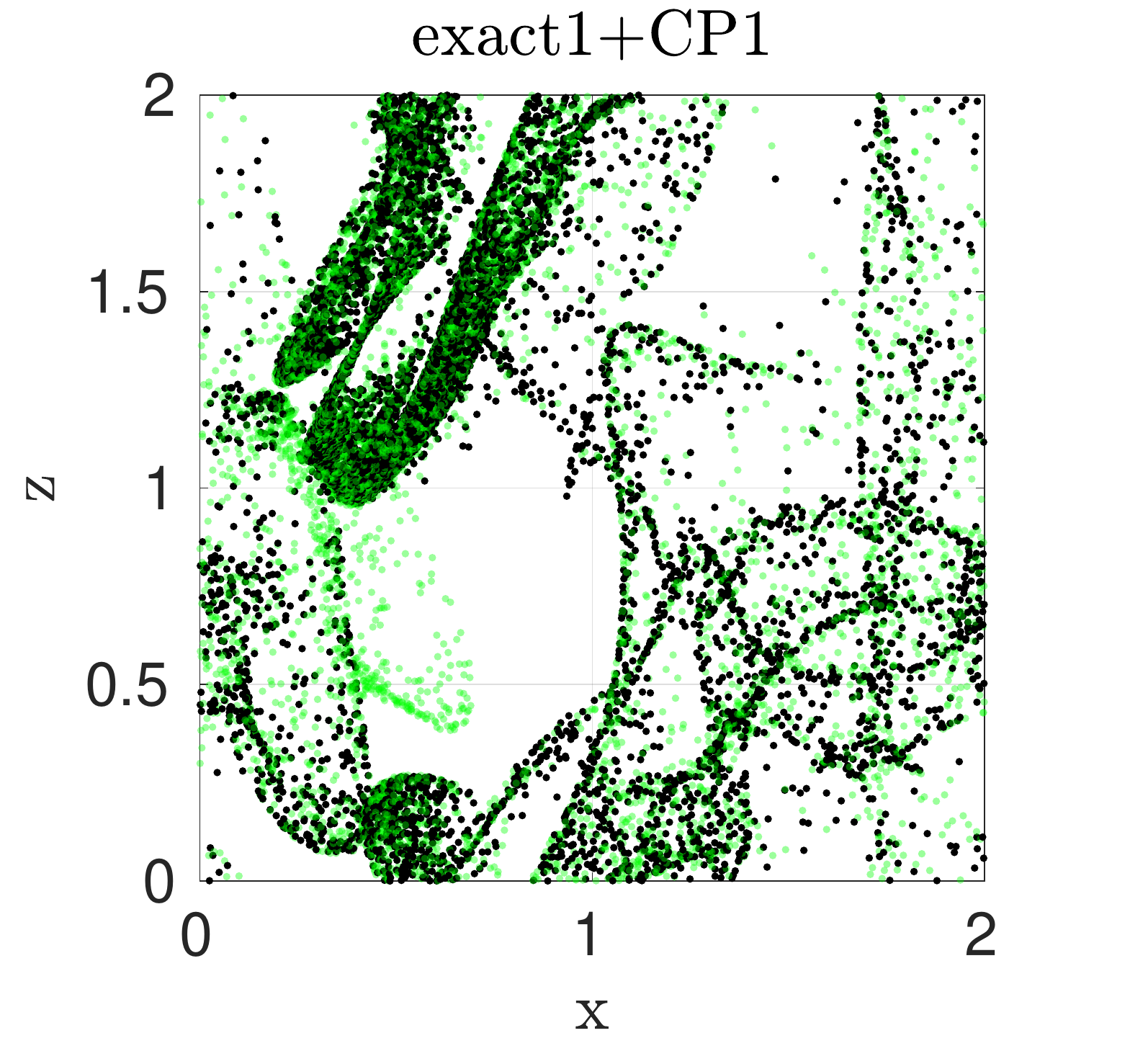}
		\caption{}\label{is1}
	\end{subfigure}
	\begin{subfigure}{0.32\textwidth}
		\includegraphics[width=\linewidth]{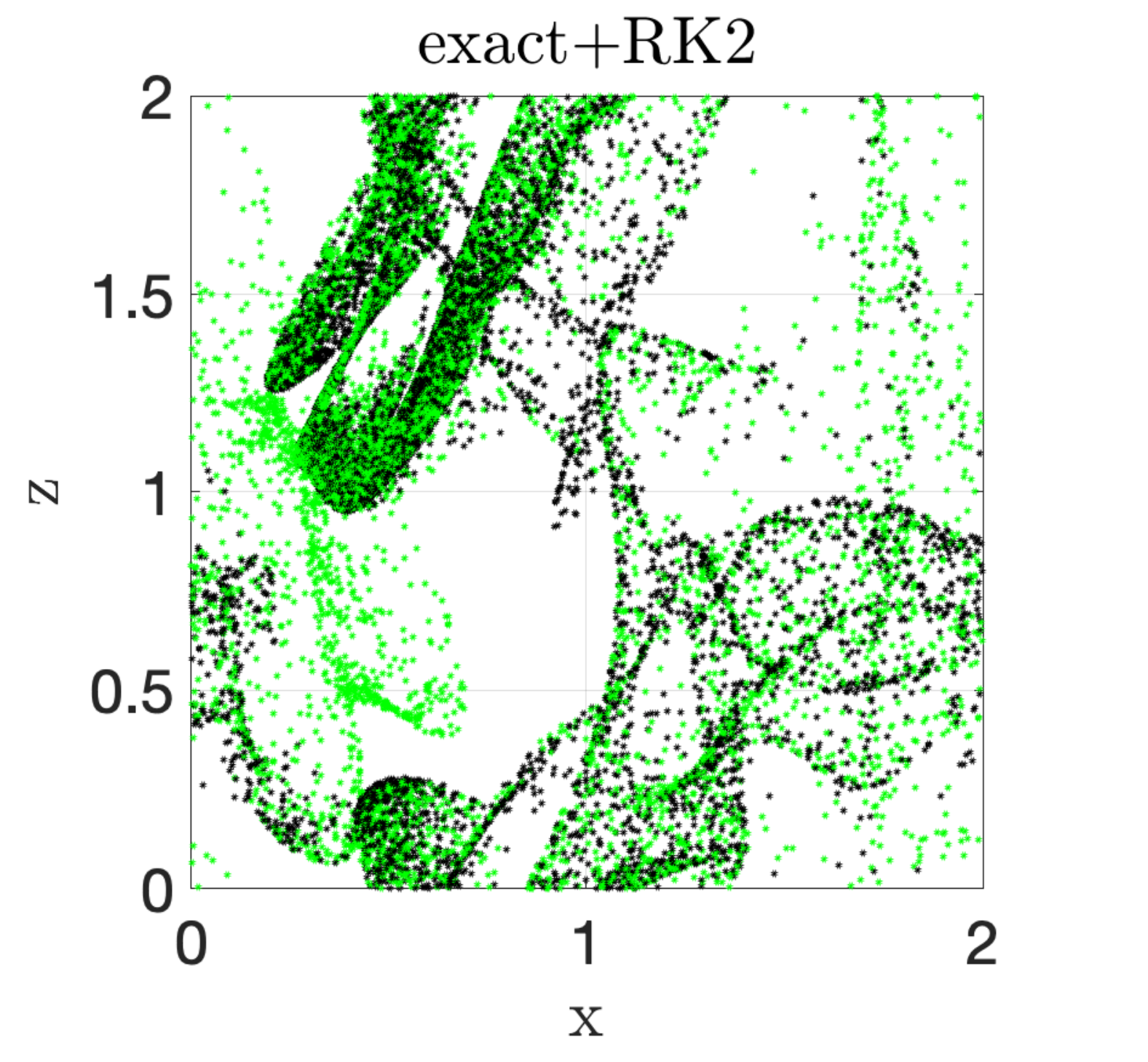}
		\caption{}\label{is2}
	\end{subfigure}
	\begin{subfigure}{0.32\textwidth}
		\includegraphics[width=\linewidth]{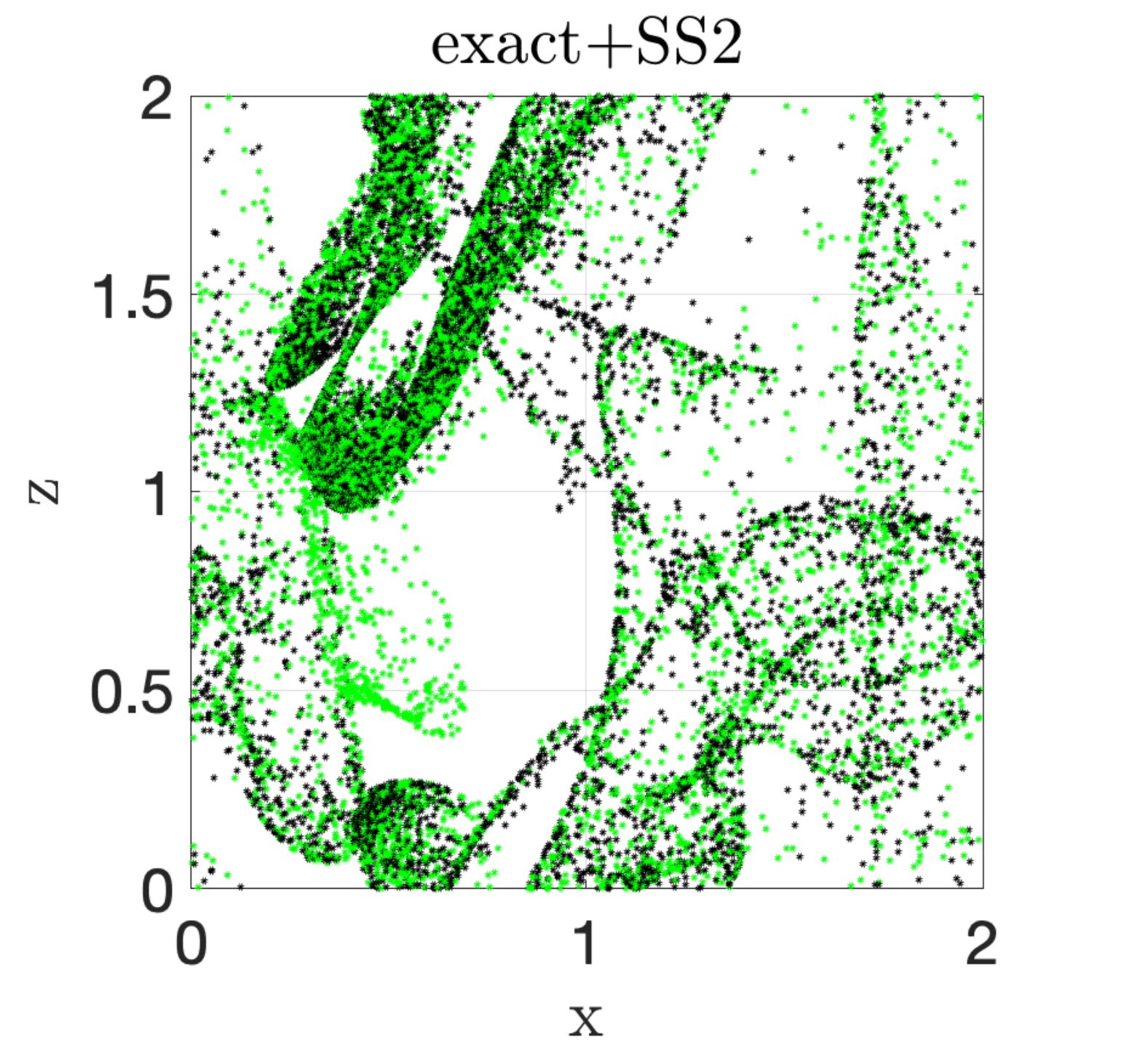}
		\caption{}\label{is3}
	\end{subfigure}

	\begin{subfigure}{0.32\textwidth}
		\includegraphics[width=\linewidth]{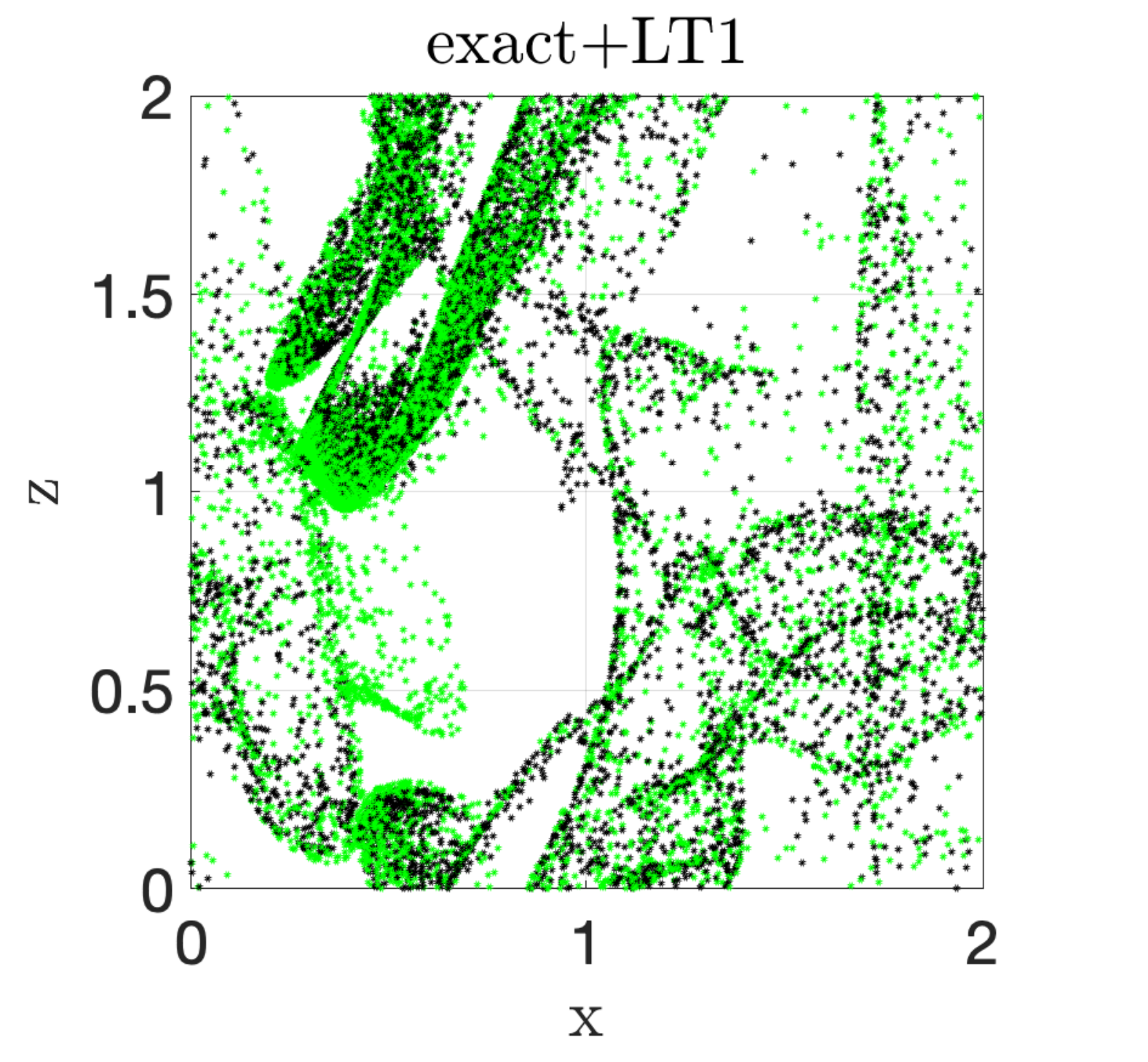}
		\caption{}\label{is4}
	\end{subfigure}
	\begin{subfigure}{0.32\textwidth}
		\includegraphics[width=\linewidth]{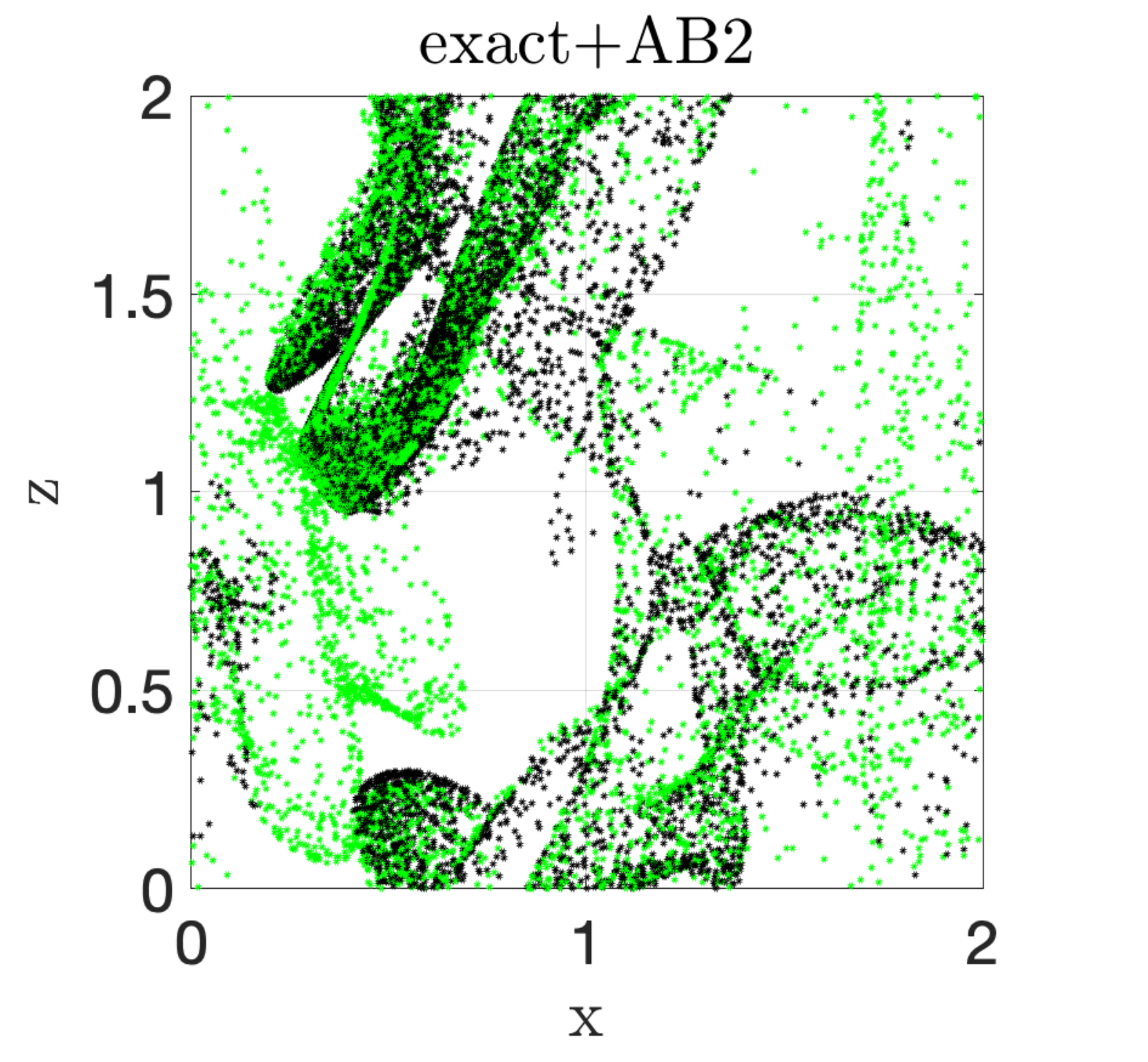}
		\caption{}\label{is5}
	\end{subfigure}
	\begin{subfigure}{0.32\textwidth}
		\includegraphics[width=\linewidth]{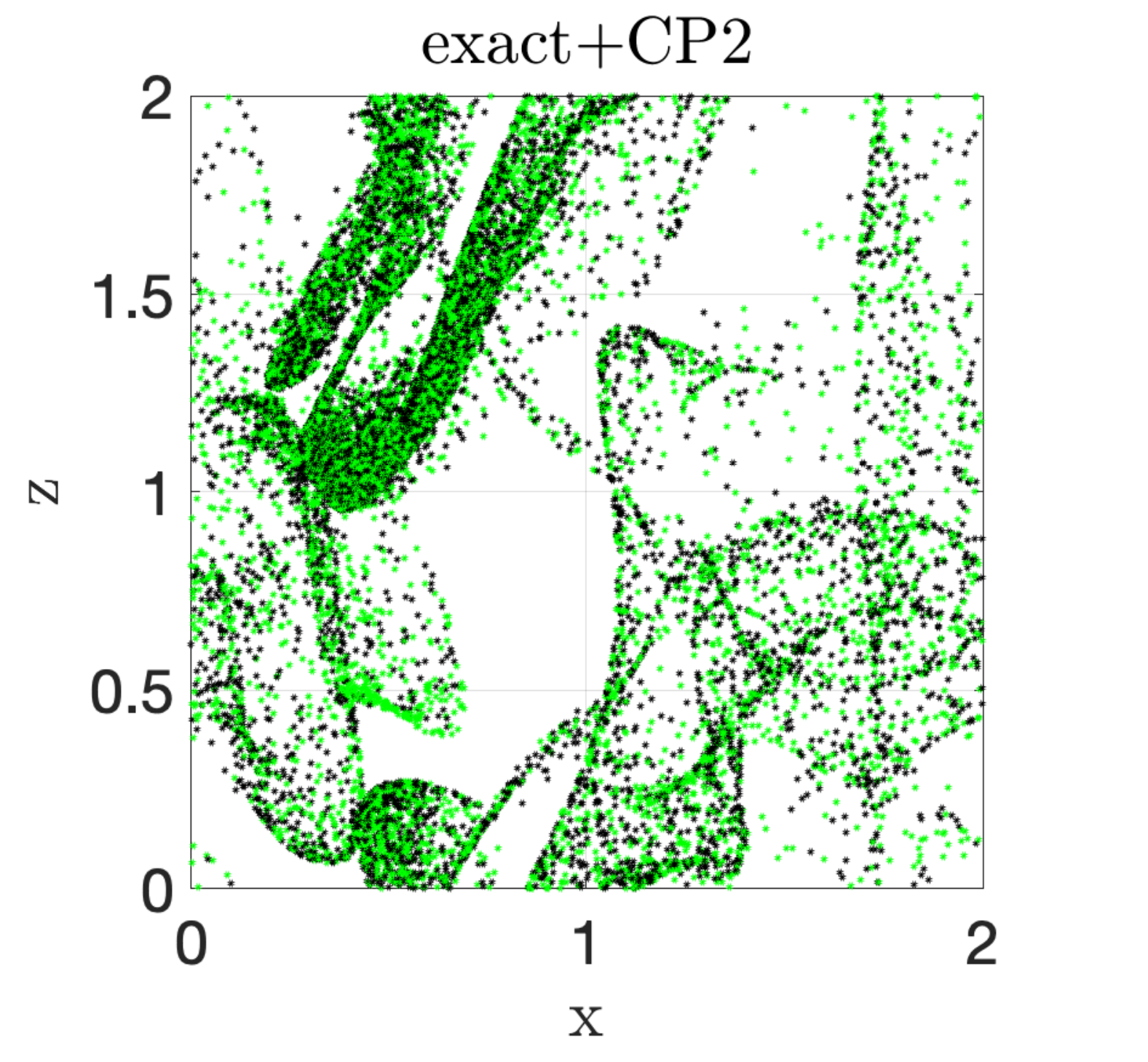}
		\caption{}\label{is6}
	\end{subfigure}
	
	\begin{subfigure}{0.32\textwidth}
		\includegraphics[width=\linewidth]{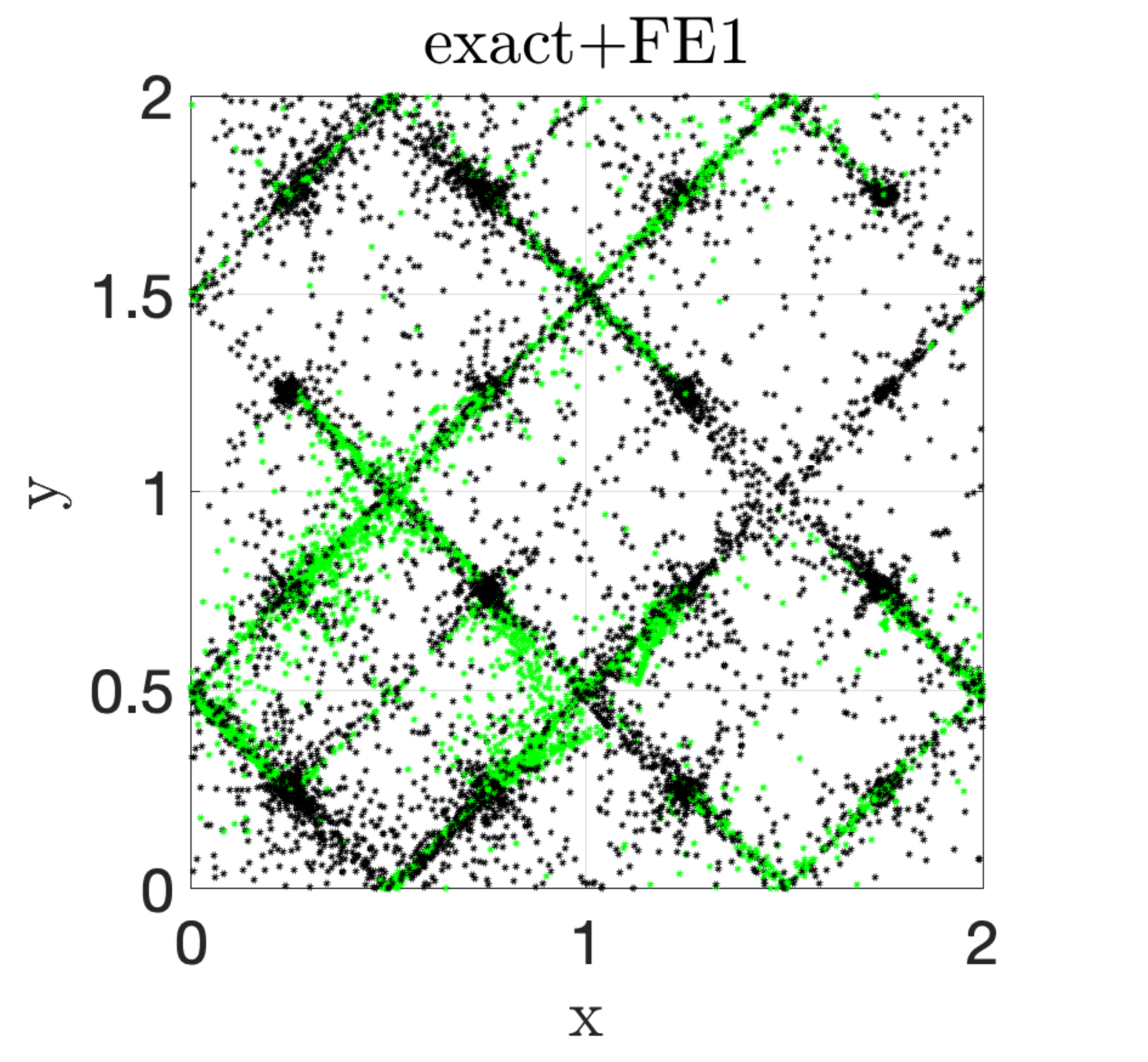}
		\caption{}\label{is7}
	\end{subfigure}
	\begin{subfigure}{0.32\textwidth}
		\includegraphics[width=\linewidth]{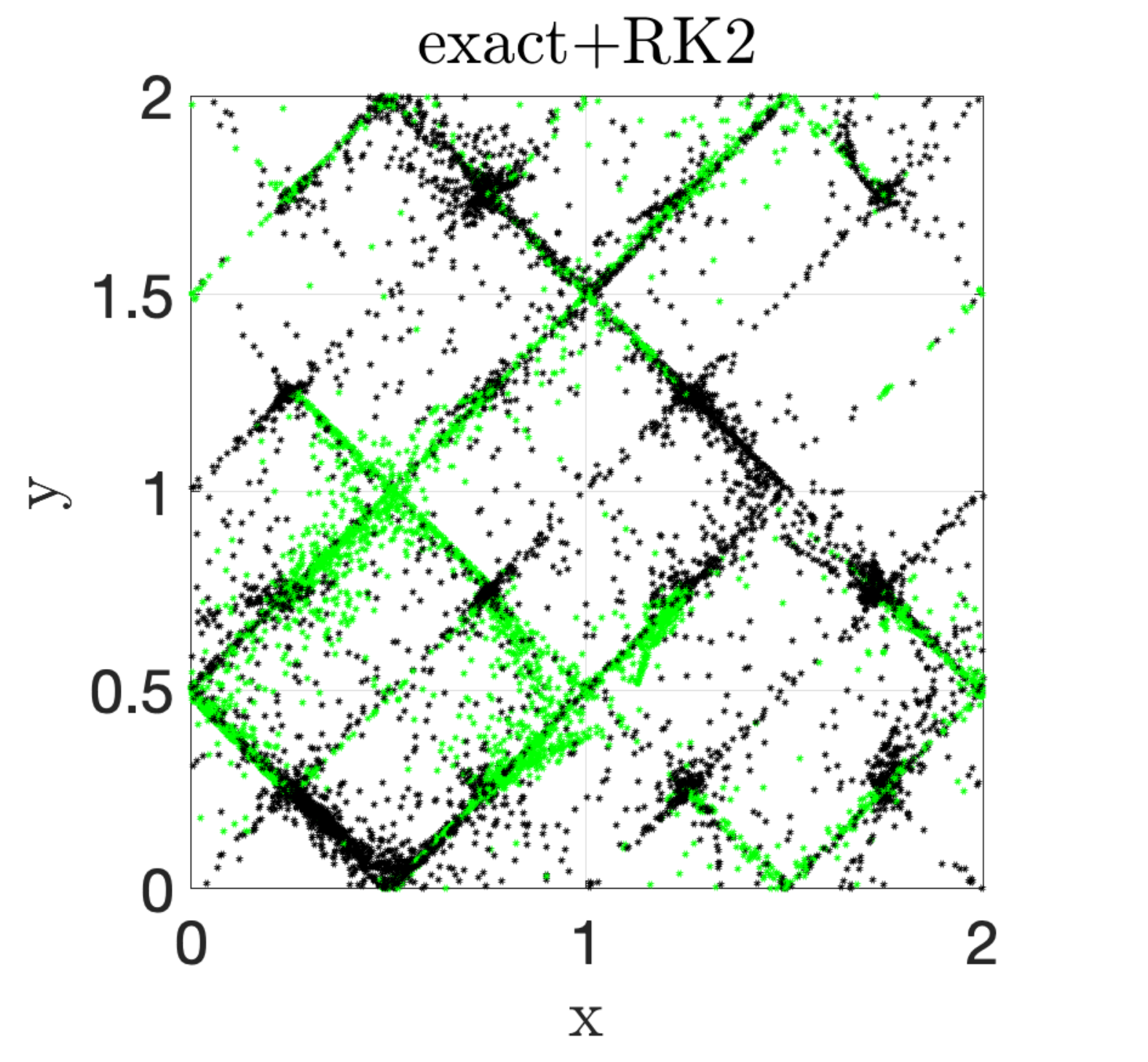}
		\caption{}\label{is8}
	\end{subfigure}
	\begin{subfigure}{0.32\textwidth}
		\includegraphics[width=\linewidth]{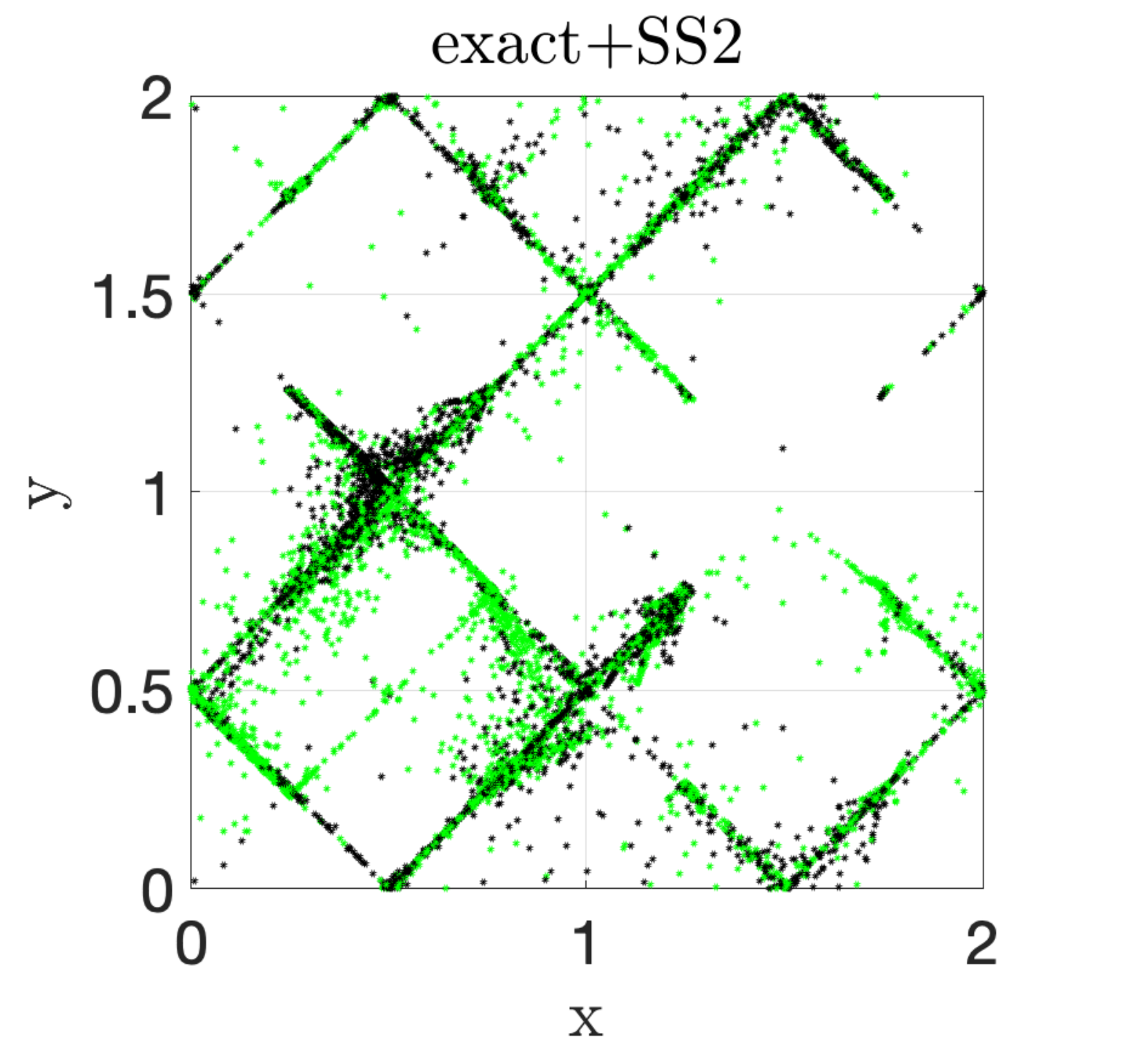}
		\caption{}\label{is9}
	\end{subfigure}

	\begin{subfigure}{0.32\textwidth}
		\includegraphics[width=\linewidth]{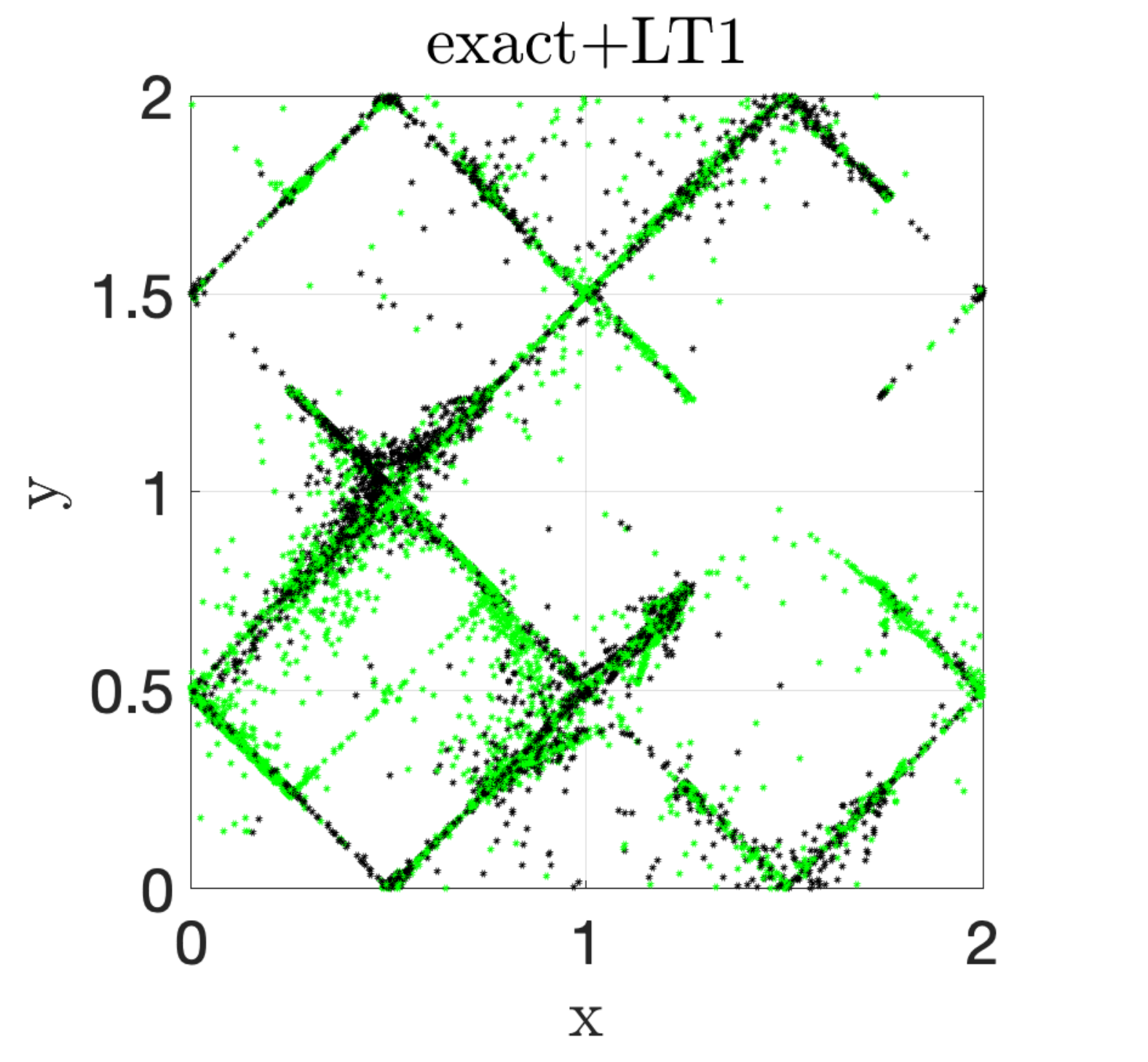}
		\caption{}\label{is10}
	\end{subfigure}
	\begin{subfigure}{0.32\textwidth}
		\includegraphics[width=\linewidth]{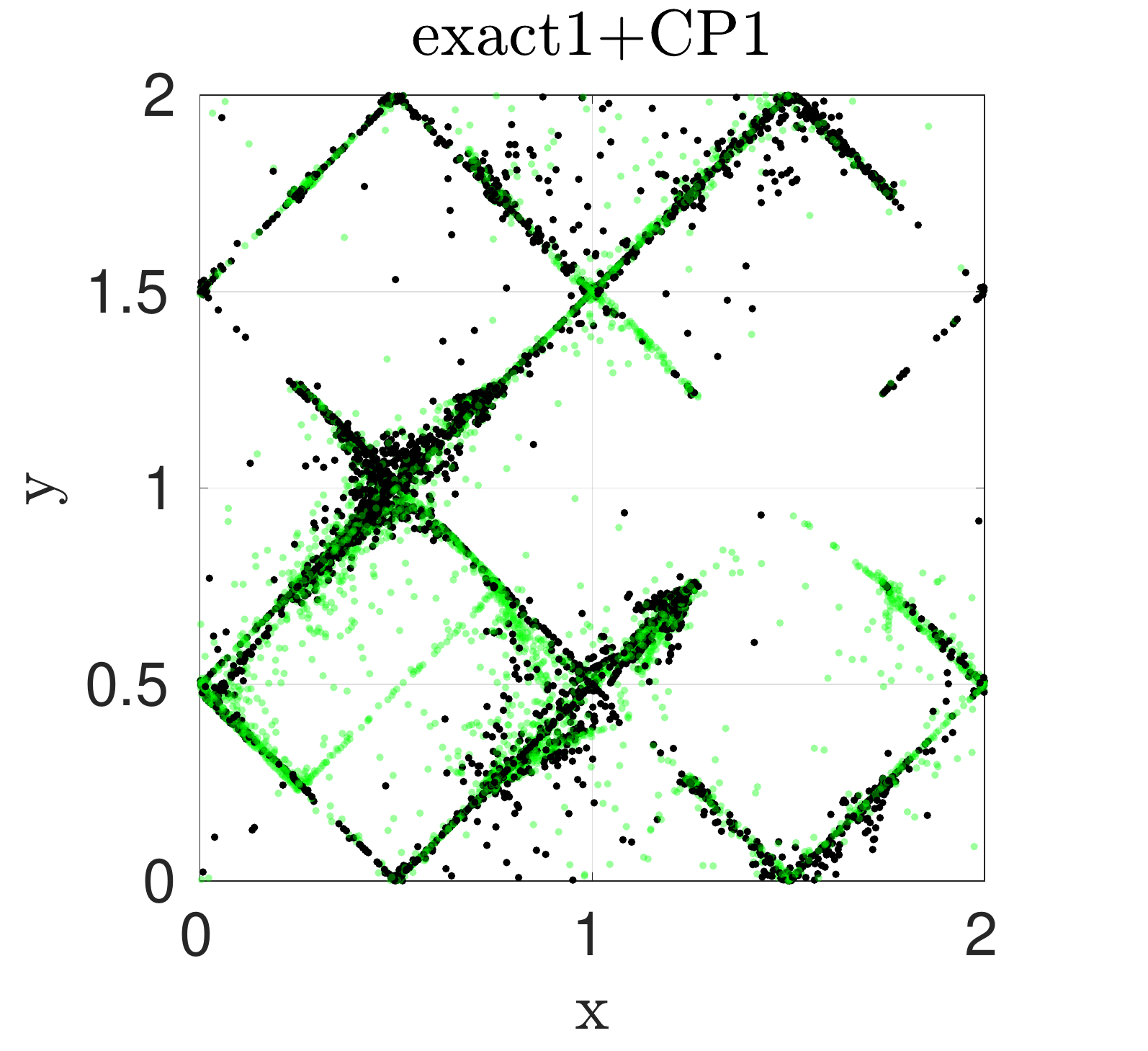}
		\caption{}\label{is11}
	\end{subfigure}
	\begin{subfigure}{0.32\textwidth}
		\includegraphics[width=\linewidth]{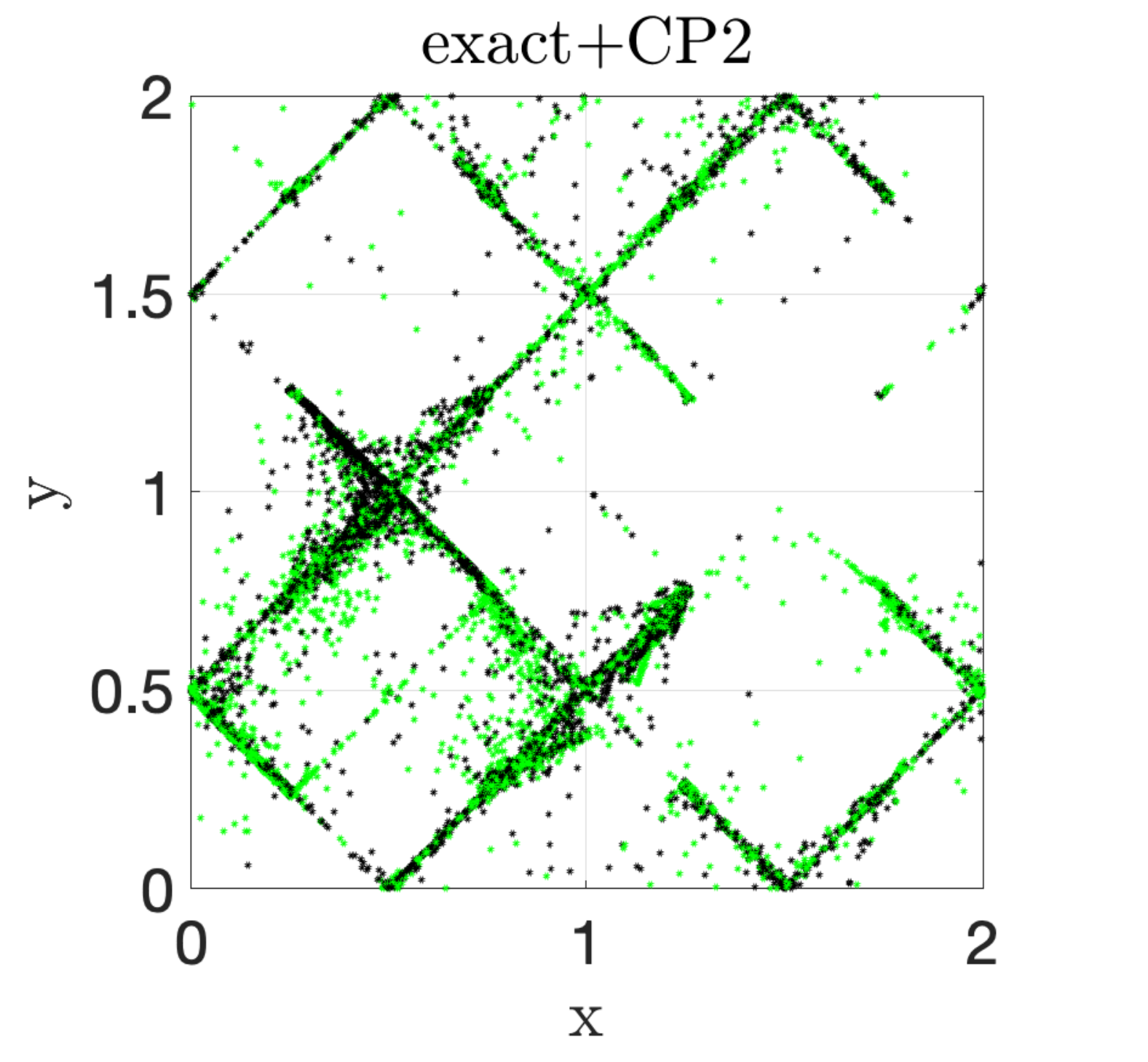}
		\caption{}\label{is12}
	\end{subfigure}

	\caption{Figures (a) through (f) show the spatial distribution of the particles in the $x-z$ plane for the $St = 10$, $h = 1/5$, $T = 20$, $\lambda = 1$ simulation from table \ref{table:integration} (The exact+FE1 is not shown). Figures (g) through (l) show the spatial distribution of the particles in the $x-y$ plane for the $St = 1$, $h = 1/20$, $T = 8$, $\lambda = 1/10$ simulation (the exact+AB2 solution is not shown). The reference solution is plotted in green in all figures.}
	\label{integration_spheres}
\end{figure}

In this experiment we use the seven integration methods outlined in table \ref{table:integration properties} to simulate a suspension of particles evolving in the Taylor-Green flow with exact interpolation. The methods are each tested in six separate simulations, three with Stokes numbers of $St = 1/5,1,10$ for spherical particles ($\lambda = 1$) and three with the same Stokes numbers for non-spherical particles ($\lambda = 1/10$). At the end of the simulation the relative entropy $E(P_n)$, first Wasserstein distance $W(P_n)$ and the average spatial error $\overline{\Delta \bx_n}$ between the numerical distribution and the reference distribution are calculated and presented in table \ref{table:integration}. The time step $h$ and total simulation time $T$ are also presented in this table. We start by discussing some qualitative features of the final distributions, examples of which are given in figure \ref{integration_spheres}. We then discuss the results of table \ref{table:integration} in detail. 

Figure \ref{integration_spheres} depicts the final distribution of the particles for the various integration methods. The particle positions are plotted modulo 2 for presentation purposes and represented by black dots, while the reference solution is plotted using green dots. Figures \ref{is1} to \ref{is6} correspond to the $St = 10$, $\lambda = 1$ simulation and is viewed along the $y$ direction. We see here that the CP2 solution is able to predict the correct clustering in all the regions that are predicted by the green reference solution. The LT1, CP1 and SS2 solutions are visually similar to each other, however do not correctly predict clustering of particles in some regions, given by regions of green dots that are void of black dots. The RK2 and AB2 solutions do a worse job as seen, again, by even more regions with a higher concentration of green dots compared to black dots. Similar observations are again seen in figures \ref{is7} to \ref{is12}, which correspond to the $St = 1$, $\lambda = 1/10$ simulation, viewed along the $z$ direction. In this simulation, the particles more closely follow the streamlines of the fluid field and more quickly concentrate in regions of high strain as seen by the regions of dense green dots. In these figures, it is even more easily seen that the four contractivity preserving methods do a good job at correctly clustering particles in regions where the reference solution does, while we see with the FE1 and RK2 solutions multiple regions exhibiting an erroneous concentration of black dots that are void of green dots. The AB2 solution is unstable for these parameters. 

To quantify the above observations, which have up until now been visual, we turn our attention to table \ref{table:integration}. We start by outlining some general observations that are common to all six simulations. Looking first at the order one methods, we observe that the LT1 and CP1 methods, which are contractivity preserving, outperform the FE1 method in almost all measures in each simulation despite the fact that their order of accuracy is the same. What is striking here is that in most simulations the LT1 and CP1 methods generally have a lower $\overline{\Delta \bx_n}$, $W(P_n)$ and $E(P_n)$ compared to the conventional RK2 and AB2 methods despite being of lower order and computational cost. Similar observations are made if we turn our attention towards the order two methods. That is, the SS2 method in most cases has lower $\overline{\Delta \bx_n}$, $W(P_n)$ and $E(P_n)$ than the RK2 and AB2 methods, and better still is the CP2 method. The advantage of the CP2 method over the SS2 is more pronounced than the advantage of the CP1 method over the LT1. The CP2 method has the lowest $\overline{\Delta \bx_n}$, $W(P_n)$ and $E(P_n)$ in all six simulations and is clearly the best method here in all three metrics. 

\begin{table}[h]
	\centering
	\begin{tabular}{|c|c|ccc|ccc|}
		\hline
		& & \multicolumn{3}{c|}{$\lambda = 1$} & \multicolumn{3}{c|}{$\lambda = 1/10$} \\
		\hline
		& $P_n$	  & $E(P_n)$ & $W(P_n)$ & $\overline{\Delta \bx_n}$ & $E(P_n)$ & $W(P_n)$ & $\overline{\Delta \bx_n}$ \\			
		\hline		
		\multirow{2}{*}{$St = \frac{1}{5}$} 
		& exact+FE1 & 7.0143 & 0.4466 & 0.5534 & --     & --     &    --  \\ 
		& exact+LT1 & 0.4596 & 0.0070 & 0.0098 & 2.3929 & 0.1793 & 0.2086 \\
		\multirow{2}{*}{$h = \frac{1}{50}$}
		& exact+CP1 & 0.1212 & 0.0049 & 0.0065 & 2.3100 & 0.1750 & 0.2048 \\
		& exact+AB2 & 4.0952 & 0.0697 & 0.0585 & --     & --     &    --  \\ 
		\multirow{2}{*}{$T = 4$}
		& exact+RK2 & 1.6791 & 0.0234 & 0.0215 & --     & --     &    --  \\
		
		& exact+CP2 & 0.0581 & 0.0022 & 0.0027 & 0.5339 & 0.0666 & 0.0983 \\
		& exact+SS2 & 0.1169 & 0.0048 & 0.0062 & 2.2956 & 0.1739 & 0.2039 \\

		\hline
		\multirow{2}{*}{
			$St = 1$
		} 
		& exact+FE1 & 6.9916 & 1.6321 & 2.5224 & 0.9086 & 0.6445 & 1.7383 \\
		& exact+LT1 & 0.1538 & 0.0929 & 1.0000 & 0.6645 & 0.4788 & 1.2536 \\
		\multirow{2}{*}{
			$h = \frac{1}{20}$
		} 
		& exact+CP1 & 0.1061 & 0.1034 & 1.0084  & 0.5716 & 0.4293 & 1.2554 \\
		
		& exact+AB2 & 0.4833 & 0.2715 & 1.4977 & --     & --     &    --  \\
		\multirow{2}{*}{
			$T = 8$
		} 
		& exact+RK2 & 0.2696 & 0.1566 & 1.3083 & 1.3583 & 0.4631 & 1.4790 \\
		
		& exact+CP2 & 0.0786 & 0.0558 & 0.5512 & 0.2505 & 0.2892 & 1.0857 \\
		& exact+SS2 & 0.1435 & 0.0890 & 1.0044 & 0.6374 & 0.4858 & 1.2535 \\
		\hline
		
		\multirow{2}{*}{
			$St = 10$
		}  
		& exact+FE1 & 6.8070 & 2.7326 & 2.8528 & 1.3145 & 1.9214 & 4.8392 \\
		& exact+LT1 & 0.3531 & 0.1626 & 0.6588 & 0.1002 & 0.5384 & 5.1653 \\
		\multirow{2}{*}{
			$h = \frac{1}{5}$
		}& exact+CP1 &  0.3014 & 0.1619 & 0.6479 &0.0658 & 0.5721 & 5.1883\\ 
		& exact+AB2 & 0.7651 & 0.3095 & 0.8671 & 0.8826 & 1.9070 & 4.9994 \\
		\multirow{2}{*}{
			$T = 20$
		} 
		& exact+RK2 & 0.5804 & 0.2522 & 0.9074 & 0.1593 & 0.7891 & 5.2612 \\
		& exact+CP2 & 0.0733 & 0.0667 & 0.2919 & 0.0711 & 0.2761 & 4.9678 \\
		& exact+SS2 & 0.3157 & 0.1567 & 0.6700 & 0.0992 & 0.5433 & 5.2206 \\
		\hline 
	\end{tabular}
	\caption{The relative entropy $E(P_n)$, first Wasserstein distance $W(P_n)$ and $\overline{\Delta \bx_n}$ between the numerical distribution $P_n$ and the reference distribution. The numerical distributions are calculated by various integration methods that use exact interpolation as shown in the second column. The first column contains the Stokes number $St$, time step $h$ and simulation time $T$ used in the six simulations. The first row contains the aspect ratio $\lambda$ of the particle shape. Values with a $-$ mean that the numerical solution is unstable. } 
	\label{table:integration}
\end{table}

For the $St = 1$, $\lambda = 1/10$ simulation, the CP1 solution has a larger $\overline{\Delta \bx_n}$ than the SS2 solution, but a lower $W(P_n)$ and $E(P_n)$ which suggests that for these simulation parameters, the centrifuge-preserving property is more advantageous for producing more accurate distributions than simply reducing the accuracy of the method in the conventional sense. In this simulation, the CP1 method is the second best in all measures, the best being the CP2 method. It is also noteworthy that in other simulations the CP1 method has roughly equal, and sometimes lower $\overline{\Delta \bx_n}$, $W(P_n)$ and $E(P_n)$ than the SS2 solution, which further suggests that the centrifuge-preserving property is advantageous.

One of the most remarkable observations is made for the $St = 10$, $\lambda = 1/10$ experiment. Here, the values of $\overline{\Delta \bx_n}$ are quite severe and roughly the same for all methods, due to the fact that the time step is quite large and the non-spherical ODEs are more stiff. Despite this, the contractivity preserving methods have a much lower $W(P_n)$ and $E(P_n)$ and the centrifuge preserving methods are better still. This highlights the fact that preserving the aforementioned physical features in the numerical solution results in spatial distributions that more closely resemble the reference solution, despite having the same $\overline{\Delta \bx_n}$. 

Finally, we mention that all the splitting schemes have better stability properties and are still able to produce accurate clusters of particles for low Stokes numbers and reasonably large time steps as noted by the $\lambda = 1/10$ simulations for $St = 1/5$ and $St = 1$ where we begin to see some of the conventional methods losing stability. 

\subsubsection{Comparison of interpolation methods}

\begin{figure}
	\centering
	
	\begin{subfigure}{0.32\textwidth}
		\includegraphics[width=\linewidth]{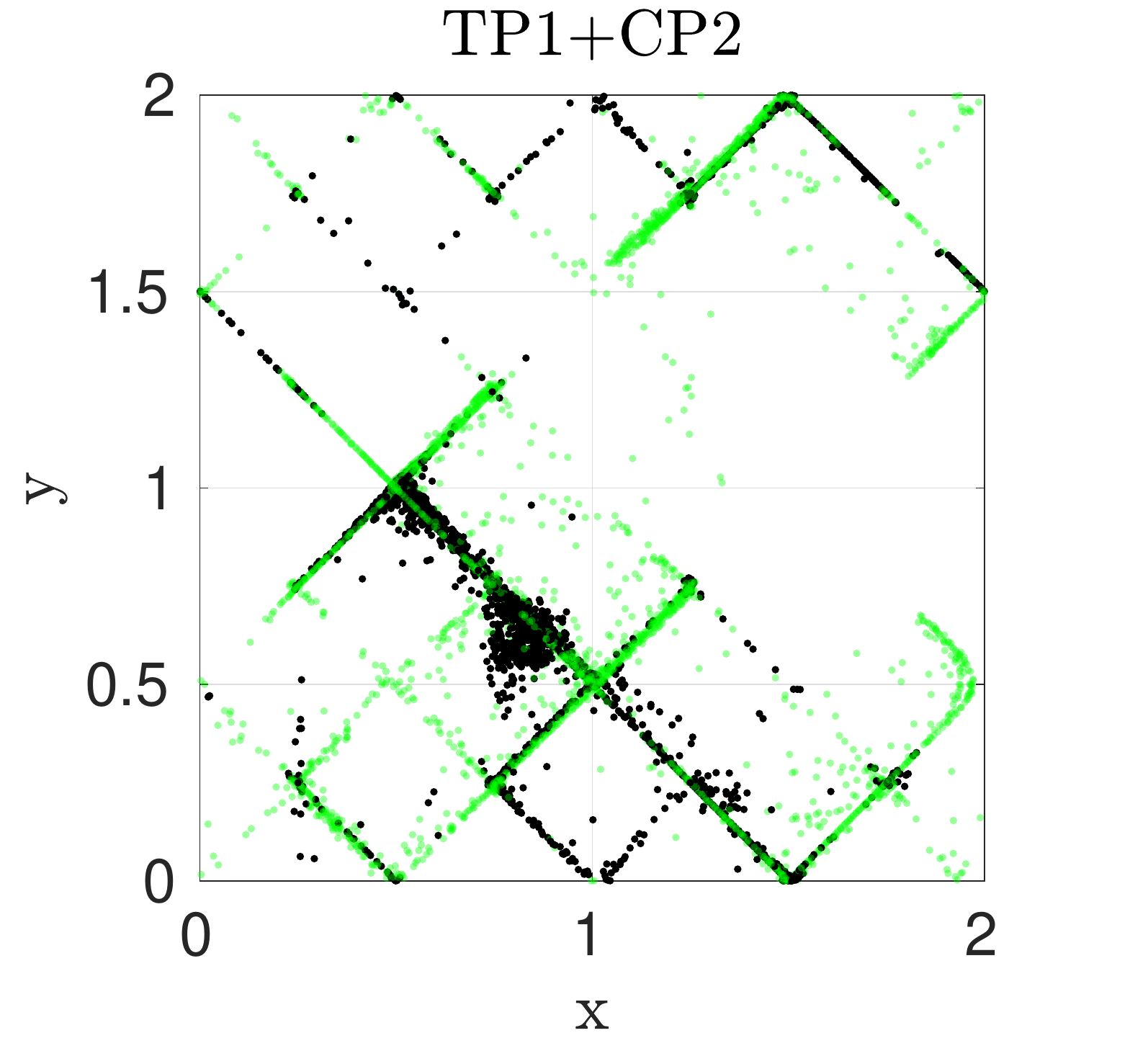}
		\caption{}\label{ii1}
	\end{subfigure}
	\begin{subfigure}{0.32\textwidth}
		\includegraphics[width=\linewidth]{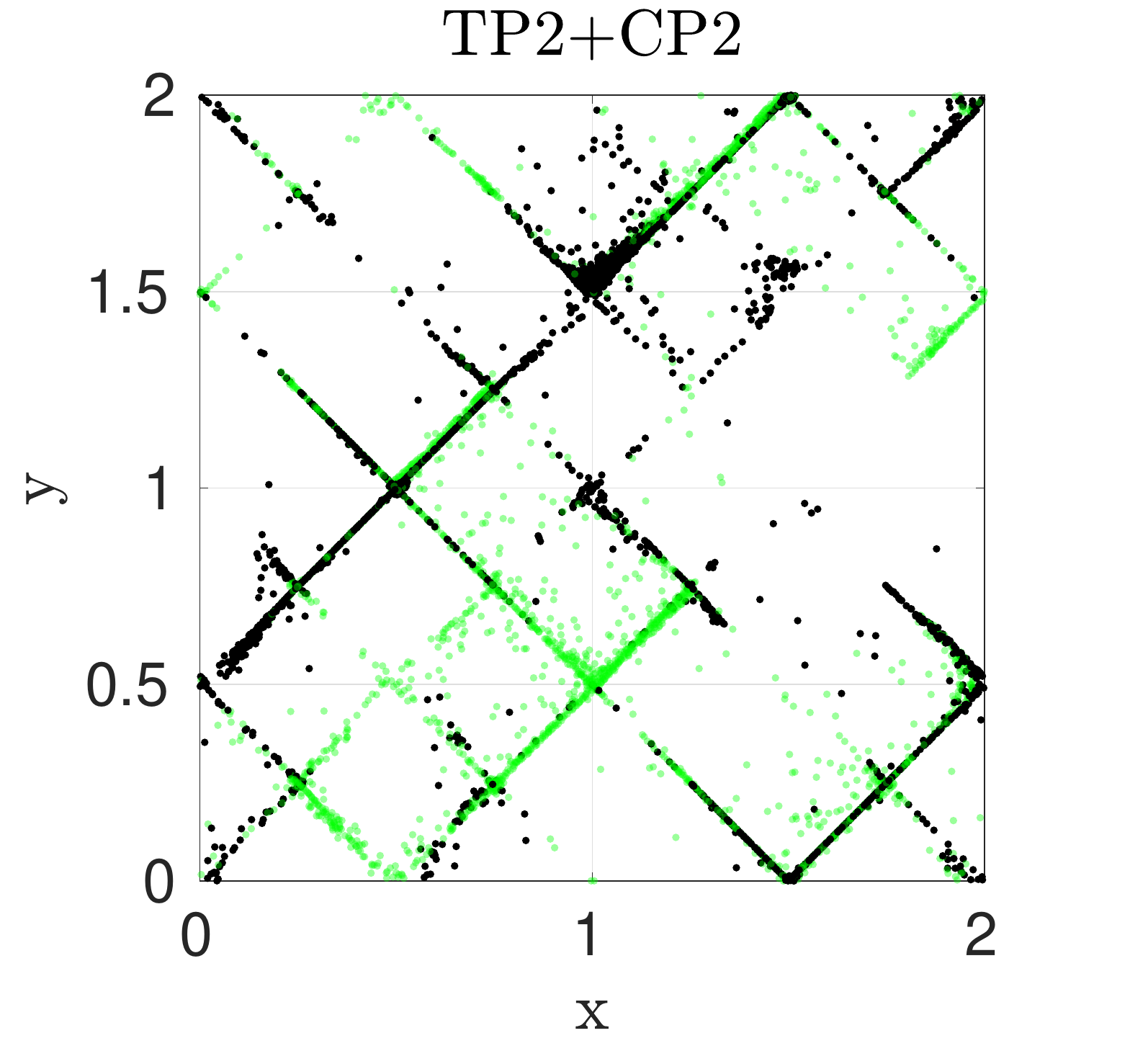}
		\caption{}\label{ii2}
	\end{subfigure}
	\begin{subfigure}{0.32\textwidth}
		\includegraphics[width=\linewidth]{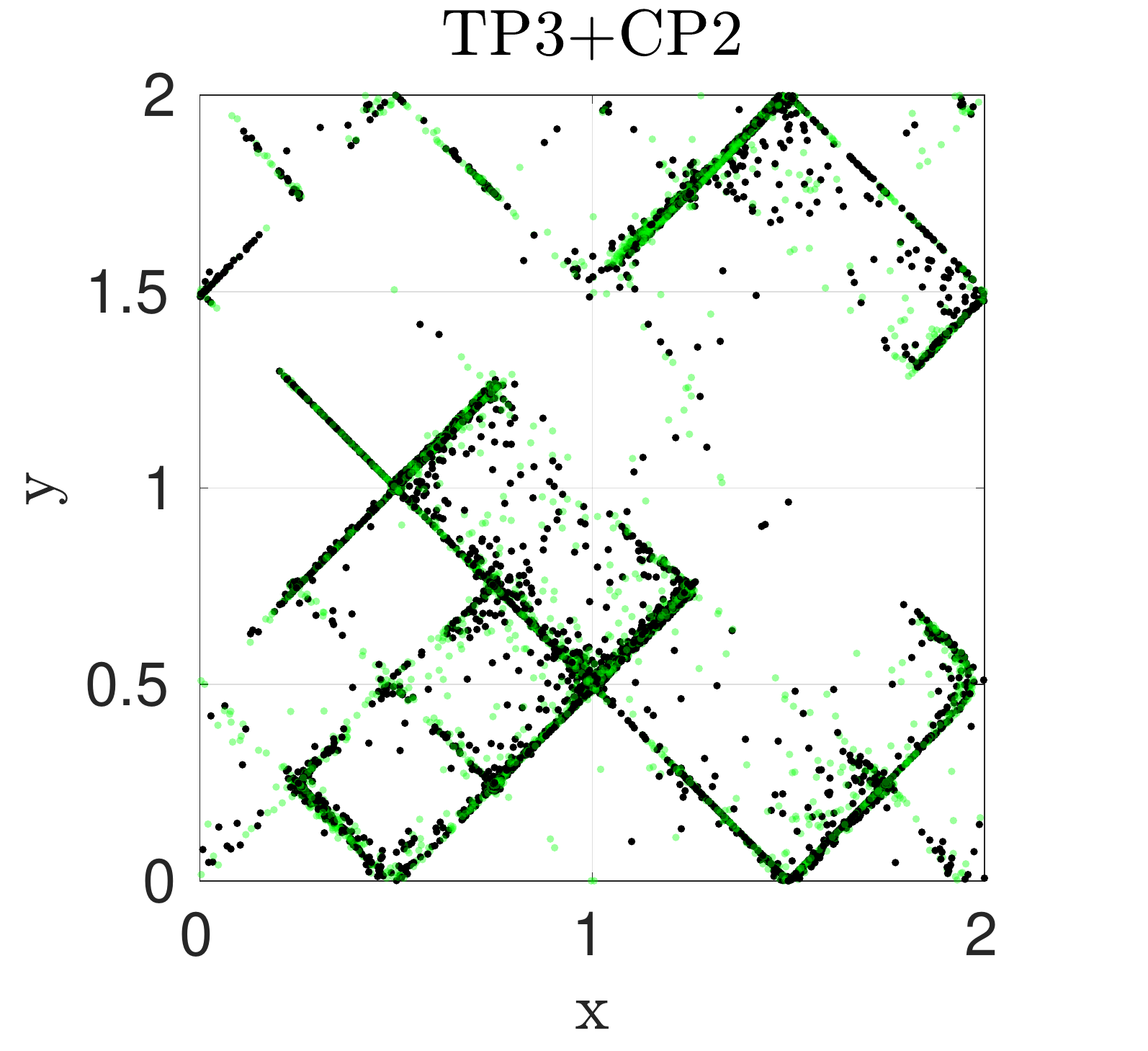}
		\caption{}\label{ii3}
	\end{subfigure}

	\begin{subfigure}{0.32\textwidth}
		\includegraphics[width=\linewidth]{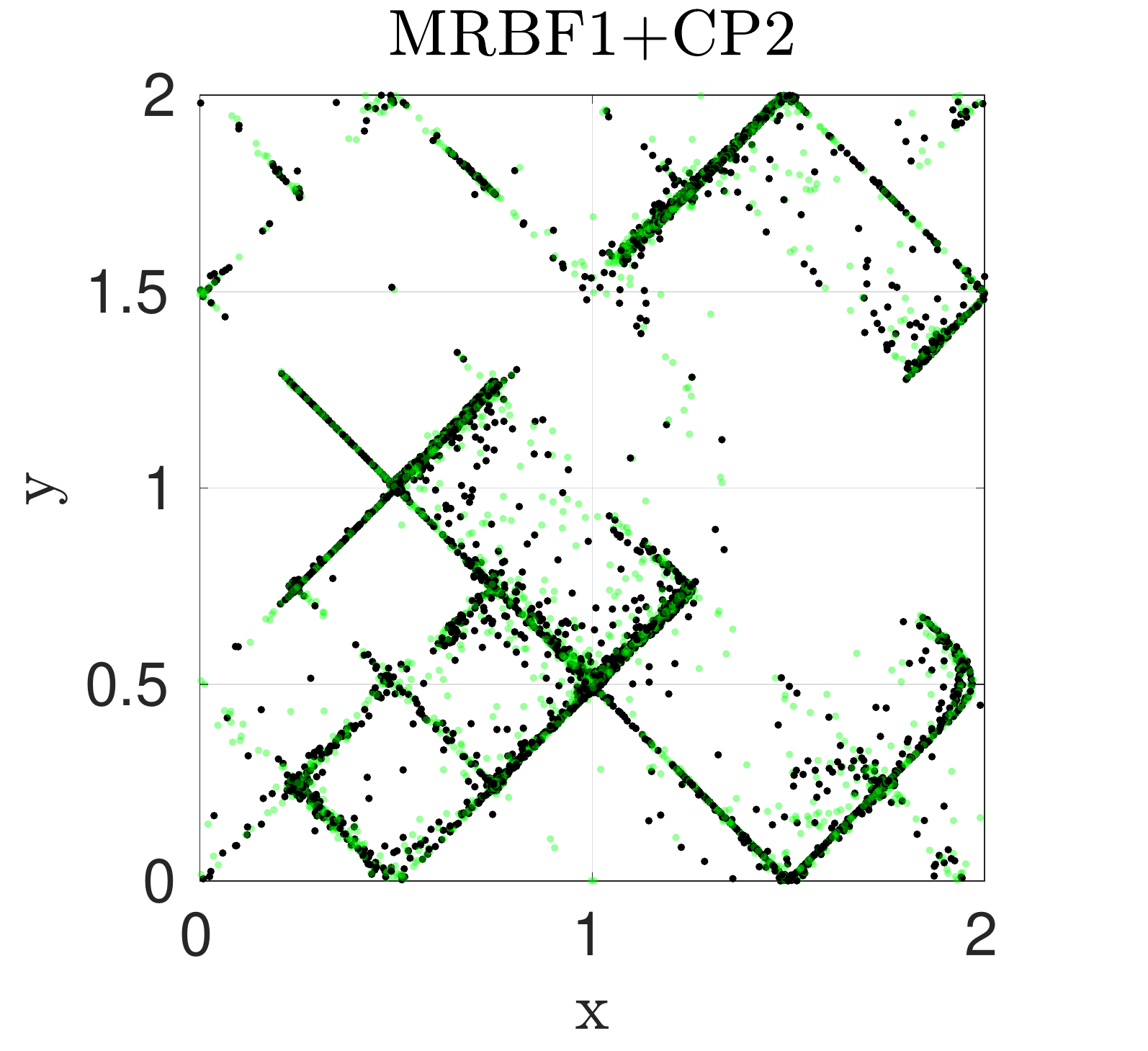}
		\caption{}\label{ii4}
	\end{subfigure}
	\begin{subfigure}{0.32\textwidth}
		\includegraphics[width=\linewidth]{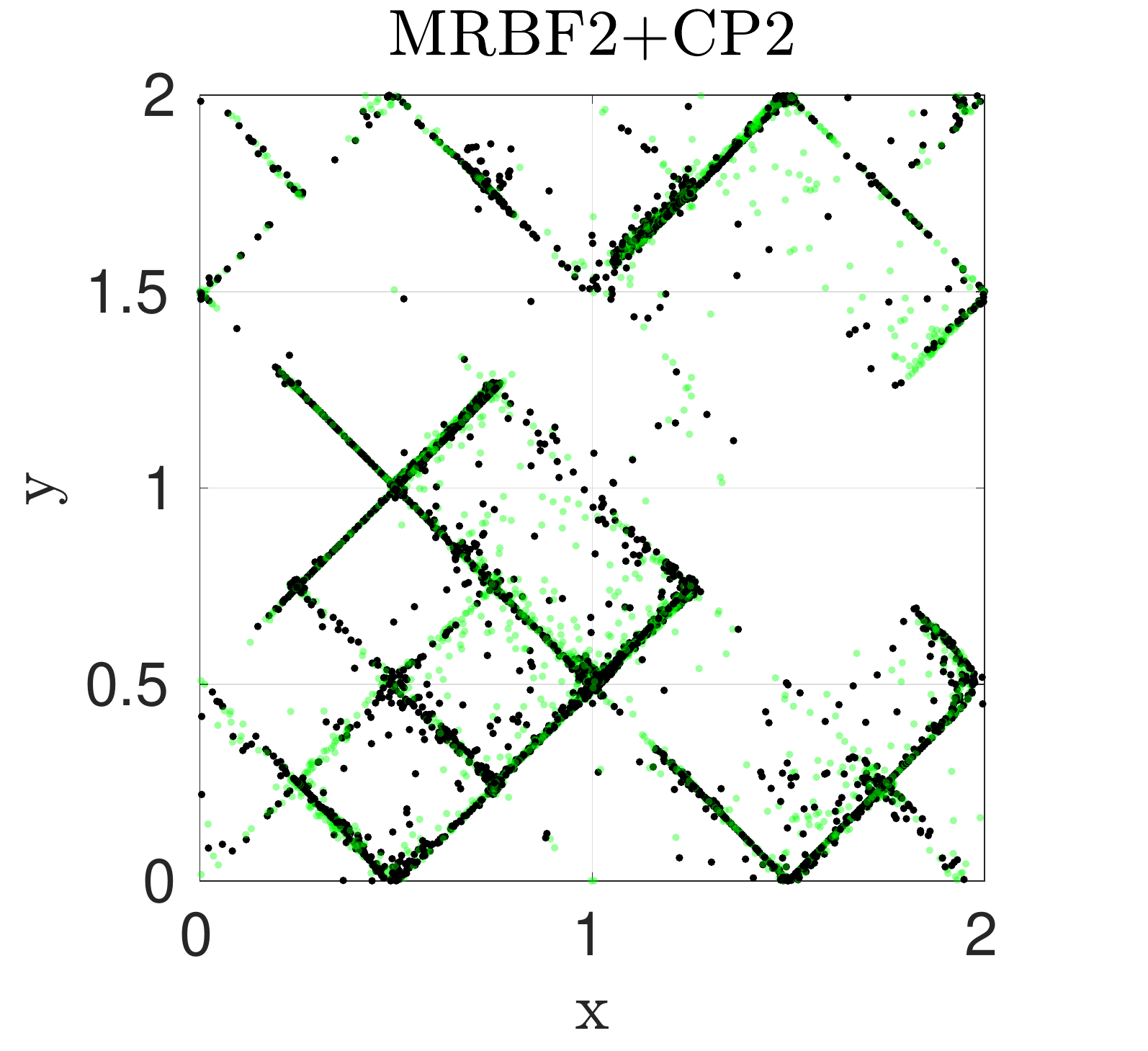}
		\caption{}\label{ii5}
	\end{subfigure}
	\begin{subfigure}{0.32\textwidth}
		\includegraphics[width=\linewidth]{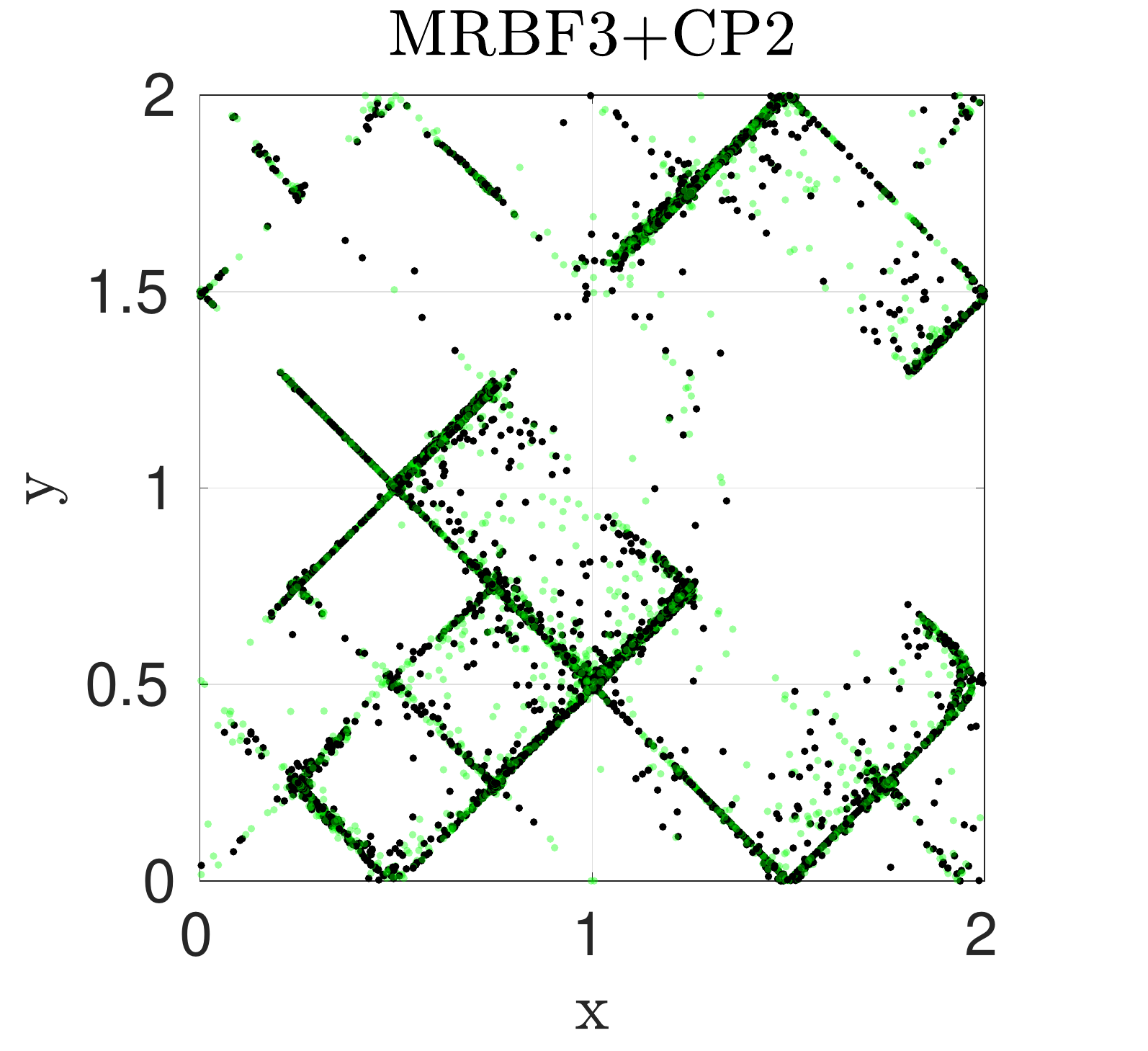}
		\caption{}\label{ii6}
	\end{subfigure}
	
	\begin{subfigure}{0.32\textwidth}
		\includegraphics[width=\linewidth]{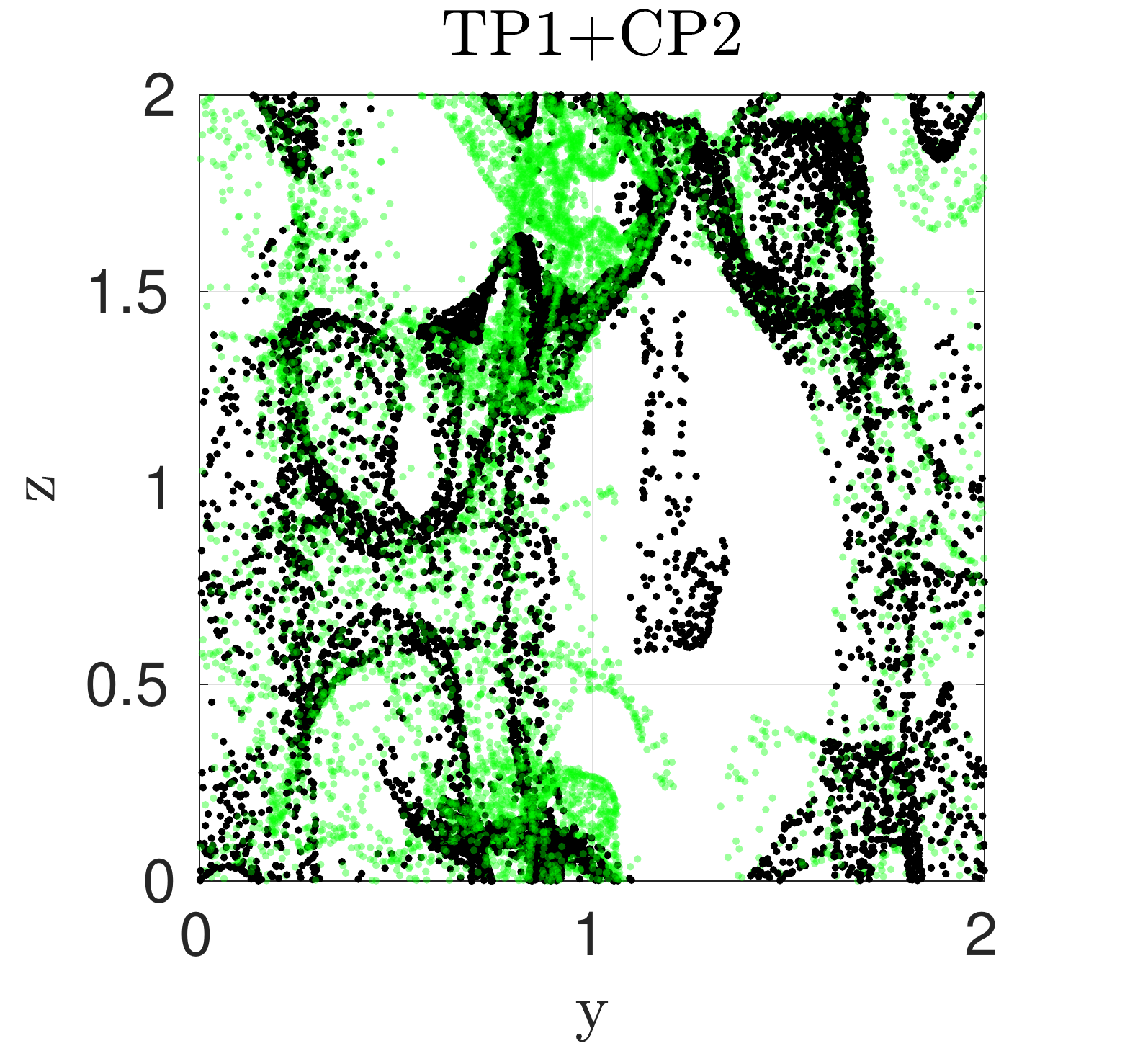}
		\caption{}\label{ii7}
	\end{subfigure}
	\begin{subfigure}{0.32\textwidth}
		\includegraphics[width=\linewidth]{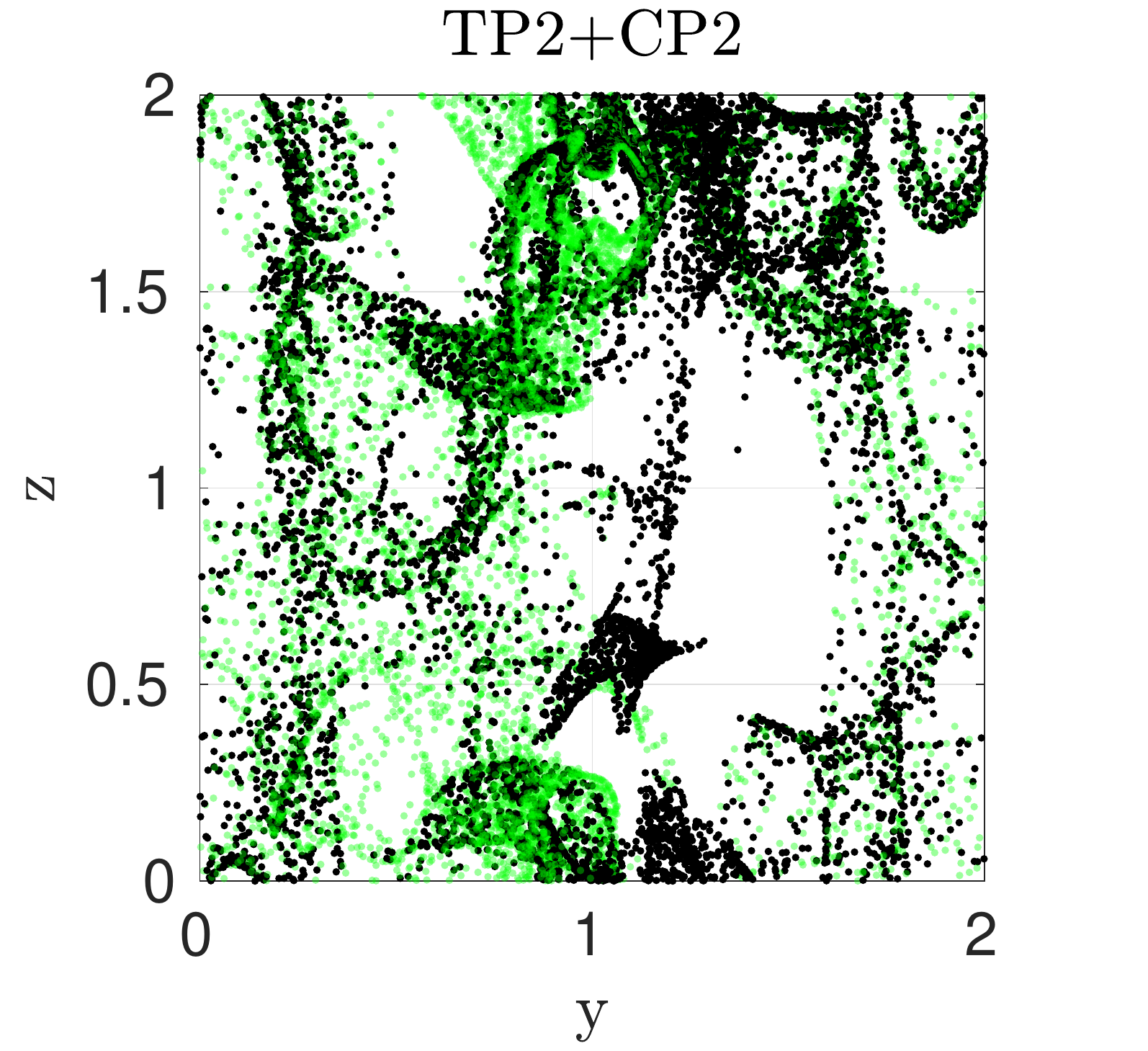}
		\caption{}\label{ii8}
	\end{subfigure}
	\begin{subfigure}{0.32\textwidth}
		\includegraphics[width=\linewidth]{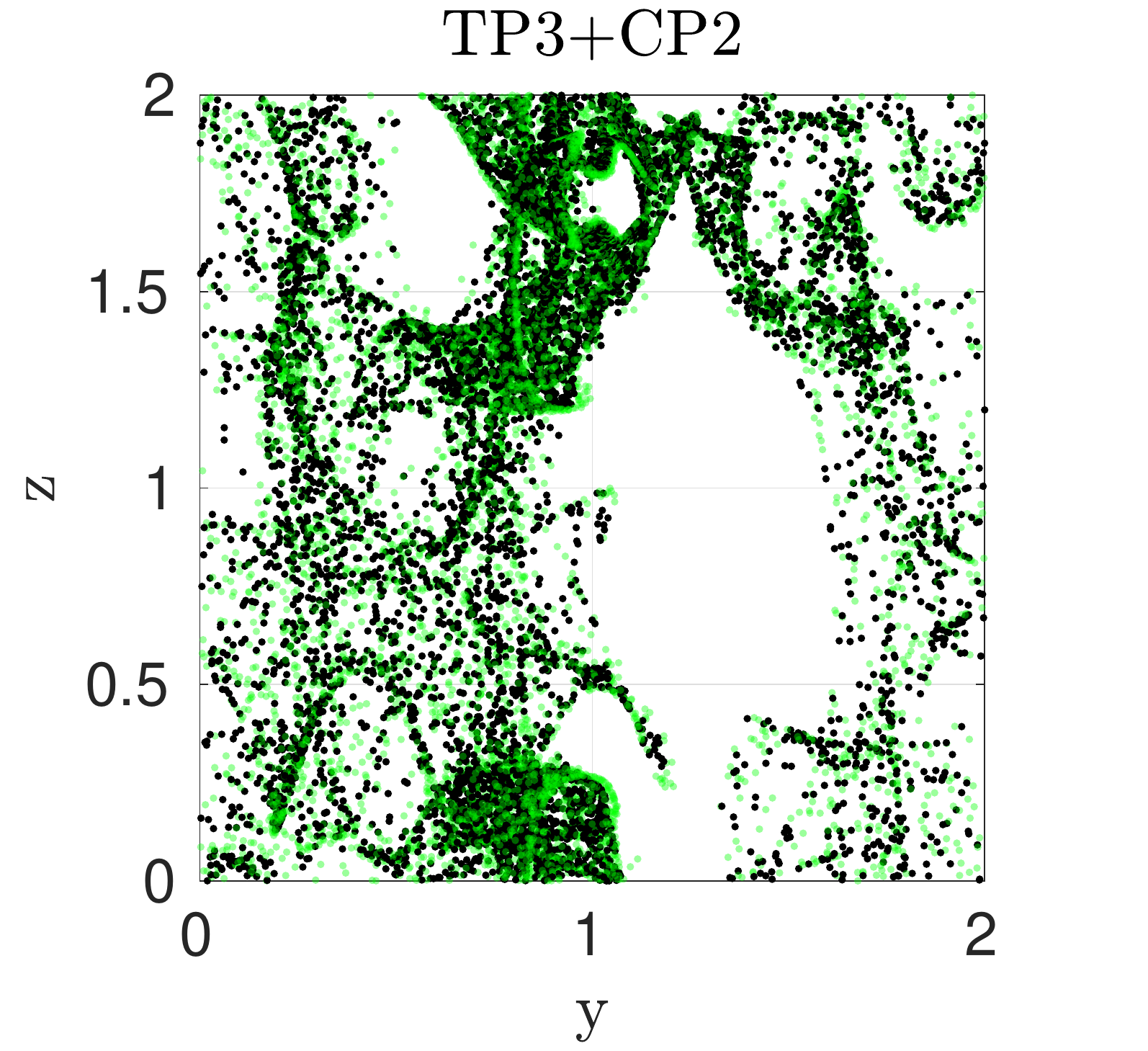}
		\caption{}\label{ii9}
	\end{subfigure}

	\begin{subfigure}{0.32\textwidth}
		\includegraphics[width=\linewidth]{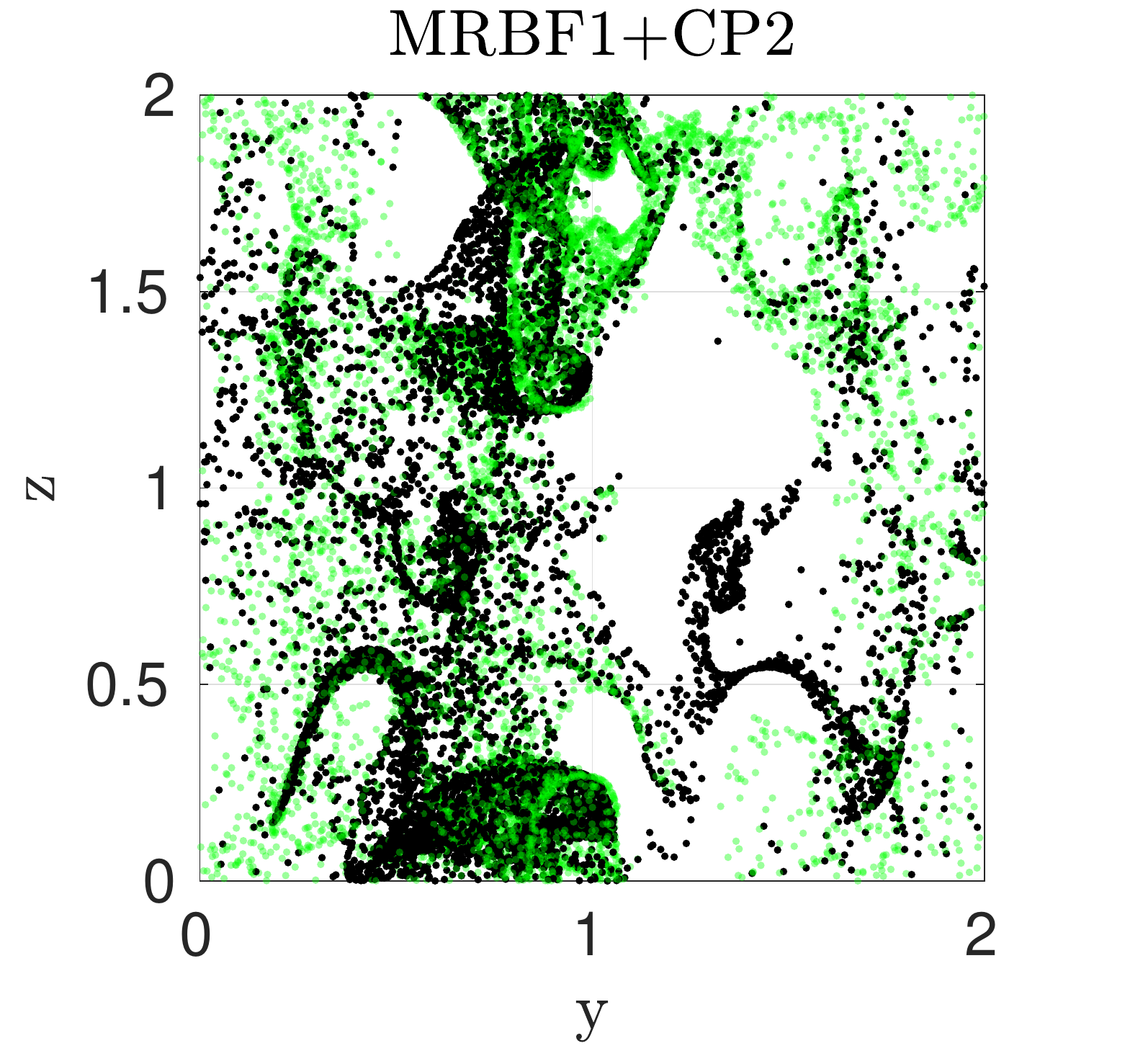}
		\caption{}\label{ii10}
	\end{subfigure}
	\begin{subfigure}{0.32\textwidth}
		\includegraphics[width=\linewidth]{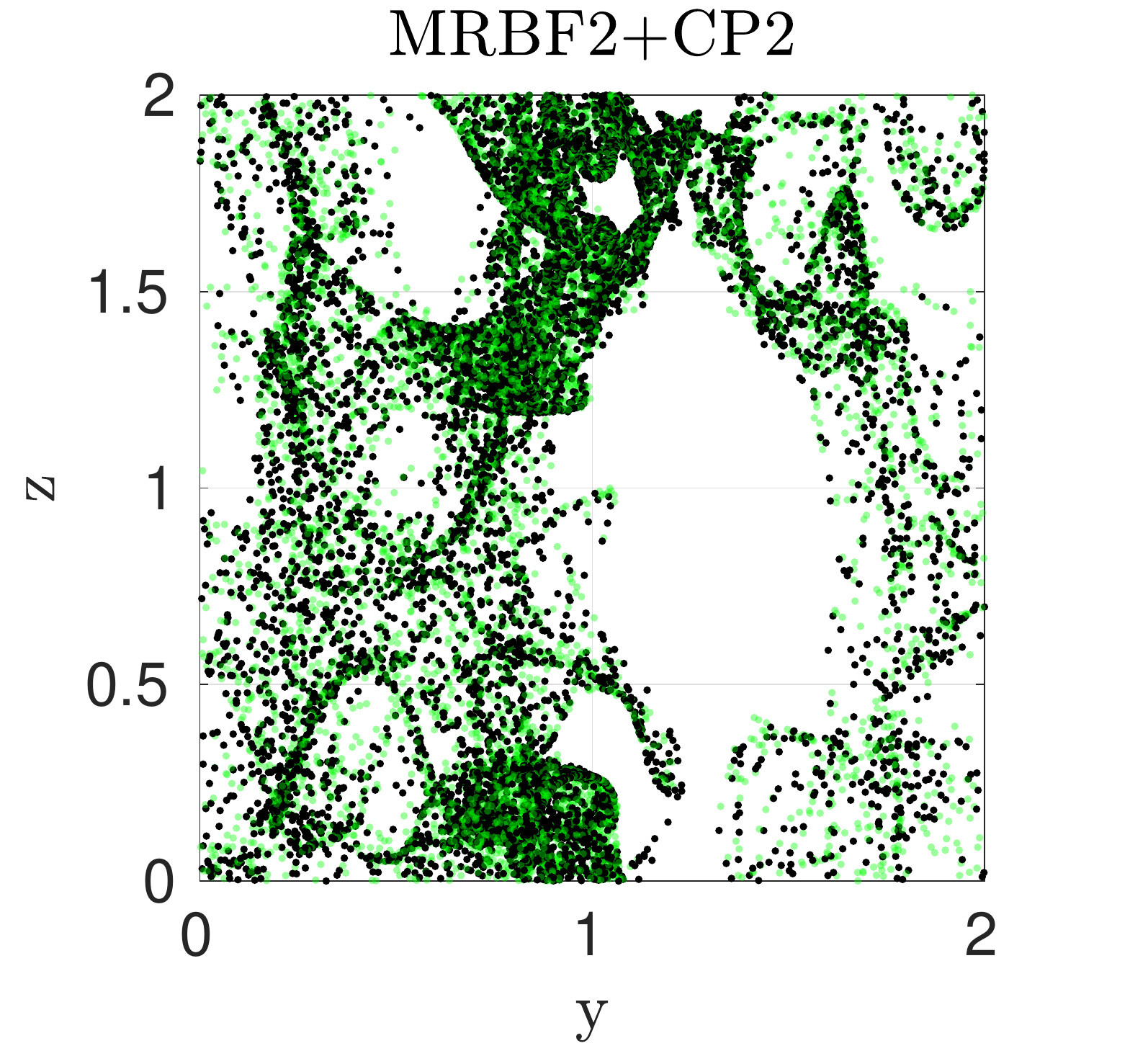}
		\caption{}\label{ii11}
	\end{subfigure}
	\begin{subfigure}{0.32\textwidth}
		\includegraphics[width=\linewidth]{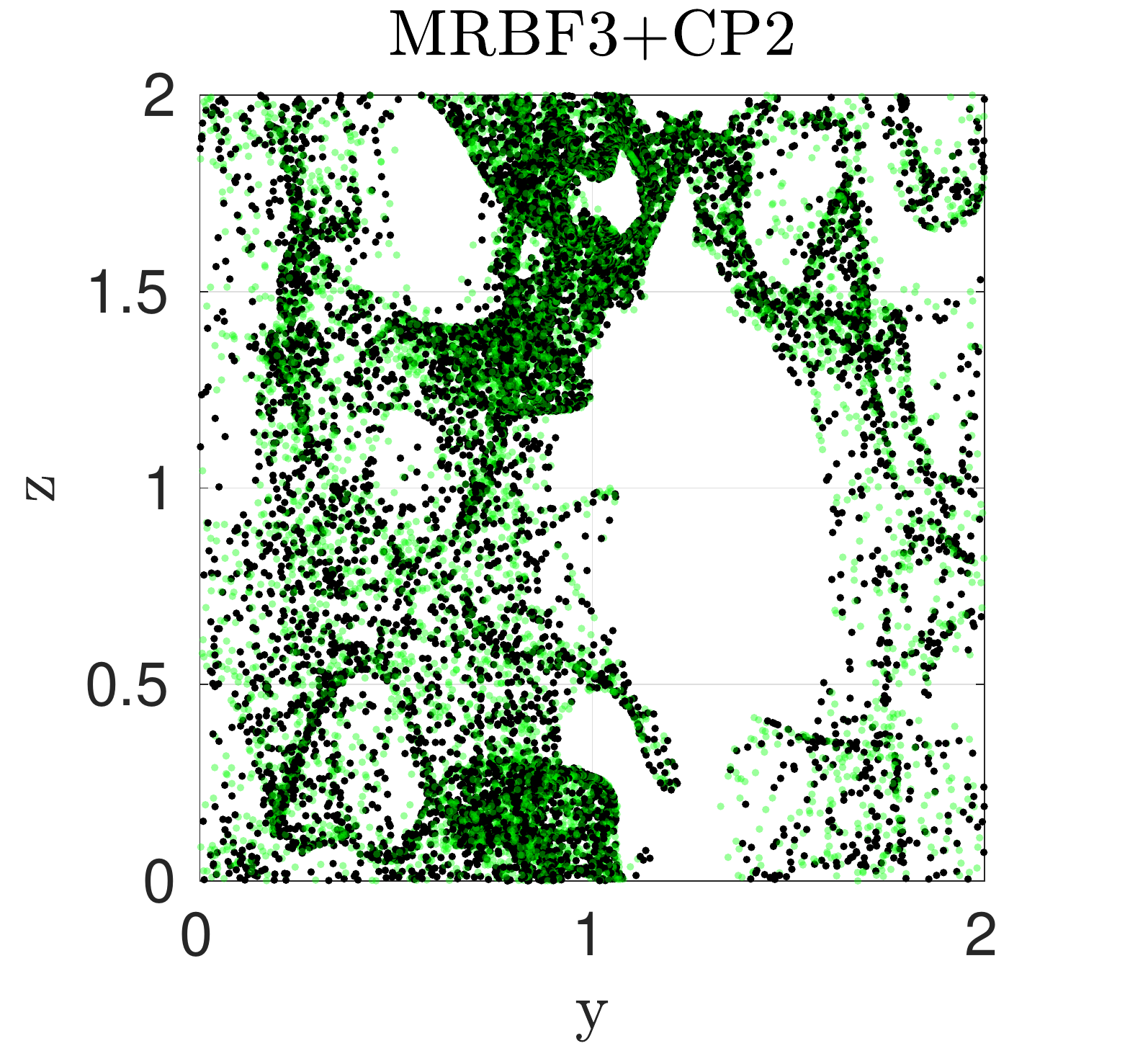}
		\caption{}\label{ii12}
	\end{subfigure}
	\caption{Figures (a) through (f) show the spatial distribution of the particles in the $x-y$ plane for the $St = 1/10$, $h = 1/100$, $T = 6$, $\lambda = 1$ simulation from table \ref{table:interpolation}. Figures (g) through (l) show the spatial distribution of the particles in the $y-z$ plane for the $St = 1$, $h = 1/40$, $T = 12$, $\lambda = 1/10$ simulation. The reference solution is plotted in green in all figures.}
	\label{interpolation_figs}
\end{figure}

In this experiment we compare the MRBF and TP interpolation methods in combination with the CP2 method to simulate a suspension of particles evolving in the Taylor-Green flow. Six separate simulations are performed, three with Stokes numbers of $St = 1/10,1,10$ for spherical particles ($\lambda = 1$) and three with the same Stokes numbers for non-spherical particles ($\lambda = 5$). At the end of each simulation the average spatial error $\overline{\Delta \bx_n}$, relative entropy $E(P_n)$ and the first Wasserstein distance $W(P_n)$ between the numerical distribution and the reference distribution are calculated and the results are presented in table \ref{table:interpolation}. The time step $h$ and total simulation time $T$ are also presented in this table. Some spatial distributions produced by the different interpolation methods are presented in figure \ref{interpolation_figs}.

Directing our attention towards figures \ref{ii1} to \ref{ii6}, which show the final distribution of the $St = 1/10$, $\lambda = 5$ simulation looking down the $z$-axis. It can be seen here that the three MRBF solutions look visually very similar to the reference solution, as does the TP3 solution. If we look towards the corresponding part of table \ref{table:interpolation}, we see that the TP3 solution has a $\overline{\Delta \bx_n}$ of 0.3468, which is lower than the MRBF1 solution, which has a $\overline{\Delta \bx_n}$ of 0.5389. Despite this, the MRBF1 solution, which we note performs exceptionally well here, has a lower $E(P_n)$ and $W(P_n)$ meaning that the final distribution is more similar to the reference distribution even though the $\overline{\Delta \bx_n}$ is greater.

Figures \ref{ii7} to \ref{ii12} show the final distribution of the $St = 1$, $\lambda = 1$ simulation looking down the $x$-axis. Here we see that the TP1, TP2 and MRBF1 solutions differ visually from the reference solution, whilst the TP3, MRBF2 and MRBF3 solutions look quite similar. From the corresponding section of table \ref{table:interpolation}, we see that the MRBF2 solution's $\overline{\Delta \bx_n}$ is 0.5004 compared to the TP3 solution, which is 0.3332, but both have a similar $E(P_n)$ and $W(P_n)$. Furthermore, there are many examples here of the MRBF solutions having higher $\overline{\Delta \bx_n}$, but lower $E(P_n)$ and $W(P_n)$. This can be seen in all three $\lambda = 1$ simulations, where the MRBF2 solution has larger $\overline{\Delta \bx_n}$ than the TP3 solution but similar or lower $E(P_n)$ and $W(P_n)$. For the $\lambda = 5$ simulations and for $St = 1$ and $St = 10$, the MRBF2 solution outperforms the TP3 solution in all three measures. Such examples indicate that preserving the divergence-free condition is important to acheive accurate spatial distributions. 

We now make some general observations about table \ref{table:interpolation}. We see that in all but one simulation, the MRBF1 solutions outperform the TP2 solution in all three measures. It is noteworthy that the MRBF1 solution is as fast as TP1 interpolation where both require only eight data points for the interpolation as opposed to 27 data points for the TP2 interpolation, which is a slower method. Additionally, in all six simulations the MRBF3 interpolation method outperforms all the TP solutions in all measures. 

To summarize, we have seen many examples of the MRBF solutions producing distributions that are more similar to the reference solution than the TP solutions, despite having worse $\overline{\Delta \bx_n}$. These observations are consistent with the fact that the CP2 method, among others, loses accuracy in $\Delta\X^{[n]}$ when evolving particles in a non-divergence-free flow field. That is, the physical volume $\X^{[n]}$ is more strongly affected when the divergence-free condition is broken, despite the fact that the order of accuracy of the method remains unaffected. 

\begin{table}[h]
	\centering
	\begin{tabular}{|c|c|ccc|ccc|}
		\hline
		& & \multicolumn{3}{c|}{$\lambda = 1$} & \multicolumn{3}{c|}{$\lambda = 5$} \\
		\hline
		& $P_n$	  & $E(P_n)$ & $W(P_n)$ & $\overline{\Delta \bx_n}$ & $E(P_n)$ & $W(P_n)$ & $\overline{\Delta \bx_n}$ \\
		
		\hline		
		\multirow{2}{*}{
			$St = \frac{1}{10}$
		} 
		& MRBF1+CP2 & 0.1285 & 0.0860 & 0.3063 & 0.0514 & 0.0478 & 0.5389  \\ 
		& MRBF2+CP2 & 0.1002 & 0.0388 & 0.1256 & 0.0967 & 0.0657 & 0.3578 \\
		\multirow{2}{*}{
			$h = \frac{1}{100}$
		}
		
		& MRBF3+CP2 & 0.0380 & 0.0164 & 0.0614 & 0.0337 & 0.0242 & 0.1732  \\ 
		& TP1+CP2 & 7.9636 & 0.7427 & 0.7957 & 3.1871 & 0.5657 & 1.1261  \\
		\multirow{2}{*}{
			$T = 4$
		}
		& TP2+CP2 & 3.7166 & 0.2975 & 0.3959 & 1.5779 & 0.3409 & 0.8653 \\
		& TP3+CP2 & 0.3349 & 0.0625 & 0.1197 & 0.0778 & 0.0704 & 0.3468 \\

		\hline
		\multirow{2}{*}{
			$St = 1$
		} 
		& MRBF1+CP2 & 0.9861 & 0.5802 & 1.4746 & 0.0402 & 0.0965 & 1.3596 \\
		& MRBF2+CP2 & 0.0444 & 0.0439 & 0.5004 & 0.0351 & 0.0698 & 0.7334 \\
		\multirow{2}{*}{
			$h = \frac{1}{40}$
		} 
		& MRBF3+CP2 & 0.0367 & 0.0353 & 0.2442 & 0.0258 & 0.0564 & 0.3755  \\
		& TP1+CP2 & 1.6501 & 0.7003 & 1.5487 & 0.4281 & 0.3222 & 1.8162 \\
		\multirow{2}{*}{
			$T = 8$
		} 
		& TP2+CP2 & 1.5784 & 1.2164 & 1.8284 & 0.0585 & 0.1871 & 1.7269 \\
		& TP3+CP2 & 0.0404 & 0.0440 & 0.3332 & 0.0310 & 0.0714 & 0.8375 \\
		
		\hline
		
		\multirow{2}{*}{
			$St = 10$
		}  
		& MRBF1+CP2 & 1.5452 & 0.0979 & 0.1223 & 0.1401 & 0.0726 & 0.1998 \\
		& MRBF2+CP2 & 0.1071 & 0.0109 & 0.0154 & 0.0178 & 0.0183 & 0.0548 \\
		\multirow{2}{*}{
			$h = \frac{1}{10}$
		} 
		& MRBF3+CP2 & 0.0416 & 0.0041 & 0.0065 & 0.0112 & 0.0084 & 0.0216 \\
		& TP1+CP2  & 5.8812 & 0.1545 & 0.1674 & 0.6419 & 0.1957 & 0.5617 \\
		\multirow{2}{*}{
			$T = 12$
		} 
		& TP2+CP2 & 4.0693 & 0.1540 & 0.2427 & 0.0955 & 0.0537 & 0.2222 \\
		& TP3+CP2 & 0.7464 & 0.0135 & 0.0125 & 0.0339 & 0.0242 & 0.0946 \\
		\hline 
	\end{tabular}
	\caption{The relative entropy $E(P_n)$, first Wasserstein distance $W(P_n)$ and average error per particle $\overline{\Delta \bx_n}$ between the numerical distribution $P_n$ and the reference distribution. The numerical distributions are calculated by various interpolation methods that use CP2 integration as shown in the second column. The first column contains the Stokes number $St$, time step $h$ and simulation time $T$ used in the six simulations. The first row contains the aspect ratio $\lambda$ of the particle shape.} 
	\label{table:interpolation}
\end{table}

\subsection{Comparison of interpolation and integration methods}

Our final experiment explores the benefit that is gained by combining MRBF interpolation with the centrifuge- and contractivity-preserving methods compared to the standard methods used in the literature. We will compare the methods TP1+FE1, MRBF1+CP1, TP2+AB2, TP3+AB2, MRBF2+CP2 and TP2+CP2. The first two methods are the cheapest and are of roughly equal cost. The method TP2+AB$n$ are used in, for example \cite{Portela, challabotla2015orientation, van2000dynamics, PanBanerjee,wang1996large} and subsequent studies. We include TP3+AB2 to test whether increasing the interpolation accuracy is worthwhile use of computational resources. We also consider the MRBF2+CP2 solution, which is an accurate and economical combination of our proposed geometric methods. Finally, the TP2+CP2 method is considered to emphasize the negative implications of using a non-divergence-free interpolation method with the CP2 method. 

Six simulations are performed, three with Stokes numbers of $St = 1/10,1,10$ for spherical particles ($\lambda = 1$) and three with the same Stokes numbers for non-spherical particles ($\lambda = 10$). At the end of the simulation, the average spatial error $\overline{\Delta \bx_n}$, relative entropy $E(P_n)$ and the first Wasserstein distance $W(P_n)$ between the numerical distribution and the reference distribution are computed and presented in table \ref{table:interpolation} along with the time step and total simulation times used. 

\begin{figure}
	\centering
	
	\begin{subfigure}{0.32\textwidth}
		\includegraphics[width=\linewidth]{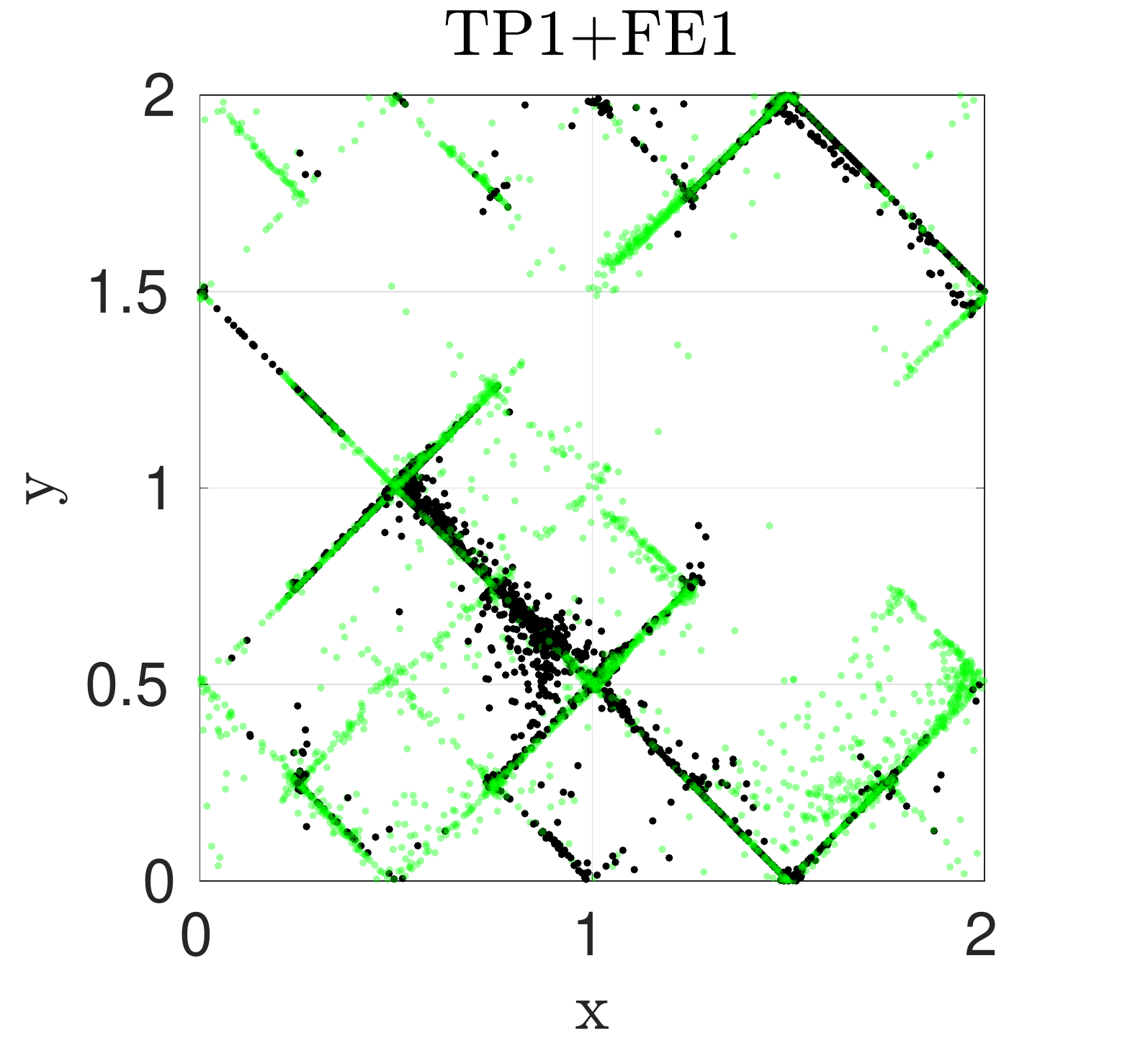}
		\caption{}\label{c1}
	\end{subfigure}
	\begin{subfigure}{0.32\textwidth}
		\includegraphics[width=\linewidth]{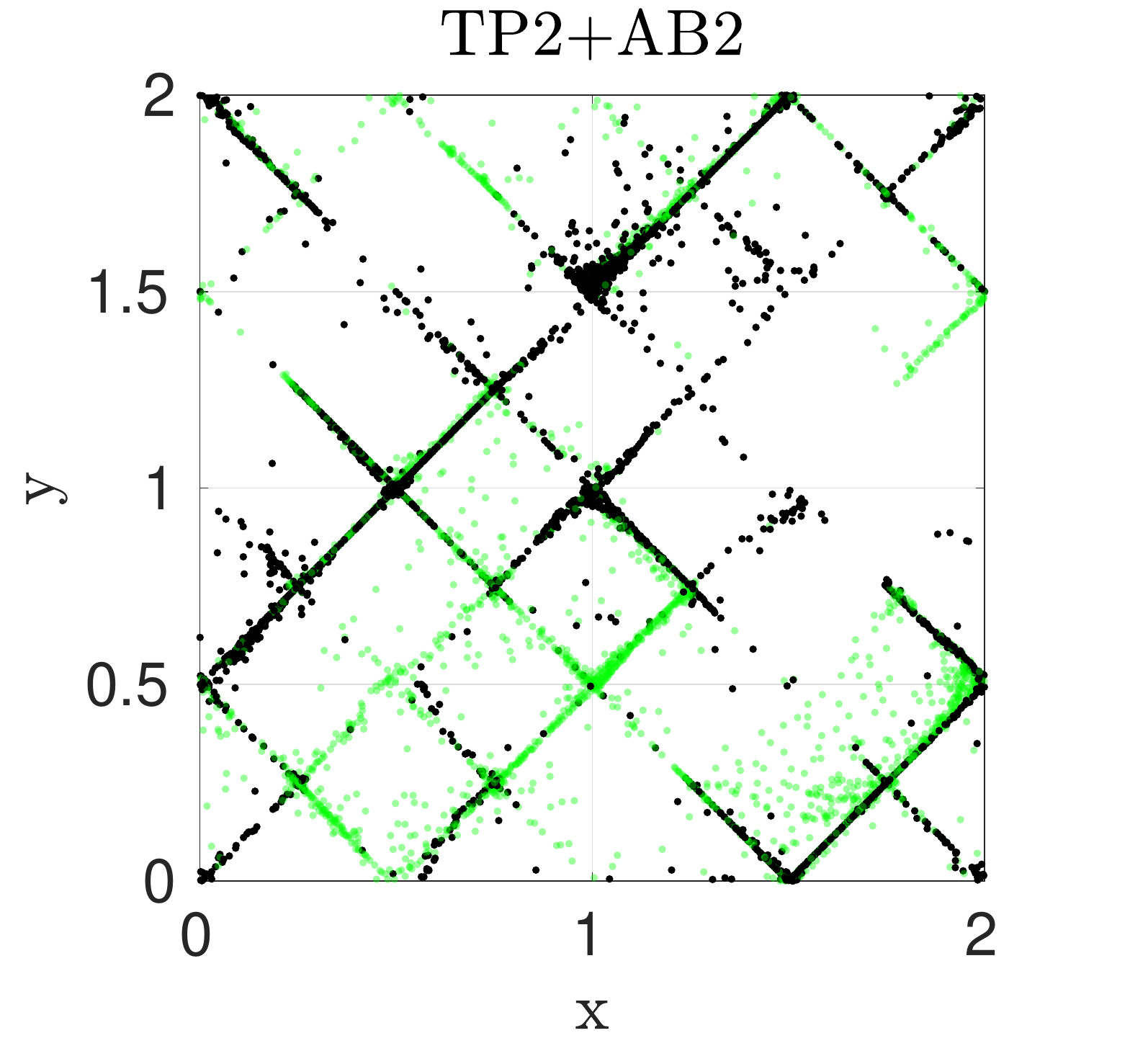}
		\caption{}
	\end{subfigure}
	\begin{subfigure}{0.32\textwidth}
		\includegraphics[width=\linewidth]{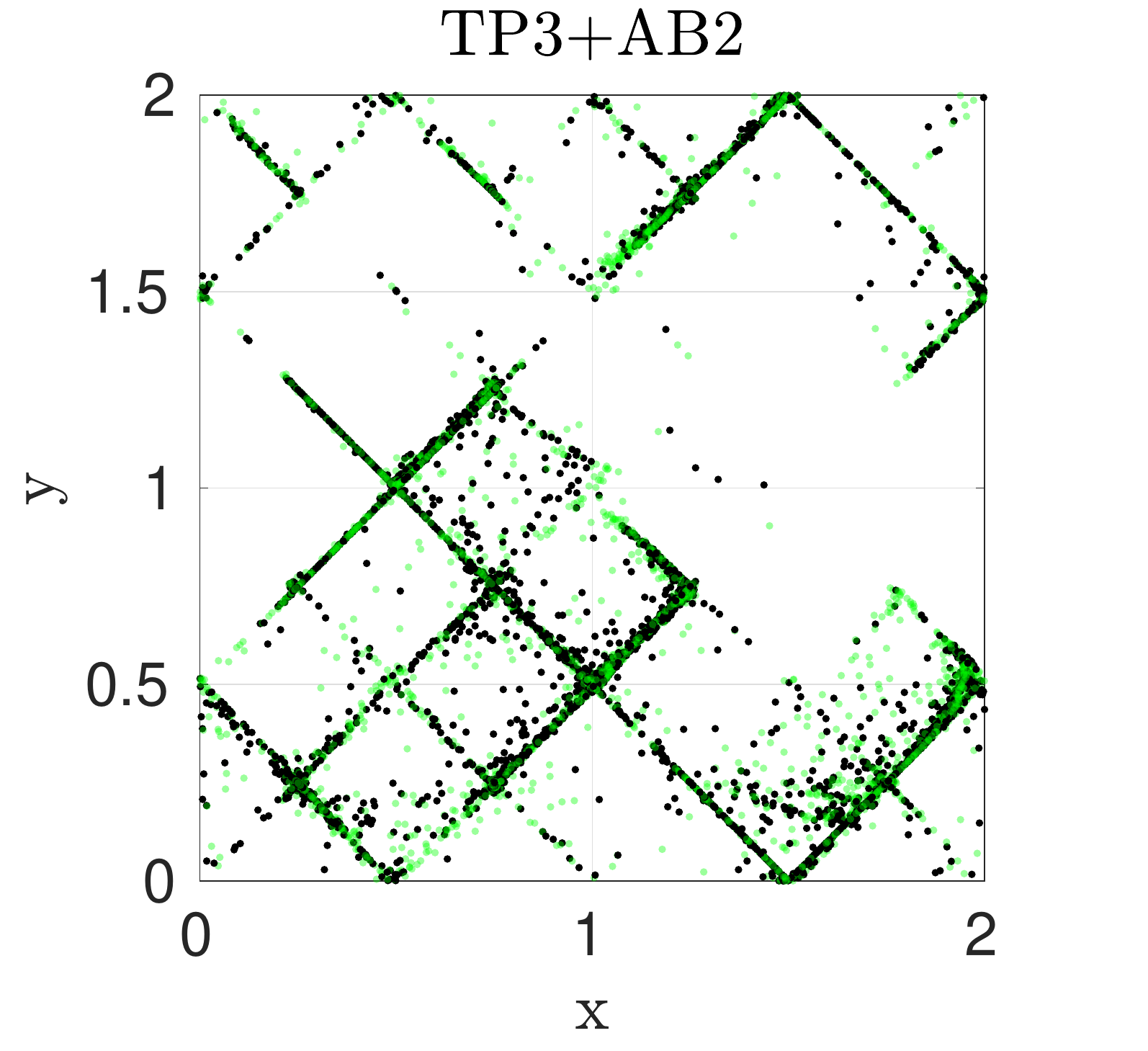}
		\caption{}
	\end{subfigure}

	\begin{subfigure}{0.32\textwidth}
		\includegraphics[width=\linewidth]{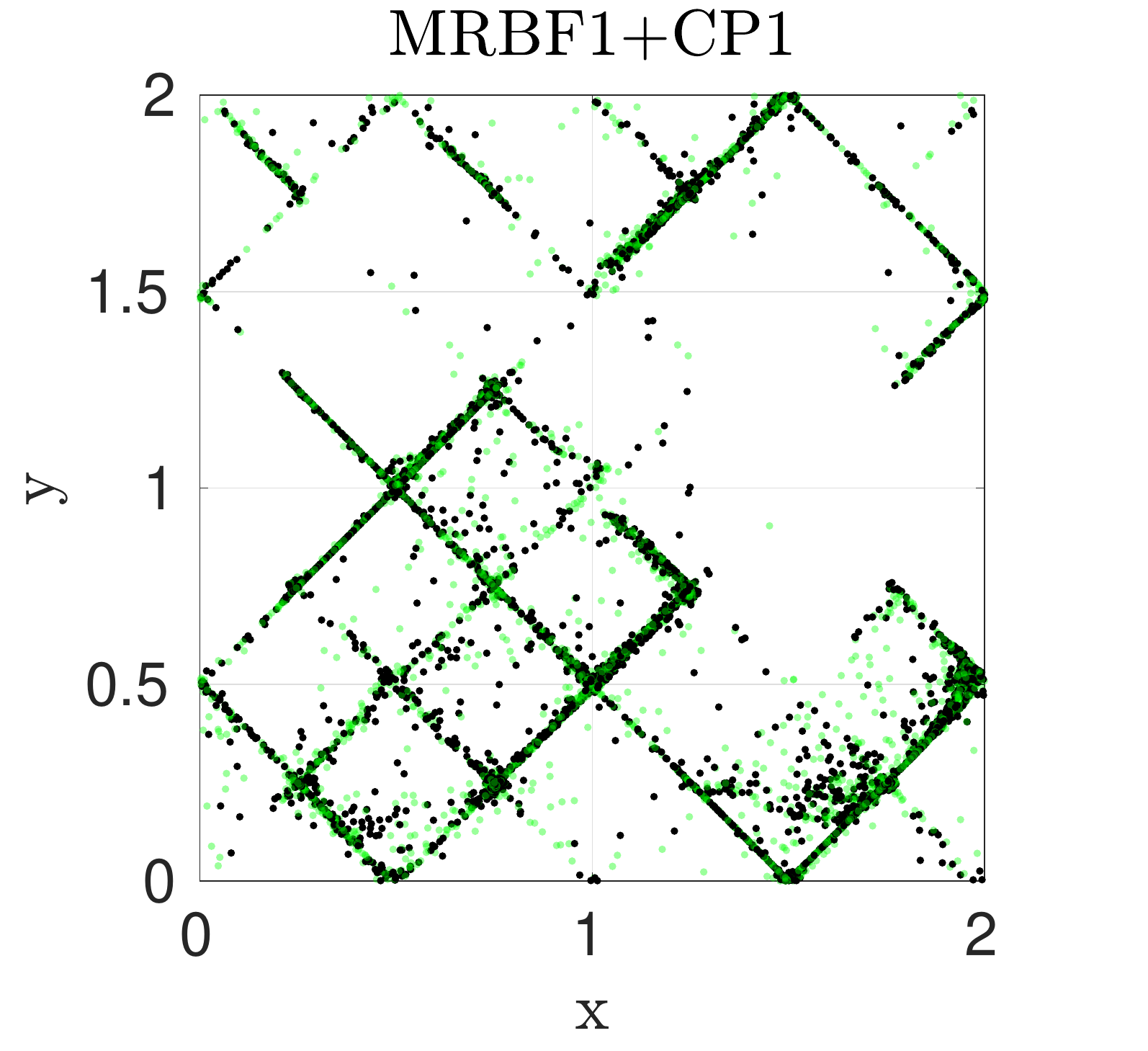}
		\caption{}
	\end{subfigure}
	\begin{subfigure}{0.32\textwidth}
		\includegraphics[width=\linewidth]{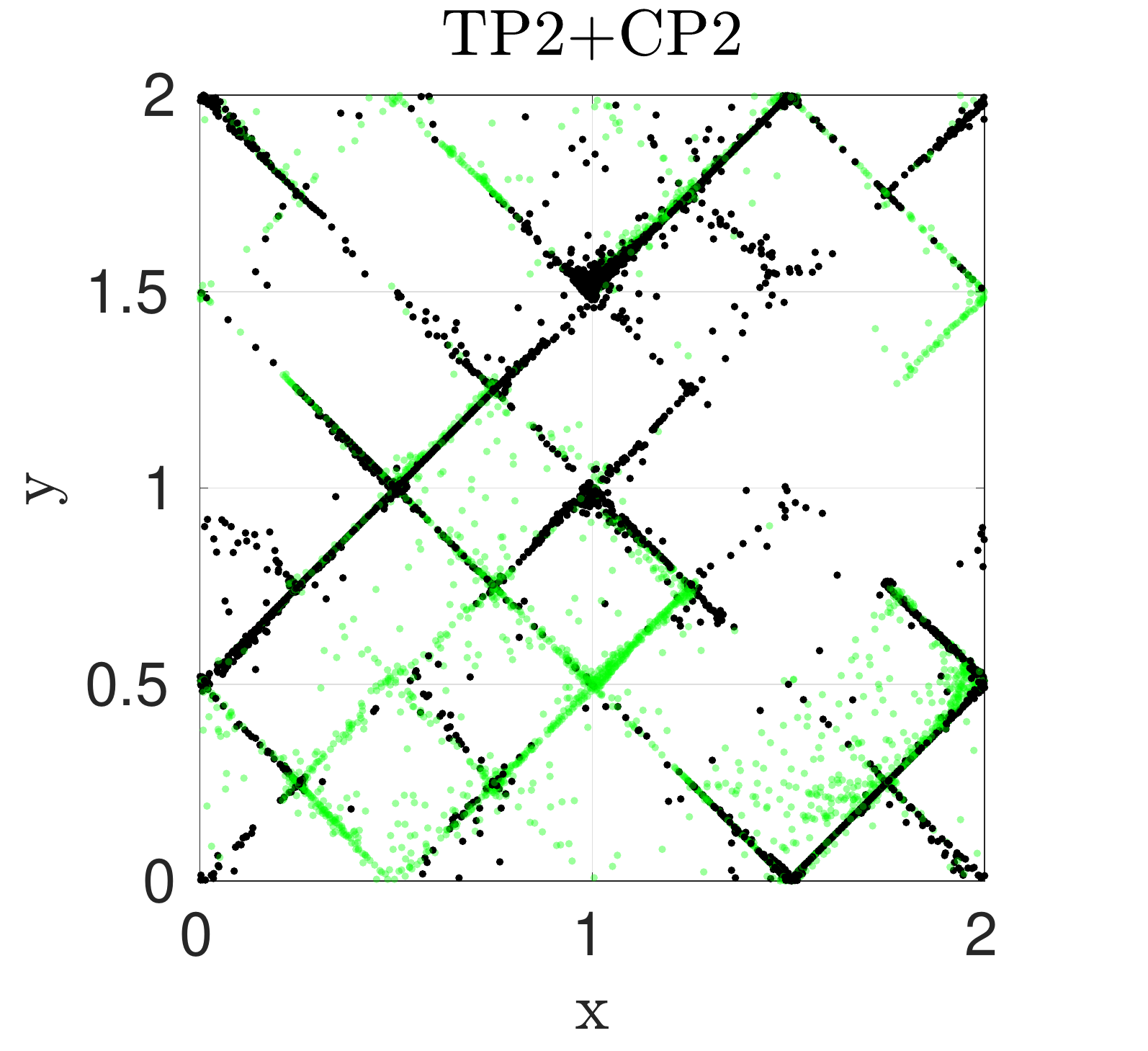}
		\caption{}
	\end{subfigure}
	\begin{subfigure}{0.32\textwidth}
		\includegraphics[width=\linewidth]{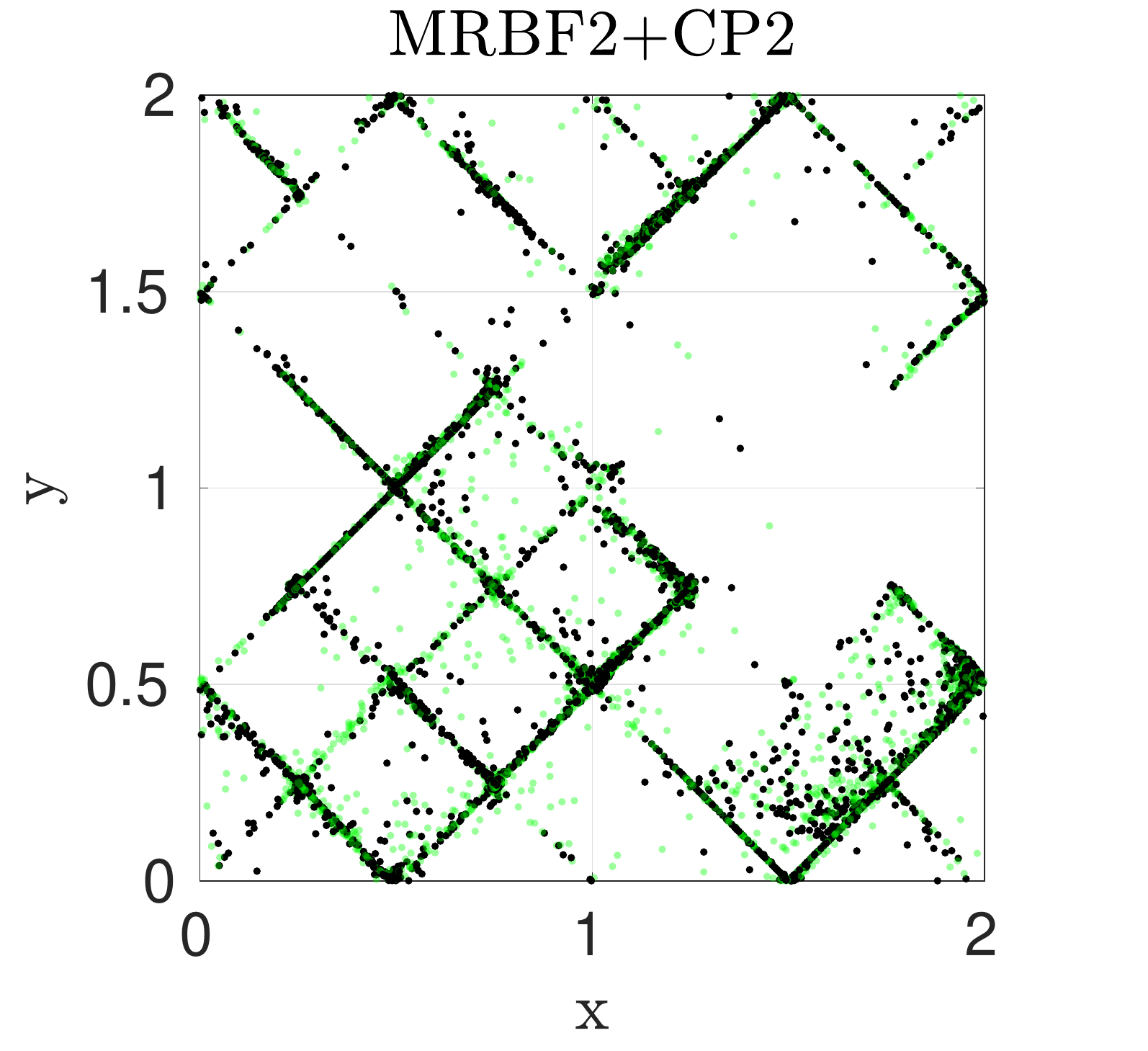}
		\caption{}\label{c6}
	\end{subfigure}
	
	\begin{subfigure}{0.32\textwidth}
		\includegraphics[width=\linewidth]{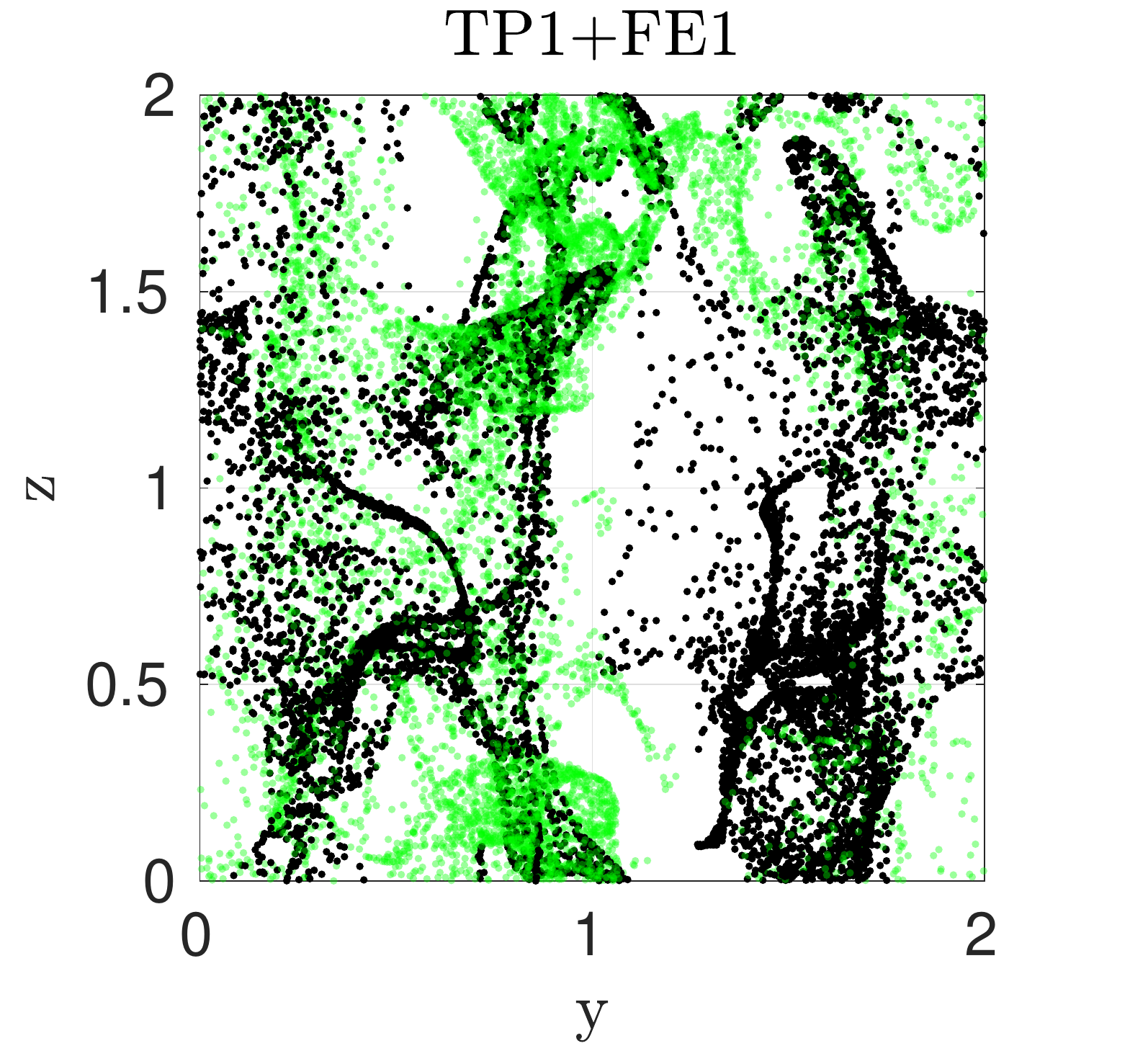}
		\caption{}\label{c7}
	\end{subfigure}
	\begin{subfigure}{0.32\textwidth}
		\includegraphics[width=\linewidth]{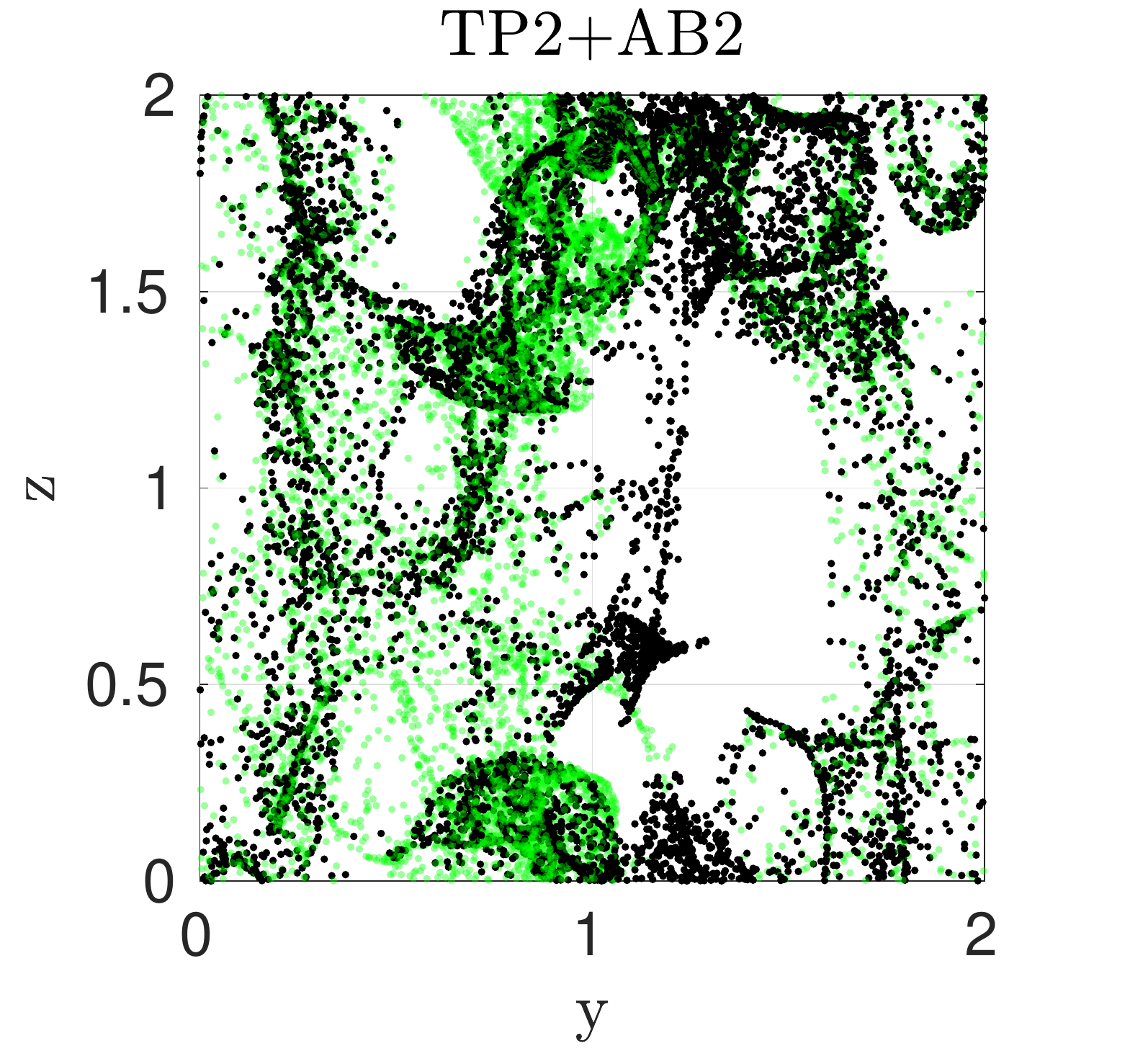}
		\caption{}
	\end{subfigure}
	\begin{subfigure}{0.32\textwidth}
		\includegraphics[width=\linewidth]{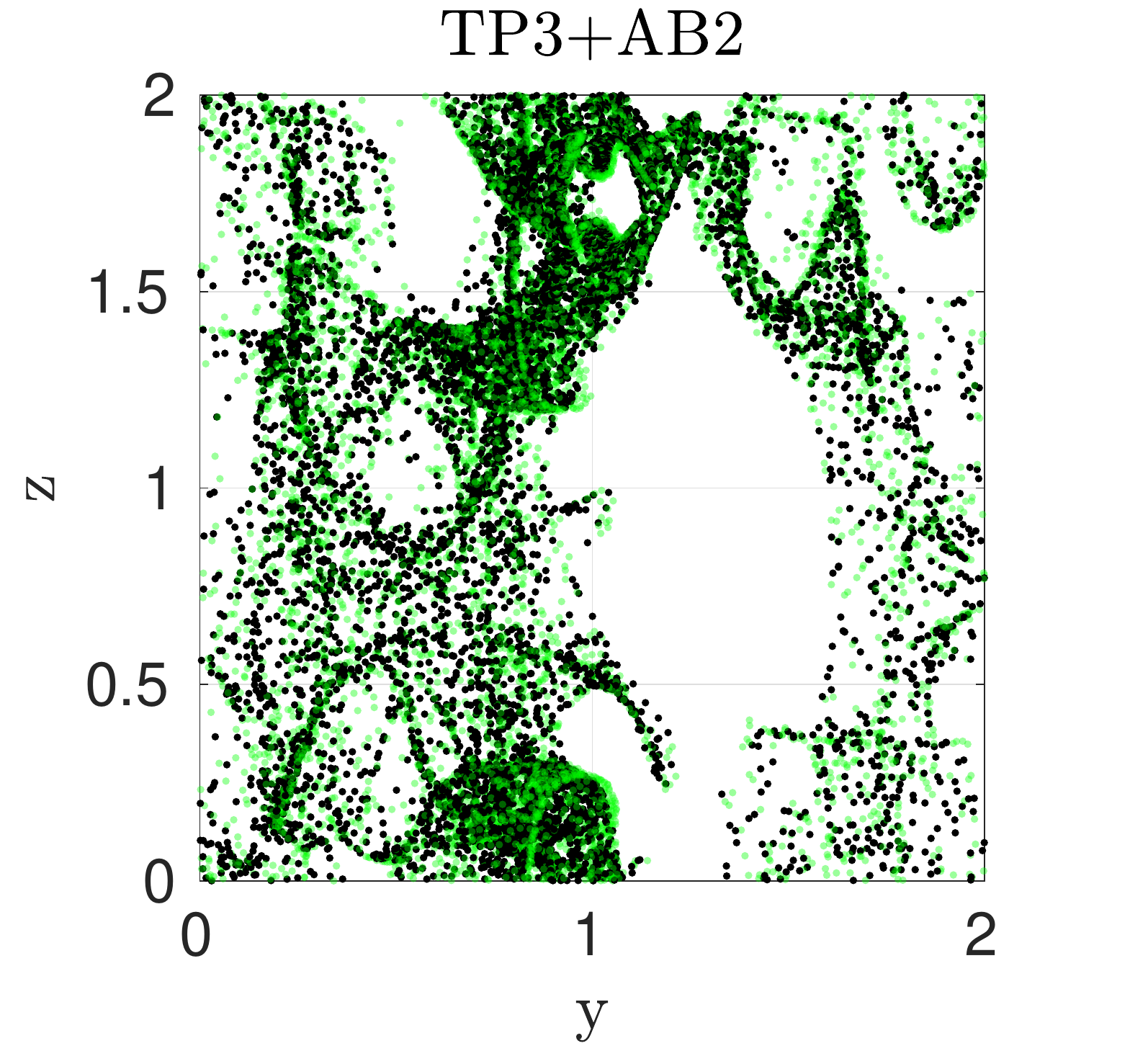}
		\caption{}
	\end{subfigure}

	\begin{subfigure}{0.32\textwidth}
		\includegraphics[width=\linewidth]{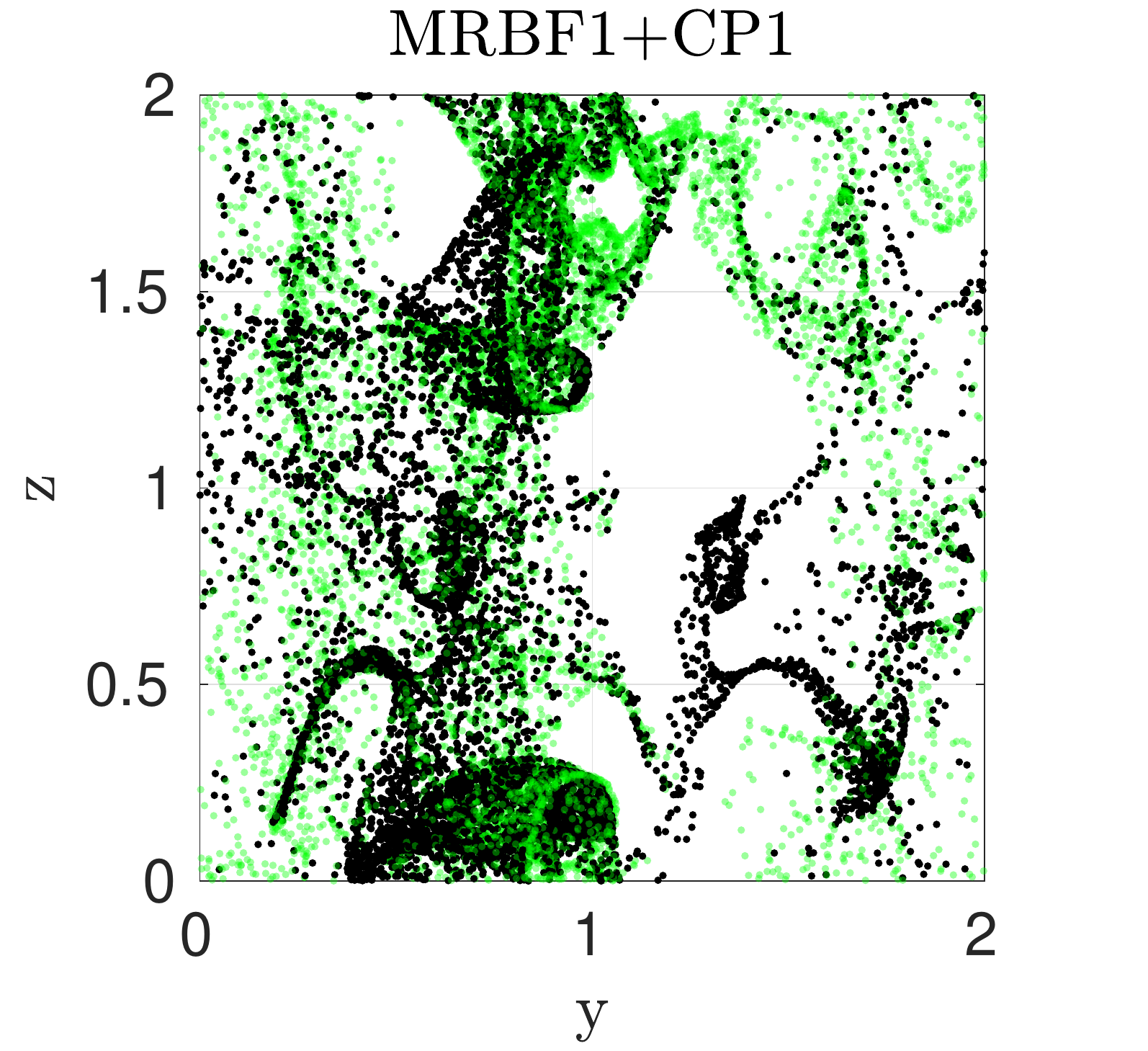}
		\caption{}
	\end{subfigure}
	\begin{subfigure}{0.32\textwidth}
		\includegraphics[width=\linewidth]{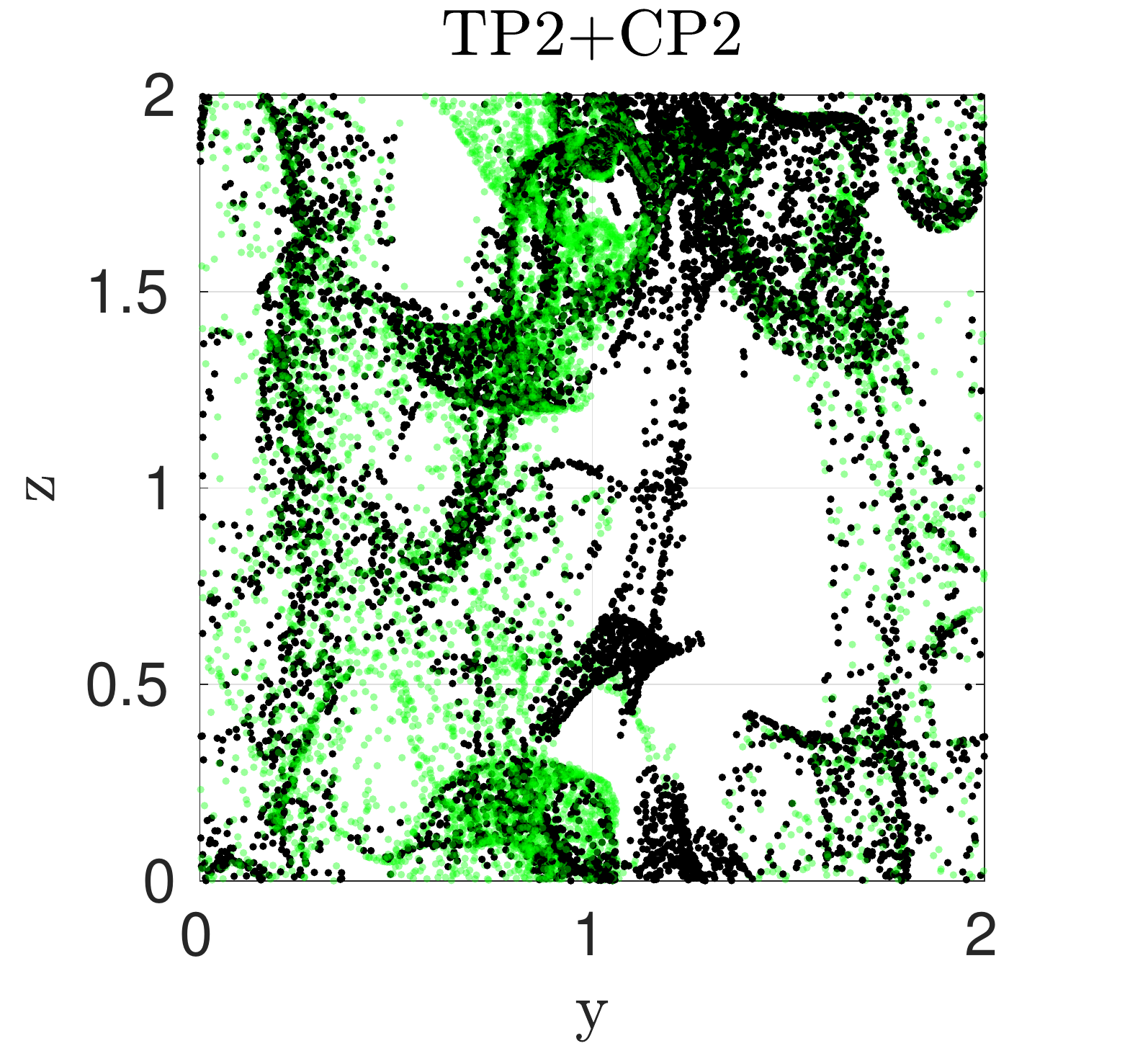}
		\caption{}
	\end{subfigure}
	\begin{subfigure}{0.32\textwidth}
		\includegraphics[width=\linewidth]{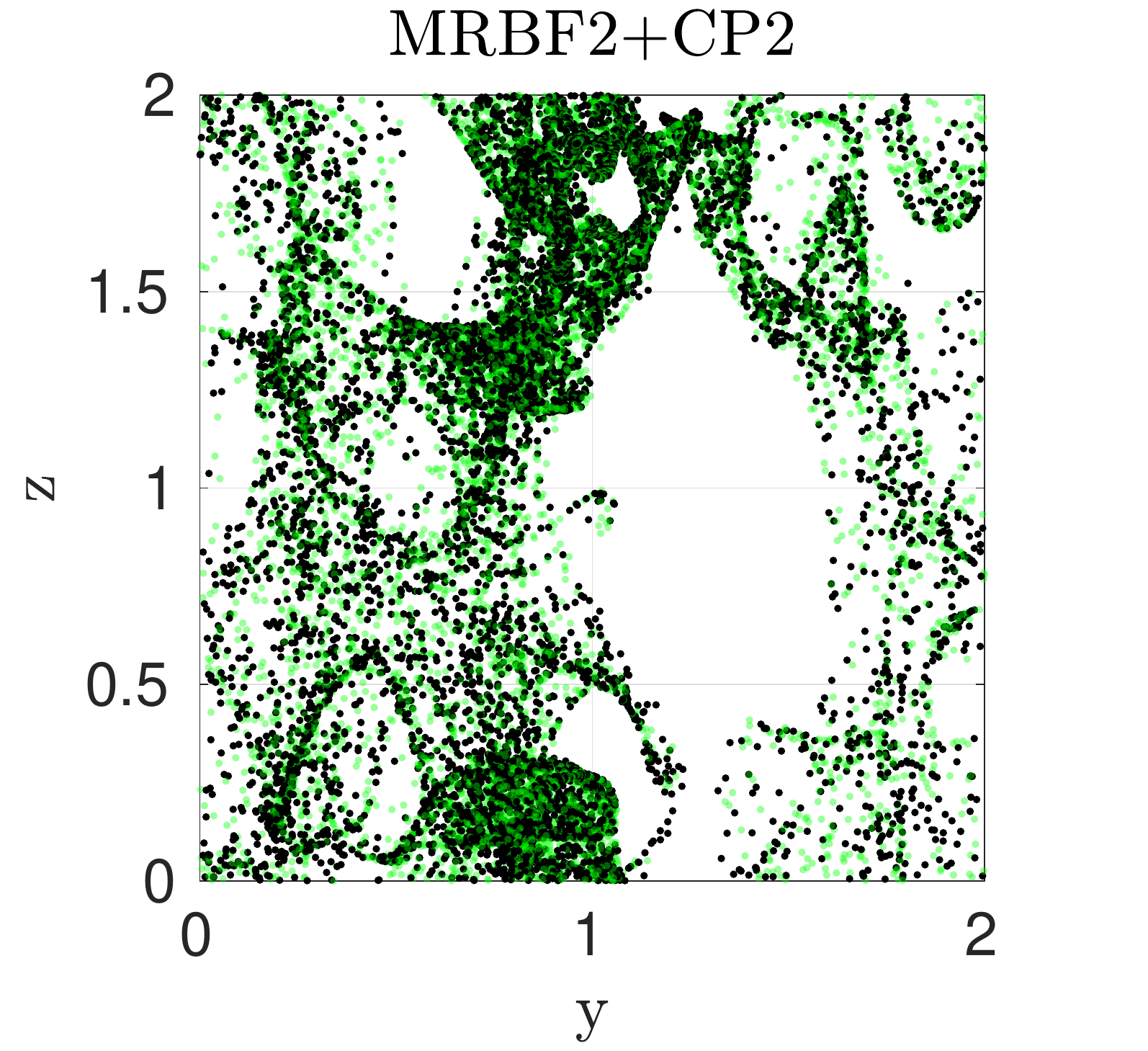}
		\caption{}\label{c12}
	\end{subfigure}
	\caption{
		Figures (a) through (f) show the spatial distribution of the particles in the $x-y$ plane for the $St = 1/10$, $h = 1/100$, $T = 6$, $\lambda = 10$ simulation from table \ref{table:comparison}. Figures (g) through (l) show the spatial distribution of the particles in the $z-y$ plane for the $St = 1$, $h = 1/40$, $T = 8$, $\lambda = 1$ simulation. The reference solution is plotted in green in all figures.
	}
	\label{comparison_figs}
\end{figure}

Figures \ref{c1} to \ref{c6} show the final distribution of the $St = 1/10$, $\lambda = 10$ simulation looking down the $z$-axis. We see that the only methods that look similar to the green reference solution are the MRBF1+CP1, MRBF2+CP2 and TP3+AB2 solutions. That is, both of our geometric methods and the most costly conventional method. It is worth noting that the TP2+CP2 and the TP2+AB2 solutions look very similar, despite one being generated by the CP2 method. This is likely due to the fact that the CP2 method loses its centrifuge-preserving properties when the fluid field is not divergence-free as seen in table \ref{table:vol errors} as well as the interpolation method being the dominant source of error. Turning our attention towards the corresponding part of table \ref{table:comparison} (i.e., the top-right) we make a few remarks. The most striking one here is that the MRBF1+CP1 solution has a larger $\overline{\Delta \bx_n}$ than the TP3+AB2 solution, but its $E(P_n)$ and $W(P_n)$ are both far lower. This is in agreement with the fact that the TP3+AB2 method is better in reducing error in the conventional sense, (i.e., the average 2-norm of the particle position errors $\overline{\Delta \bx_n}$), but does a poorer job at reproducing the mechanisms that are responsible for the preferential distribution of particles (i.e., the centrifuge effect and the sum of the Lyapunov spectrum). We can make similar remarks for the  $St = 1/10$, $\lambda = 1$ experiment where, despite having higher $\overline{\Delta \bx_n}$, the MRBF1+CP1 method has a lower $E(P_n)$ and $W(P_n)$ than the TP3+AB2 method. These significant observations are prevalent in both $St<1$ experiments, which corresponds to a more stiff vector field with greater relative influence of fluid inertia. 

\begin{table}[h]
	\centering
	\begin{tabular}{|c|c|ccc|ccc|}
		\hline
		& & \multicolumn{3}{c|}{$\lambda = 1$} & \multicolumn{3}{c|}{$\lambda = 10$} \\
		\hline
		& $P_n$	  & $E(P_n)$ & $W(P_n)$ & $\overline{\Delta \bx_n}$ & $E(P_n)$ & $W(P_n)$ & $\overline{\Delta \bx_n}$ \\
		
		\hline		
		\multirow{2}{*}{
			$St = \frac{1}{10}$
		} 
		& MRBF1+CP1 & 0.1305 & 0.1050 & 0.3351 & 0.0423 & 0.0431 & 0.5453  \\ 
		& TP1+FE1 & 7.9359 & 0.6340 & 0.7866 & 3.2075 & 0.7961 & 1.2645 \\
		\multirow{2}{*}{
			$h = \frac{1}{100}$
		}
		
		& MRBF2+CP2 & 0.1249 & 0.0414 & 0.1242 & 0.0706 & 0.0456 & 0.3513  \\ 
		& TP2+CP2 & 2.7166 & 0.2998 & 0.3990 & 1.2931 & 0.3131 & 0.8785  \\
		\multirow{2}{*}{
			$T = 6$
		}
		& TP2+AB2 & 2.6957 & 0.2836 & 0.3805 & 1.2585 & 0.2852 & 0.8833 \\
		& TP3+AB2 & 2.9179 & 0.1339 & 0.2512 & 0.1326 & 0.0804 & 0.4506 \\

		\hline
		\multirow{2}{*}{
			$St = 1$
		} 
		& MRBF1+CP1 & 0.9463 & 0.5888 & 1.4858 & 0.0395 & 0.1000 & 1.3961 \\
		& TP1+FE1 & 4.3703 & 1.4721 & 2.1060 & 0.4507 & 0.5446 & 1.9263 \\
		\multirow{2}{*}{
			$h = \frac{1}{40}$
		} 
		& MRBF2+CP2 & 0.0507 & 0.0525 & 0.5055 & 0.0306 & 0.0753 & 0.7845  \\
		& TP2+CP2 & 1.6212 & 1.2123 & 1.8251 & 0.0534 & 0.1630 & 1.7856 \\
		\multirow{2}{*}{
			$T = 8$
		} 
		& TP2+AB2 & 1.1982 & 1.0943 & 1.7759 & 0.0532 & 0.1595 & 1.7693 \\
		& TP3+AB2 & 0.0589 & 0.0532 & 0.4793 & 0.0376 & 0.0865 & 1.0213 \\

		\hline
		\multirow{2}{*}{
			$St = 10$
		}  
		& MRBF1+CP1 & 1.2748 & 0.3379 & 0.5122 & 0.0727 & 0.1190 & 0.7322 \\
		& TP1+FE1 & 6.1882 & 0.7767 & 0.8750 & 1.9416 & 0.6882 & 1.5169 \\
		\multirow{2}{*}{
			$h = \frac{1}{10}$
		} 
		& MRBF2+CP2 & 0.0979 & 0.0348 & 0.0663 & 0.0262 & 0.0438 & 0.2688 \\
		& TP2+CP2 & 3.2931 & 0.4742 & 0.7083 & 0.0754 & 0.1176 & 0.8004 \\
		\multirow{2}{*}{
			$T = 16$
		} 
		& TP2+AB2 & 3.2219 & 0.4063 & 0.6304 & 0.0774 & 0.1194 & 0.7933 \\
		& TP3+AB2 & 0.2631 & 0.0575 & 0.0865 & 0.0397 & 0.0594 & 0.4864 \\
		\hline 
		
	\end{tabular}
	\caption{The relative entropy $E(P_n)$, first Wasserstein distance $W(P_n)$ and average error per particle $\overline{\Delta \bx_n}$ between the numerical distribution $P_n$ and the reference distribution. The numerical distributions are calculated by various combinations of integration and interpolation methods as shown in the second column. The first column contains the Stokes number $St$, time step $h$ and simulation time $T$ used in the six simulations. The first row contains the aspect ratio $\lambda$ of the particle shape.} 
	\label{table:comparison}
\end{table}

We finish with some general observations. First, the MRBF1+CP2 solutions outperform the TP1+FE1, TP2+AB2 and TP2+CP2 methods in all three measures. Second, the TP2+AB2 and TP2+CP2 methods perform about the same in each simulation, suggesting there is little advantage from using the CP2 method in conjunction with the TP2 solution. Finally, we note that the MRBF2+CP2 method is more accurate than the TP2+AB2 and the TP3+AB3 in all cases. The only exception is the $St = 1$, $\lambda = 1$ simulation where the MRBF2+CP2 method has worse $\overline{\Delta \bx_n}$ than the TP3+AB3 solution.

\section{Conclusions}\label{sec:conclusions}

A novel combination of geometric numerical methods for calculating accurate distributions of inertial particles in viscous flows is proposed. The algorithm consists of MRBFs to construct a divergence-free approximation of the background flow field and a geometric splitting method for the time integration. 

The splitting method is shown to preserve the sum of the Lyapunov spectrum and hence the contractivity of phase space volume. By expanding the exact solution we derive an expression for how a physical volume of particles change over a small time step $h$, with which we recover the centrifuge effect at $O(h^4)$. We show that when a divergence-free interpolation method is used, one can implement a so-called centrifuge-preserving splitting method that preserves not only the qualitative but also the quantitative behavior of this centrifuge effect. Moreover, it is shown that errors to the divergence of the fluid field can overshadow this effect when a conventional polynomial interpolation method is used, for example.  

It is shown through numerical experiments that MRBF interpolation yields particle distributions that are more similar to the exact solution than standard TP interpolation. In many examples, this is observed even when the MRBF solution has higher error per particle. This is, in part, explained by the fact that: (1) MRBF interpolation is divergence-free meaning that the numerical time integration methods mimic the qualitative centrifuge effect; and (2) MRBF interpolation produces a vector field that solves the Stokes equations, meaning that the background flow field more physically resembles that of the exact solution (e.g., the flow field is related to the gradient of a scalar pressure function). 

Furthermore, we see that the proposed centrifuge-preserving methods are superior to the standard methods in terms of error per particle and how closely the particle distribution resembles the exact distribution. This is true, remarkably so, even when comparing the order-one CP1 method to the order-two RK2 and AB2 methods. In particular, for experiments with low particle inertia, the MRBF1+CP1 method produces more accurate distributions of particles than the expensive TP3+AB2 solutions, despite having slightly worse error and being an order one integration method that uses far less data points for the interpolation step. 

These observations strongly suggest that preserving certain physical features of ODEs under study in the numerical solution is of importance when simulating inertial particles in discrete flow fields. Of particular interest for future studies would be to implement the proposed methods in a physically realistic flow fields generated by a direct numerical simulation of homogeneous isotropic turbulence or turbulent channel flow, for example. 

\section{Acknowledgments}
This work has received funding from the European Unions Horizon 2020 research and innovation programme under the Marie Sklodowska-Curie grant agreement (No. 691070). B. K. Tapley, E. Celledoni and B. Owren would like to thank the Isaac Newton Institute for Mathematical Sciences, Cambridge, for support and hospitality during the programme {\it Geometry, compatibility and structure preservation in computational differential equations} (2019) where part of the work on this paper was undertaken.

\appendix
\section{Non-spherical particle model}\label{model}
Here, we give details of the specific rigid spheroid model that is used in the numerical experiments. The surface of a spheroid is defined by the equation
\begin{equation}\label{eqn:spheroid}
\frac{x^2}{a^2}+\frac{y^2}{a^2}+\frac{z^2}{c^2} = 1,
\end{equation}
where $a$ and $c$ are the distinct semi-axis lengths. The particle shape is characterised by the dimensionless aspect ratio $\lambda = c/a>0$, which distinguishes between spherical ($\lambda = 1$), prolate ($\lambda > 1$) and oblate ($\lambda < 1$) particles (the latter two shapes are also called as rods and disks).

An inertial particle immersed in a fluid will experience forces on its surface that have magnitude governed by many parameters such as the particles density $\rho_p$, length $a$, fluid density $\rho_f$, kinematic viscosity $\nu$ and fluid time scale $\tau_f$. The particle Stokes number is formally defined as the ratio of the particle and fluid time scales $St=\tau_p / \tau_f$. For a spherical particle the Stokes number is
\begin{equation}
St_0 = \frac{2 D a^2}{9 \nu \tau_f},
\end{equation}
where $D = \rho_p/\rho_f$ is the particle-fluid density ratio. Note that this definition only depends on the particle size and inertia. For spheroidal particles, the following shape dependent Stokes numbers are used, which are derived by Shapiro and Goldenberg \cite{shapiro1993deposition} and Zhao, et al. \cite{zhao2015rotation}

\begin{equation}
St = \left\{
\begin{array}{cc}
St_0 \, \lambda \, \log(\lambda+\sqrt{\lambda^2-1})/\sqrt{\lambda^2-1} & \mathrm{for} \quad \lambda > 1\\
St_0 \, (\pi-k_0)/(2\sqrt{1-\lambda^2})  & \mathrm{for} \quad \lambda < 1 \\
\end{array}\right.
\end{equation}
where $k_0 = \log((\lambda-\sqrt{\lambda^2-1})/(\lambda+\sqrt{\lambda^2-1}))$. Note that $St\rightarrow St_0$ as $\lambda\rightarrow 1$ from above or below. All the following equations are implemented in their non-dimensional form and all parameters have dimension equal to $1$.

The particle experiences a hydrodynamic drag force due to Brenner \cite{brenner1963stokes},
\begin{equation}
\mathbf{F} = QK_bQ\trans(\mathbf{u}-\mathbf{v}),
\end{equation}
where $\bu=\bu(\bx,t)$ is the fluid velocity evaluated at the particle center of mass $\bx$ and $\bv=\bp/m$ is the particle velocity. The body frame resistance tensor $K_b$ was calculated by Oberbeck \cite{oberbeck1876uber}, is diagonal, positive definite and given by

\begin{equation}
K_b=16\pi\lambda~\mathrm{diag}\left( \frac{1}{\chi_0+\alpha_0},\frac{1}{\chi_0+\beta_0} ,\frac{1}{\chi_0+\lambda^2\gamma_0}\right)
\end{equation}

where the constants $\chi_0$, $\alpha_0$, $\beta_0$ and $\gamma_0$ were calculated for ellipsoidal particles by Siewert et al. \cite{siewert2014orientation} and are presented in table \ref{table:constants}

\begin{table}[h]
	\begin{center}
		\renewcommand{\arraystretch}{1.8}
		\begin{tabular}{c|ccc}
			& $\lambda<1$ & $\lambda=1$ & $\lambda>1$ \\
			\hline
			
			$\chi_0$ 		& $\frac{\lambda^2(\pi-\kappa_0)}{\sqrt{1-\lambda^2}}$ & $2$ &  $\frac{-\kappa_0\lambda}{\sqrt{\lambda^2-1}}$\\
			
			$\alpha_0 = \beta_0$ 	& $\frac{-\lambda \left(\kappa_0-\pi+2\lambda\sqrt{1-\lambda^2}\right)}{2(1-\lambda^2)^{3/2}}$ & $\frac{2}{3}$ & $\frac{\lambda^2}{\lambda^2-1}+\frac{\lambda\kappa_0}{2(\lambda^2-1)^{3/2}}$\\
			
			$\gamma_0$	& $\frac{\left(\lambda(\kappa_0-\pi)+2\sqrt{1-\lambda^2}\right)}{(1-\lambda^2)^{3/2}}$ & $\frac{2}{3}$ & $\frac{-2}{\lambda^2-1}-\frac{\lambda\kappa_0}{(\lambda^2-1)^{3/2}}$\\
			
			$\kappa_0$ 		& $2\arctan\left(\frac{\lambda}{\sqrt{1-\lambda^2}}\right)$ & $1$ & $\ln\left(\frac{\lambda-\sqrt{\lambda^2-1}}{\lambda+\sqrt{\lambda^2-1}}\right)$ \\
		\end{tabular} 
	\end{center}
	\caption{The expressions for the constants $\chi_0$, $\alpha_0$, $\beta_0$ and $\gamma_0$ for $\lambda<1$, $\lambda=1$ and $\lambda>1$. }
	\label{table:constants}
\end{table}

The torque vector $\mathbf{T}$ depends on the particle shape and the local fluid velocity derivatives, and is given in non-dimensional form by \cite{jeffery1922motion}
\begin{align}
T_x = & \frac{16\pi \lambda}{3(\beta_0+\lambda^2\gamma_0)}\left[(1-\lambda^2)S_{yz}+(1+\lambda^2)(\Omega_{x}-\omega_y)\right],\label{eq:JT1} \\
T_y = & \frac{16\pi \lambda}{3(\alpha_0+\lambda^2\gamma_0)}\left[(\lambda^2-1)S_{zx}+(1+\lambda^2)(\Omega_{y}-\omega_z)\right],\label{eq:JT2} \\
T_z = & \frac{32\pi \lambda}{3(\alpha_0+\beta_0)}(\Omega_{z}-\omega_z).\label{eq:JT3}
\end{align}

\bibliography{bibliography}
\end{document}